\newcommand{\cnotenum}{\texorpdfstring{CNOT}{CNOT}}

\documentclass[times,twocolumn,final]{elsarticle}

\usepackage{medima}
\usepackage{framed,multirow}
\usepackage{stackengine}

\usepackage{epsfig}
\usepackage{amsmath}
\usepackage{lipsum}
\usepackage{enumitem}
\usepackage[toc, page]{appendix}

\usepackage{amssymb}
\usepackage{latexsym}
\usepackage{array, makecell}
\usepackage{bm}
\usepackage{xspace}
\usepackage{nicematrix}
\usepackage{subfigure}
\usepackage[T1]{fontenc}
\usepackage{graphicx}
\usepackage{multirow}
\usepackage{math}
\usepackage{float}
\usepackage{tikz}
\usepackage{booktabs}
\usepackage{fontawesome5}
\usepackage{url}
\usepackage{xcolor}
\usepackage[colorlinks = true,
            linkcolor = tealblue,
            urlcolor  = tealblue,
            citecolor = tealblue,
            anchorcolor = tealblue]{hyperref}

\definecolor{newcolor1}{rgb}{.0, .502, .675}

\makeatletter
\AtBeginDocument{\def\@citecolor{newcolor1}}
\AtBeginDocument{\def\@linkcolor{newcolor1}}
\AtBeginDocument{\def\@anchorcolor{newcolor1}}
\AtBeginDocument{\def\@filecolor{newcolor1}}
\AtBeginDocument{\def\@urlcolor{newcolor1}}
\AtBeginDocument{\def\@menucolor{newcolor1}}
\AtBeginDocument{\def\@pagecolor{newcolor1}}
\makeatother

\newcolumntype{?}{!{\vrule width 1pt}}
\usepackage{color, colortbl}
\definecolor{Gray}{gray}{0.9}
\definecolor{newcolor}{rgb}{.8,.349,.1}

\usepackage{lipsum}
\definecolor{candyAppleRed}{RGB}{255 8 0}
\definecolor{Gray}{gray}{0.9}
\definecolor{bailiang}{RGB}{148,170,178}
\definecolor{Bailiang}{RGB}{148,170,178}
\definecolor{next-gen-nn}{RGB}{230,210,130}
\definecolor{honkamj}{RGB}{123,86,160}
\definecolor{LoRA-FT}{RGB}{187,100,75}
\definecolor{MadeForLife}{RGB}{190,160,220}
\definecolor{lukasf}{RGB}{233,100,160}
\definecolor{LYU1}{RGB}{130,190,80}
\definecolor{TimH}{RGB}{240,49,49}
\definecolor{VROC}{RGB}{80,217,120}
\definecolor{DutchMasters}{RGB}{220,80,230}
\definecolor{zhuoyuanw210}{RGB}{255,120,0}

\definecolor{ANTsSyN}{RGB}{220,220,220}
\definecolor{deedsBCV}{RGB}{190,190,190}
\definecolor{DeedsBCV}{RGB}{190,190,190}
\definecolor{FireANTsGreedy}{RGB}{160,160,160}
\definecolor{FireANTsSyN}{RGB}{130,130,130}
\definecolor{SynthMorph}{RGB}{100,100,100}
\definecolor{TransMorph}{RGB}{0,0,255}
\definecolor{uniGradICON}{RGB}{0,75,255}
\definecolor{uniGradICONiso}{RGB}{0,150,255}
\definecolor{VFA}{RGB}{0,220,255}
\definecolor{VoxelMorph}{RGB}{140,255,255}
\definecolor{ZeroDisplacement}{RGB}{40, 40, 40}
\definecolor{FireANTsSyN$^\dagger$}{RGB}{130,130,130}
\definecolor{ANTsSyN$^\dagger$}{RGB}{220,220,220}

\definecolor{cream}{RGB}{255 253 208}

\newcommand{\gold}{\includegraphics[height = 2em]{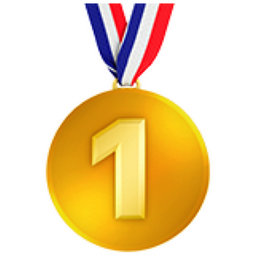}}
\newcommand{\silver}{\includegraphics[height = 2em]{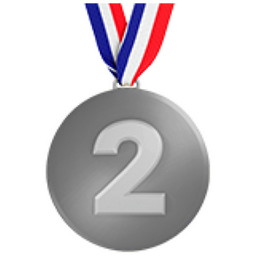}}
\newcommand{\bronze}{\includegraphics[height = 2em]{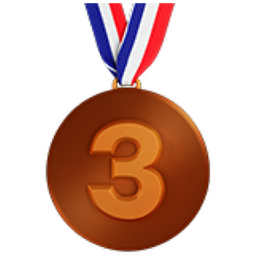}}
\newif\ifcb
\cbtrue

\definecolor{R1}{RGB}{254,0,0}
\definecolor{R2}{RGB}{242,0,8}
\definecolor{R3}{RGB}{232,0,22}
\definecolor{R4}{RGB}{219,0,35}
\definecolor{R5}{RGB}{206,0,48}
\definecolor{R6}{RGB}{195,0,58}
\definecolor{R7}{RGB}{184,0,72}
\definecolor{R8}{RGB}{169,0,83}
\definecolor{R9}{RGB}{156,0,95}
\definecolor{R10}{RGB}{150,0,104}
\definecolor{R11}{RGB}{135,0,120}
\definecolor{R12}{RGB}{124,0,130}
\definecolor{R13}{RGB}{106,0,148}
\definecolor{R14}{RGB}{99,0,160}
\definecolor{R15}{RGB}{86,0,167}
\definecolor{R16}{RGB}{71,0,182}
\definecolor{R17}{RGB}{64,0,189}
\definecolor{R18}{RGB}{52,0,204}
\definecolor{R19}{RGB}{37,0,217}
\definecolor{R20}{RGB}{26,0,229}
\definecolor{R21}{RGB}{9,0,244}
\definecolor{R22}{RGB}{0,0,252}

\newcommand{\mycomment}[1]{}
\newcommand{\pBox}[2]{\raisebox{1.5pt}{\fcolorbox{#1}{#2}{\rule{0pt}{2pt}\rule{2pt}{0pt}}}}
\newcommand{\mBox}[2]{\textbf{#2}~\raisebox{2.8pt}{\fcolorbox{#1}{#2}{\rule{0pt}{2pt}\rule{2pt}{0pt}}}}
\newcommand{\tbBox}[1]{\texttt{#1}~\raisebox{2.8pt}{\fcolorbox{black}{#1}{\rule{0pt}{2pt}\rule{2pt}{0pt}}}}
\newcommand{\twBox}[1]{\texttt{#1}~\raisebox{2.8pt}{\fcolorbox{white}{#1}{\rule{0pt}{2pt}\rule{2pt}{0pt}}}}

\colorlet{color max}[rgb]{red}
\colorlet{color min}[rgb]{blue}
\def\min{1}
\def\max{22}

\newcommand{\rCirc}[1]{%
\pgfmathtruncatemacro\lambda{(#1 - \min)/(\max - \min)*100}%
\colorlet{lambda}[rgb]{color min!\lambda!color max}
\raisebox{0.1ex}{%
    \tikz[baseline=(char.base)]{
      \node[
        draw=lambda,
        fill=lambda,
        circle,
        minimum size=1.2em,  
        inner sep=0pt,       
        font=\scriptsize,          
        align=center         
      ] (char) {\textcolor{cream}{#1}};
    }%
}%
}

\newcommand{\desc}[4]{%
\vspace{2em plus 1em minus 1em}
{\noindent\nopagebreak\pBox{#1}{#2}~\textbf{#2}\nopagebreak\\\nopagebreak\textit{#3}\nopagebreak\\\nopagebreak(#4)\nopagebreak}%
}

\newcommand{\doubleDesc}[6]{%
\vspace*{\baselineskip}
\vspace{\baselineskip}
{\noindent\pBox{#1}{#2}~\textbf{#2}~~\textbf{/}~~\pBox{#3}{#4}~\textbf{#4}\nopagebreak\\\nopagebreak\textit{#5}\nopagebreak\\\nopagebreak(#6)\nopagebreak}%
}

\newcommand\sbullet[1][.5]{\mathbin{\vcenter{\hbox{\scalebox{#1}{$\bullet$}}}}}

\newcommand{\hbullet}[1][0.5ex]{%
  \mathbin{%
    \raisebox{0.55ex}{
    \tikz[baseline=(a.base)]{
      \node[draw=black, fill=white, circle, inner sep=#1, line width=0.7pt] (a) {};
      }
    }%
  }%
}

\journal{}

\begin{document}

\verso{Chen \textit{et~al.}}

\begin{frontmatter}

\title{Beyond the LUMIR challenge: The pathway to foundational registration models}

\author[1]{Junyu \snm{Chen}\corref{cor1}}
\author[2]{Shuwen \snm{Wei}}
\author[3]{Joel \snm{Honkamaa}}
\author[3]{Pekka \snm{Marttinen}}
\author[4]{Hang \snm{Zhang}}
\author[5]{Min \snm{Liu}}
\author[6]{Yichao \snm{Zhou}}
\author[6]{Zuopeng \snm{Tan}}
\author[7]{Zhuoyuan \snm{Wang}}
\author[7]{Yi \snm{Wang}}
\author[8]{Hongchao \snm{Zhou}}
\author[8]{Shunbo \snm{Hu}}
\author[9]{Yi \snm{Zhang}}
\author[9]{Qian \snm{Tao}}
\author[10]{Lukas \snm{F\"{o}rner}}
\author[10]{Thomas \snm{Wendler}}
\author[11]{Bailiang \snm{Jian}}
\author[11]{Benedikt \snm{Wiestler}}
\author[12]{Tim \snm{Hable}}
\author[13]{Jin \snm{Kim}}
\author[13]{Dan \snm{Ruan}}
\author[14]{Frederic \snm{Madesta}}
\author[14]{Thilo \snm{Sentker}}
\author[12]{Wiebke \snm{Heyer}}
\author[15]{Lianrui \snm{Zuo}}
\author[16]{Yuwei \snm{Dai}}
\author[17]{Jing \snm{Wu}}
\author[2]{Jerry L. \snm{Prince}}
\author[1]{Harrison \snm{Bai}}
\author[1]{Yong \snm{Du}}
\author[15]{Yihao \snm{Liu}}
\author[18]{Alessa \snm{Hering}}
\author[19,20]{Reuben \snm{Dorent}}
\author[21]{Lasse \snm{Hansen}}
\author[12]{Mattias P. \snm{Heinrich}}
\author[2]{Aaron \snm{Carass}}

\address[1]{The Russell H. Morgan Department of Radiology and Radiological Science, Johns Hopkins Medical School, Baltimore, MD, USA}
\address[2]{Image Analysis and Communications Laboratory, Department of Electrical and Computer Engineering, Johns Hopkins University, Baltimore, MD, USA}
\address[3]{Department of Computer Science, Aalto University, Espoo, Uusimaa, Finland}
\address[4]{Cornell University, New York, NY, USA}
\address[5]{College of Electrical and Information Engineering, Hunan University, Changsha, Hunan, China}
\address[6]{Canon Medical Systems (China) Co. Ltd., Beijing, China}
\address[7]{School of Biomedical Engineering, Shenzhen University Medical School, Shenzhen, Guangdong, China}
\address[8]{School of Information Science and Engineering, Linyi University, Linyi, Shandong, China}
\address[9]{Department of Imaging Physics, Delft University of Technology, Delft, South Holland, Netherlands}
\address[10]{Department of Diagnostic and Interventional Radiology and Neuroradiology, University Hospital Augsburg and Institute of Digital Medicine, Augsburg, Bavaria, Germany}
\address[11]{Technical University of Munich and Klinikum Rechts der Isar, Munich, Bavaria, Germany}
\address[12]{Institute of Medical Informatics, University of L\"{u}beck, L\"{u}beck, Schleswig-Holstein, Germany}
\address[13]{Department of Radiology and Center for Computer Vision and Imaging Biomarkers, University of California Los Angeles, Los Angeles, CA, USA}
\address[14]{Institute for Applied Medical Informatics and Institute of Computational Neuroscience, University Medical Center Hamburg, Hamburg, Germany}
\address[15]{Department of Electrical and Computer Engineering, Vanderbilt University, Nashville, TN, USA}
\address[16]{Radboud University Medical Center, Nijmegen, Gelderland, Netherlands}
\address[17]{Department of Radiology and Radiological Science, Johns Hopkins Medical Institutions, Baltimore, MD, USA}
\address[18]{Radboud University Medical Center, Nijmegen, Gelderland, Netherlands}
\address[19]{Inria, Paris, France}
\address[20]{Department of Neurosurgery, Brigham \& Women's Hospital and Harvard Medical School, Boston, MA, USA}
\address[21]{EchoScout GmbH, L\"{u}beck, Schleswig-Holstein, Germany}

\cortext[cor1]{Corresponding author. E-mail: \texttt{jchen245@jhmi.edu}}
\received{xxxx}
\finalform{xxxx}
\accepted{xxxx}
\availableonline{xxxx}
\communicated{xxxx}

\begin{abstract}
Medical image challenges have played a transformative role in advancing the field, catalyzing innovation and establishing new performance benchmarks. 
Image registration, a foundational task in neuroimaging, has similarly advanced through the Learn2Reg initiative.
Building on this, we introduce the \textbf{L}arge-scale \textbf{U}nsupervised Brain \textbf{M}RI \textbf{I}mage \textbf{R}egistration~(LUMIR) challenge, a next-generation benchmark for unsupervised brain MRI registration.
%
%
\textcolor{black}{Previous challenges relied upon anatomical label maps, however LUMIR provides 4,014 unlabeled T1-weighted MRIs for training, encouraging biologically plausible deformation modeling through self-supervision.}
Evaluation includes 590 in-domain test subjects and extensive zero-shot tasks across disease populations, imaging protocols, and species.
Deep learning methods consistently achieved state-of-the-art performance and produced anatomically plausible, diffeomorphic deformation fields. 
They outperformed several leading optimization-based methods and remained robust to most domain shifts.
These findings highlight the growing maturity of deep learning in neuroimaging registration and its potential to serve as a foundation model for general-purpose medical image registration.
\end{abstract}


\end{frontmatter}



\section{Introduction}
\label{s:introduction}
Medical image registration is a cornerstone of medical image analysis, enabling voxel correspondence between images to support a wide range of critical clinical applications.
%
\textcolor{black}{For brain magnetic resonance imaging~(MRI) in particular, accurate registration plays a pivotal role in longitudinal monitoring~\citep{xue2006ni, holland2011mia}, assessing treatment effects~\citep{vaarkamp2000ijrobp}, planning interventions~\citep{rosenman1998ijrobp}, and conducting population-level studies of brain structure~\citep{gerber2010mia} and function~\citep{toga2001ivc, bilgel2015ni}.}
Deformable brain image registration is widely recognized as a fundamental tool in neuroimaging for applications ranging from Alzheimer's disease research to functional mapping for surgical planning.
Although the brain is encased within the rigid skull, which limits large deformations and constrains anatomical variability both within and between subjects after affine pre-alignment, this apparent simplicity belies significant challenges.
Structures like the hippocampus, putamen, and amygdala, often lack strong intensity-based contrast, complicating voxel-based alignment.
In addition, neurodegenerative conditions introduce cortical thinning, atrophy, and other subtle deformations that demand highly sensitive and robust registration algorithms to capture clinically meaningful changes.

Medical image challenges have played a crucial role in advancing the field by providing rigorous evaluation frameworks and standardized benchmarks, fostering collaboration, driving innovation, and encouraging the development of robust and generalizable algorithms for various imaging tasks~\citep{murphy2011tmi, sirinukunwattana2017mia, maierhein2018nc, halabi2019rad, antonelli2022natc, hering2023tmi, hansen2025melba}. However, despite the fundamental role of image registration in many clinical and research applications, it has been relatively underrepresented in medical image challenges, particularly those focused on brain MRI.
While existing challenges have extensively addressed segmentation tasks involving tumors~\citep{menze2015tmi}, lesions~\citep{menze2015tmi, maier2017mia, carass2017ni, commowick2018nsr, kuijf2019tmi}, brain tissue~\citep{mendrik2015cin, kuijf2019tmi}, and the parcellation of fine brain structures~\citep{carass2018ni}, fewer efforts have been dedicated to systematically evaluating registration algorithms for brain MRI.
Notable exceptions, such as Hippocampus-MR and OASIS from previous editions of the Learn2Reg challenge~\citep{hering2023tmi}, have offered valuable contributions but were limited in scope.
For example, they focus on small regions of interest or rely on label maps that can inadvertently bias models towards over-optimizing label overlap ~\citep{rohlfing2011image} at the expense of generating smooth and physically plausible deformations~\citep{rohlfing2011image, chen2024survey}.
Moreover, their evaluations were restricted to in-domain datasets, without assessing generalizability.
To address these limitations, we introduce the \textbf{L}arge-scale \textbf{U}nsupervised Brain \textbf{M}RI \textbf{I}mage \textbf{R}egistration (LUMIR) challenge.
LUMIR leverages data from publicly available brain MRI datasets, which have been carefully prepared and preprocessed to support unsupervised learning of inter-subject registration.
It emphasizes both anatomical accuracy and deformation smoothness without reliance on label map supervision.
By providing gold-standard evaluation metrics alongside this curated dataset, the LUMIR challenge establishes a critical platform for benchmarking and advancing the next generation of brain image registration models, with the ultimate goal of enhancing clinical workflows and supporting high-impact neuroimaging research.

\begin{figure*}[!tb]
\begin{center}
\includegraphics[width=0.8\textwidth]{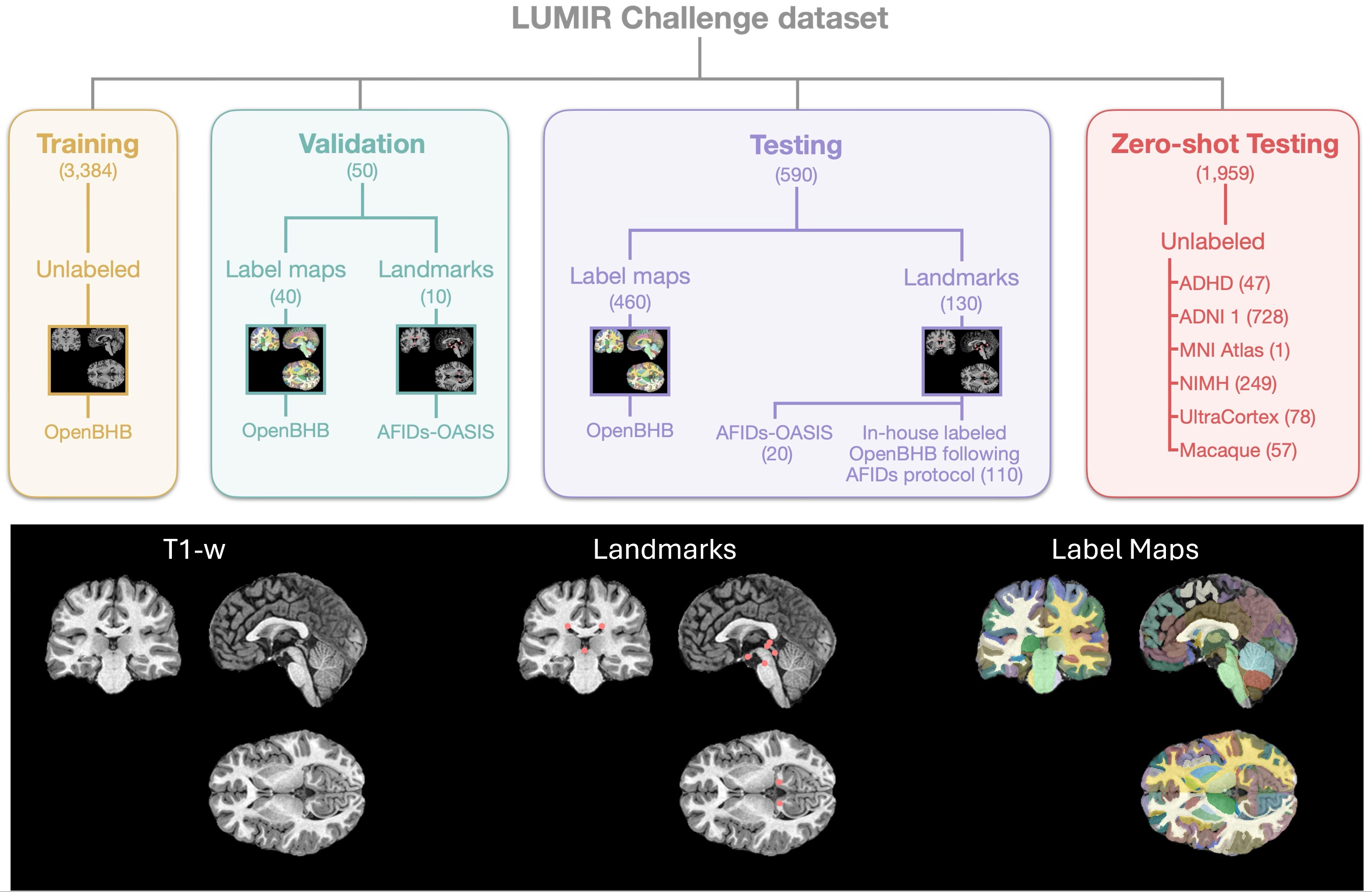}
\end{center}
   \caption{The top panel provides a breakdown of the training, validation, and testing data used in the LUMIR Challenge; with each source listed and the quantity of data in parentheses. The bottom panel shows from left to right, a tri-planar view of a typical T1-weighted MRI after our preprocessing, the manually identified landmarks overlaid on the same T1-weighted MRI, and finally the corresponding SLANT labels of the T1-weighted MRI. See Sec.~\ref{sec:datasets} for complete details.}
\label{fig:lumir_samples}
\end{figure*}

The challenge features 4,014 T1-weighted~(T1-w) brain MRI images, with 3,384 training images that generate more than 5.7 million unique image pairs.
This unprecedented scale promotes the development of foundational registration models while enabling fair benchmarking of traditional optimization-based and deep learning-based approaches.
Furthermore, to complement the label maps used to assess accuracy, the LUMIR challenge also provides a subset of 130 test images, each manually annotated with 32 anatomical landmarks.
These landmarks enable gold-standard evaluations of registration accuracy and allow analyses of correlations between landmark-based and label map-based measures.
Detailed descriptions of the LUMIR challenge dataset are provided in Fig.~\ref{fig:lumir_samples}.
Through extensive evaluation, the results reveal that \textbf{deep learning-based methods achieve superior performance and efficiency}, with the top-performing approaches employing a single-step, optimization-free strategy while producing smooth and even diffeomorphic deformations, \textcolor{black}{suggesting that effective model design can be sufficient for strong performance, though instance optimization may offer further gains in some settings}.
These findings underscore the advances and maturity of deep learning methods for brain image registration.

Beyond the in-domain test set, the LUMIR challenge provides the opportunity to explore a broader and more ambitious question: \textbf{Can deep learning models trained on this large-scale dataset effectively register brain images across various modalities, acquisition protocols, and even species?}
This exploration is grounded in the unique nature of image registration, which focuses on identifying spatial correspondences between image pairs.
Unlike tasks that rely on the semantic interpretation of individual images, registration models prioritize inter-image relationships.
This suggests that such models may be inherently robust to out-of-distribution samples, as previous studies have shown that those trained on random shapes can still perform well in brain MRI registration~\citep{hoffmann2021synthmorph, chen2025pretraining}.
To further investigate this, we evaluated challenge submissions and baseline registration models on extensive out-of-distribution brain MRI datasets.
The results are striking: \textbf{deep learning methods trained solely on T1-w brain MRI images of healthy subjects showed exceptional generalization capability.}
Moreover, the top-performing deep learning methods outperformed traditional pairwise optimization methods, despite the latter being theoretically immune to domain shift.
Our results directly address concerns that deep learning-based registration methods may struggle to generalize across datasets~\citep{jena2024deep}.
In contrast, \textbf{our findings demonstrate that with sufficient training data and appropriate preprocessing pipelines, learning-based registration models evaluated in this study exhibit excellent out-of-domain generalization to intra-contrast registration on unseen MRI data}, \textcolor{black}{including data from MR sequences not encountered during training,} even surpassing models designed to be robust across image contrasts, such as SynthMorph~\citep{hoffmann2021synthmorph}.
This robustness suggests that specialized \textcolor{black}{contrast}-agnostic training strategies may become less critical when large-scale datasets and consistent preprocessing pipelines are available.

The LUMIR challenge underscores the transformative potential of deep learning and large scale data in creating robust and generalizable solutions for medical image registration.
By addressing critical limitations in training data, evaluation methods, and application scope, this challenge establishes an unparalleled platform for benchmarking and advancing brain image registration models.
This work not only summarizes the challenge outcomes but also provides empirical insights to guide the development of robust and adaptable registration models, unlocking new possibilities for clinical and research advancements.
%
%
To help ensure that the LUMIR challenge has a meaningful impact, the submission website and associated evaluation of results \textcolor{black}{will remain open} for the foreseeable future; which is in keeping with the tradition of previous Learn2Reg Challenges.
This fosters reproducibility across iterations of technological innovation within the medical image registration community.

For the rest of the paper, in Sec.~\ref{s:materials}, we introduce the datasets and evaluation metrics employed in our experiments.
Sec.~\ref{s:setup} describes the design and organization of the LUMIR challenge. 
In Sec.~\ref{s:methods}, we summarize both the participating methods and the baseline methods included in the challenge.
Sec.~\ref{s:results} presents detailed quantitative and qualitative results.
We then discuss key findings and limitations in Sec.~\ref{s:discussion}, before concluding the paper in Sec.~\ref{s:conclusion}.

\section{Materials}
\label{s:materials}

\begin{figure*}[!tb]
\begin{center}
\includegraphics[width = 0.9\textwidth]{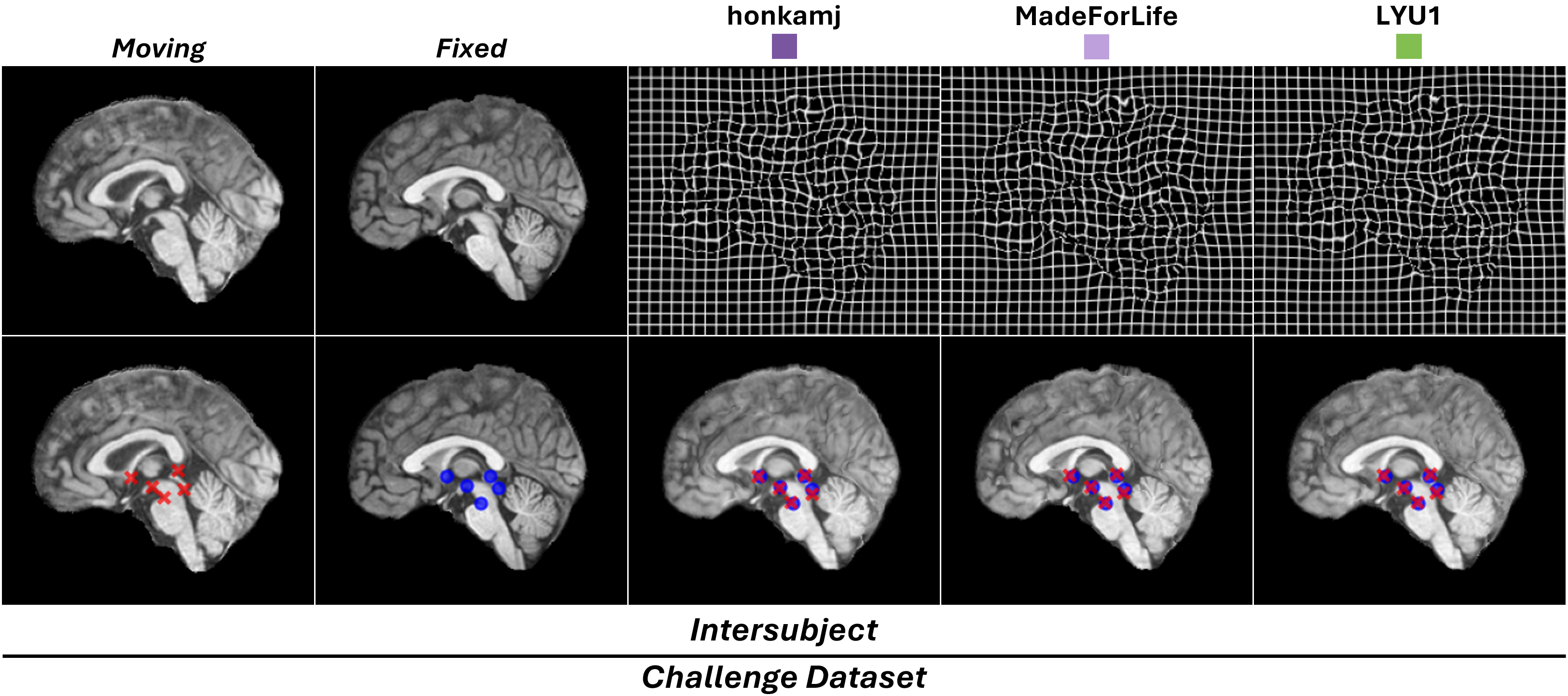}
\end{center}
   \caption{Representative registration results from the challenge evaluation task. Shown are the moving and fixed images, the deformation fields and deformed moving images for the top three methods (selected from Fig.~\ref{fig:rank_all_heatmaps} and shown left to right in rank order), and the corresponding anatomical landmarks.}
%
\label{fig:result_challenge_samples}
\end{figure*}

\begin{figure*}[!tp]
\begin{center}
\begin{tabular}{c}
\hspace*{2mm}\includegraphics[width = 0.6\textwidth]{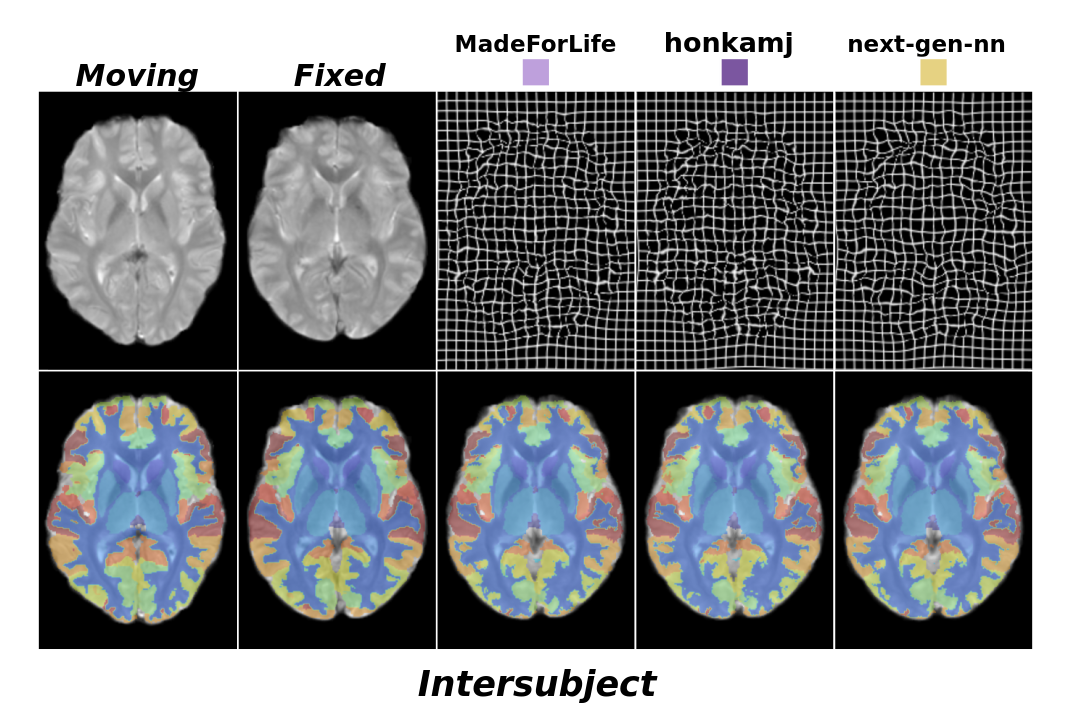}\\
\includegraphics[width = 0.6\textwidth]{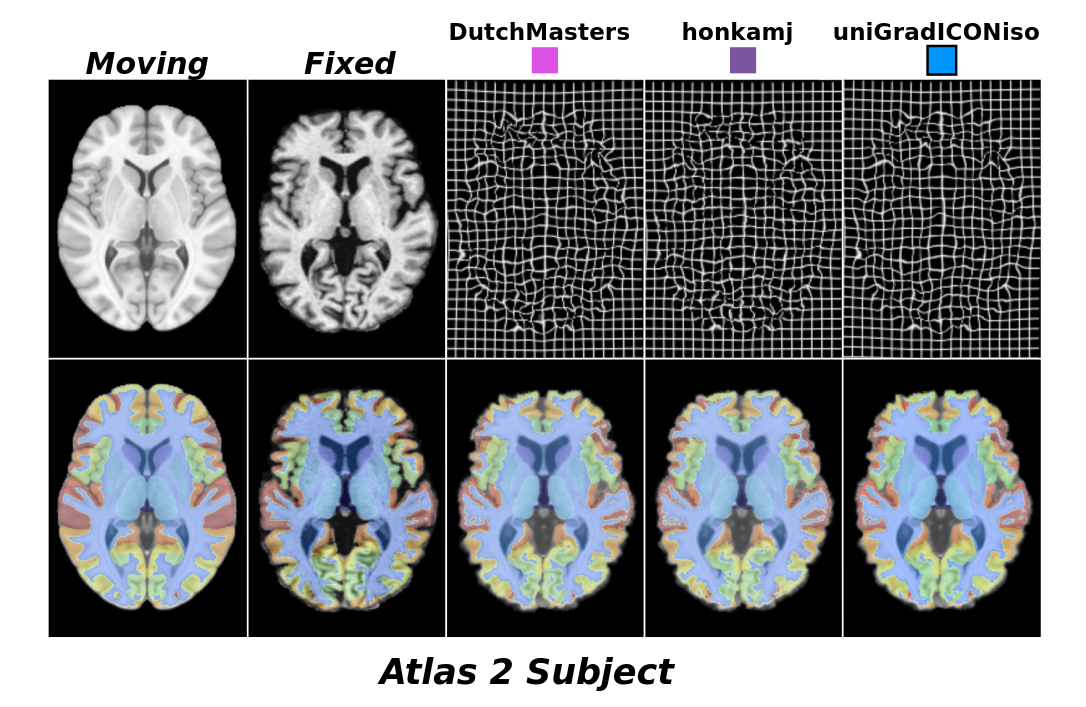}\\
\includegraphics[width = 0.6\textwidth]{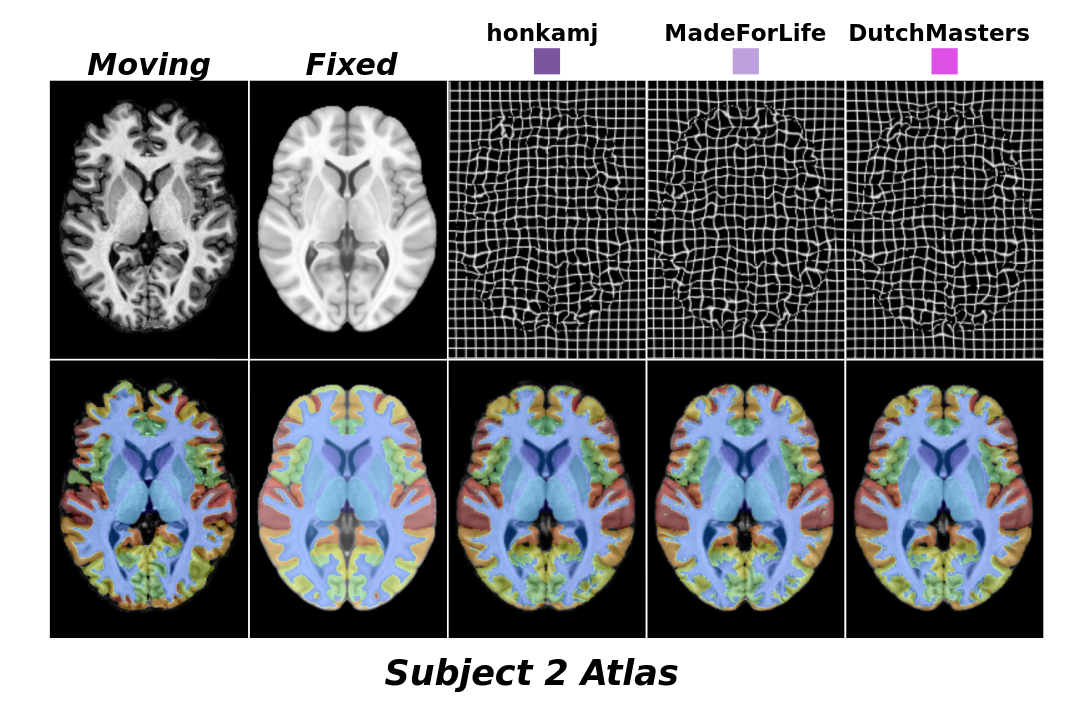}\\
\end{tabular}
\end{center}
   \caption{Representative registration results from the zero-shot evaluation tasks. From top to bottom the three panels show the Intersubject, Atlas to Subject~(Atlas2Subject) and Subject to Atlas~(Subject2Atlas) results. Each panel shows an example moving and fixed image, with the deformation fields from the top three methods (selected from Fig.~\ref{fig:rank_all_heatmaps} and shown left to right in rank order). Also shown are the corresponding anatomical label maps for the moving and fixed images, as well as the propagated anatomical label maps for the top three methods.}
\label{fig:result_zeroshot_samples}
\end{figure*}

\subsection{Datasets}
\label{sec:datasets}
\subsubsection{Challenge Dataset}
\label{sec:challenge_set}
The LUMIR challenge training, validation, and test data is built from three primary public sources, these include ten different public datasets~\citep{di2014autism, di2017enhancing, braindevelopmentDatasetx2013, zuo2014open, holmes2015sd, Nastase2020Narratives, babayan2019mind, sunavsky2020neuroimaging, li2019reading, follmer2018predicts, orfanos2017brainomics} that are part of the OpenBHB dataset~\citep{dufumier2022ni} and the OASIS dataset~\citep{marcus2007jcn}, in combination with public landmarks~\citep{taha2023sd} and our own privately held landmarks.
%
%
The OpenBHB dataset contains T1-weighted brain MRIs that have been skull-stripped and affinely aligned to the isotropic 1mm MNI template using FreeSurfer~\citep{fischl2012freesurfer}. 
Subsequently, we applied kernel-based intensity normalization~\citep{reinhold2019evaluating} to normalize the mean white matter value to 110 and used the automated segmentation tool SLANT~\citep{huo20193d} to segment the brains into 133 cortical and subcortical structures.
The AFIDs-OASIS dataset~\citep{taha2023magnetic} \textcolor{black}{is a subset of the OASIS-1 dataset~\citep{marcus2010open}, comprising 30 T1-weighted 3T MR images from cognitively intact subjects, including 17 female and 13 male participants, with a mean age of
58.0 years and a standard deviation of 17.9 years.
Each image is annotated with 32 manually identified
anatomical landmarks and underwent the same preprocessing steps.}
The final LUMIR dataset contains 4,014 images, with 3,984 from OpenBHB and 30 from AFIDs-OASIS. \textcolor{black}{The 3,984 OpenBHB images are acquired from 64 different sites and 3,984 different healthy control subjects.
From 3,984 images, 3,637 are acquired on 3T MRI and 347 are acquired on 1.5T MRI. From 3,984 subjects, 2,087 are males and 1,897 are females.
The average age of these subjects is 24.9~years, with the standard deviation of age being 14.3.}
From these, we randomly allocated 3,384 OpenBHB images for training; 40 images (30 from OpenBHB, 10 from AFIDs-OASIS) for validation via the grand-challenge.org public leaderboard; and 590 images (570 from OpenBHB, 20 from AFIDs-OASIS) for testing.
Additionally, following AFIDs protocols~\citep{taha2023magnetic}, a neurologist (Y. Dai) and two radiologists (J. Wu, H. Bai) manually annotated landmarks on 110 OpenBHB images in the test set for landmark-based evaluation.
Specifically, the final breakdown includes:
\begin{itemize}
    \item \textbf{\textit{Training}}: 3,384 unlabeled brain MRI images.
    \item \textbf{\textit{Validation}}: 50 brain MRI images, comprising 40 with segmentation label maps and 10 with landmarks.
    \item \textbf{\textit{Testing}}: Privately held 590 brain MRI images, including 460 images with segmentation labels and 130 images with landmark annotations.
\end{itemize}
In line with our focus on unsupervised image registration, only imaging data were provided for training (i.e., no label maps or deformation fields), allowing participants to freely create inter-subject pairs or deformation fields as they see fit.
%
\textcolor{black}{This setup yields 5,724,036 unique unordered image pair combinations from the 3,384 training images, or up to 11,448,072 possible ordered moving-to-fixed training pairs, where participants were free to construct training pairs according to their own training strategy from the available 3,384 training images.}
Anatomical label maps and landmarks for the training and validation sets were kept private. However, a small subset (label maps from five images) was made available to participants for preliminary sanity checks before submitting their results to the Grand Challenge platform.

\subsubsection{\textcolor{black}{Zero-shot Datasets}}

\label{sec:zeroshot_set}
\begin{itemize}
    \item \textbf{\textit{ADHD}}~\citep{lytle2020neuroimaging, lytle2021neuroimaging}: This dataset includes MRI scans from 47 children diagnosed with attention deficit hyperactivity disorder (ADHD) in two previous studies. The cohort consists of 35~males (ages 8.59--11.85) and 12~females (ages 9.8--12.3). MRI scans were acquired using 3T and 1.5T systems with the MP-RAGE sequence.
    
    \item \textbf{\textit{ADNI1 1.5T\&3T}}: Subsets of the ADNI1 dataset were used, specifically the ``Complete 1Yr 1.5T'' and ``Complete 1Yr 3T'' cohorts, comprising 639 and 89 subjects, respectively, all diagnosed with Alzheimer’s disease. Scans were acquired using 1.5T and 3T MRI systems. The age range for the 1.5T cohort is 55--91 years, and for the 3T cohort, 55--89 years. ADNI1 data was obtained from the Alzheimer’s Disease Neuroimaging Initiative~(ADNI) database (adni.loni.usc.edu). ADNI was launched in 2003 as a public-private partnership, led by Principal Investigator Michael W. Weiner,~MD.
    
    \item \textbf{\textit{NIMH}}~\citep{nugent2022nimh}: This dataset includes MRI scans of healthy subjects aged 18--72 years. The scans were categorized by imaging modality, comprising 249 subjects with T1-weighted scans, 247 with T2-weighted scans, 241 with T2*-weighted scans, and 240 with FLAIR scans. Notably, T2*-weighted and FLAIR scans typically exhibit larger voxel sizes along the slice-select (z-axis) direction, leading to significantly lower resolution along this axis.
    
    \item \textbf{\textit{UltraCortex}}~\citep{mahler2024ultracortex}: This data set consists of 9.4T ultra-high-field T1 weighted MRI scans from 78 healthy adults (28~females, 50~males) aged 20--53 years. MRI acquisition was performed using a 9.4T whole-body scanner with the MP-RAGE and MP2RAGE sequences.
    
    \item \textbf{\textit{Atlas}}: The ICBM 2009c Nonlinear Symmetric T1-weighted template with $1\times1\times1$ mm isotropic resolution was used for atlas based registration. This template is part of the ICBM 152 Nonlinear atlases~(2009 version)~\citep{fonov2009unbiased}.
    
    \item \textbf{\textit{Macaque}}~\citep{milham2018open}: This dataset includes T1-weighted MRI scans from 57 macaques, collected from five different sources: \textbf{1)~Aix-Marseille Université}: Contains 4 scans from 3~male and 1~female \textit{Macaca mulatta}, aged 7--8 years, acquired using a 3T MRI system with the MPRAGE sequence. \textbf{2) Mount Sinai School of Medicine (Philips)}: Includes 9~scans from 8 males and 1~female \textit{Macaca mulatta} and \textit{Macaca fascicularis}, aged 3.4--8 years, acquired on a 3T MRI scanner with the MPRAGE sequence. \textbf{3) McGill University}: Comprises 3 scans from 3 female \textit{Macaca mulatta} and \textit{Macaca fascicularis}, aged 10--12 years, acquired using a 3T system with the MP2RAGE sequence. \textbf{4) Stem Cell and Brain Research Institute}: includes 22 scans from 9 male and 13 female \textit{Macaca mulatta} and \textit{Macaca fascicularis}, aged 3.6--14 years, acquired using 1.5T and 3T systems with the MPRAGE sequence. \textbf{5) University of California, Davis}: includes 19 scans from 19 female \textit{Macaca mulatta}, aged 18.5--22.5 years, acquired using a 3T system with the MPRAGE sequence.
    
\end{itemize}

For consistency with the challenge dataset, a similar preprocessing pipeline was applied to the zero-shot datasets.
This included skull-stripping, affine registration to a $1\times 1\times 1$ mm MNI \textcolor{black}{template~\citep{mazziotta2001phil}}, bias field correction using ANTs~\citep{avants2009ij}, \textcolor{black}{and intensity normalization based on kernel density estimation~\citep{reinhold2019evaluating}}.
As the macaque brain images were \textcolor{black}{initially resampled to} a resolution of $1\times 1\times 1$~mm but were significantly smaller than human brains, we scaled them up by a factor of 1.5.
To ensure proper affine alignment, a macaque brain was first affinely registered to the MNI template and qualitatively evaluated; all remaining macaque brains were subsequently aligned to this initial macaque image.
T1-weighted MRI scans from human participants were segmented into 133 cortical and subcortical anatomical structures using SLANT~\citep{huo20193d}.
The resulting label maps were propagated from T1-weighted scans to T2*-weighted and FLAIR scans.
For non-human primates (ie., macaques), brain segmentation was performed using nBEST~\citep{zhong2024nbest}, an automated method that delineates nine anatomical structures.
The resulting label maps from the segmentation algorithms were reviewed by the challenge organizers, who have extensive experience in neuroimaging research.
\textcolor{black}{There is no overlap between the Challenge Training Dataset and the Zero-shot Dataset.}

\subsection{Comparisons}
\label{ss:comparisons}
We compute a range of comparisons to evaluate registration performance that assess the accuracy, robustness, plausibility, and speed of the algorithms. 
We specifically use the following:
\begin{itemize}
    \item \textbf{\textit{DSC}}: The Dice similarity coefficient~(DSC)~\citep{dice1945e} measures the overlap of two sets of segmentation labels. The DSC provides a reference for the general accuracy of the methods, although it has limitations~\citep{carass2020sr} specifically related to registration evaluation~\citep{rohlfing2011image}.
    \item \textbf{\textit{DSC30}}: \textcolor{black}{To assess robustness, DSC30 considers the 30\textsuperscript{th} percentile of DSC scores on all anatomical structures and cases. It serves to evaluate the methods' performance on challenging scenarios, representing the lower tail of the distribution.}
    \item \textbf{\textit{HD95}}: The Hausdorff distance~(HD)~\citep{hausdorff1914book} indicates the maximum distance in a metric space between two sets, in this case segmentation labels of the fixed and warped moving images.
    For a robust score, it is customary to consider the 95$^{\mbox{\tiny{th}}}$~(HD95) percentile instead of the maximum distance.
    \item \textbf{\textit{TRE}}: The target registration error~(TRE) is defined as the \textcolor{black}{mean of the} Euclidean distance \textcolor{black}{across all} corresponding landmarks in the fixed and warped moving images.
    \item \textbf{\textit{TRE30}}: Similarly to the DSC30 score, the TRE30 reports the 30\textsuperscript{th} percentile of the largest distances from the landmark.
    \item \textbf{\textit{NDV}}: The non-diffeomorphic volume~(NDV)~\citep{liu2024ijcv} quantifies the size of the non-diffeomorphic space in 3D caused by a digital transformation. It avoids the incorrect estimation arising from the central differences based Jacobian determinant computation by computing the volume of folded space in 3D.
    \textcolor{black}{Since larger brains inherently contain more voxels and are therefore more likely to exhibit higher absolute NDV values independent of algorithmic performance, we report the \%NDV relative to the total brain volume.
    Ideally, it would have a value of zero, as a non-zero value indicates that there has been a fold in the transformation between the two images, which is not physically feasible.}
    \item \textbf{\textit{ACC}}: To provide a summary statistic of registration accuracy~(ACC), we pool the rankings of DSC, HD95, and TRE (when available).
    \textcolor{black}{ACC is intended as a summary measure across complementary measures of registration accuracy.}
    \item \textbf{\textit{Runtime}}: The average runtime of each method is computed from ten repeated measurements performed on CPU, including image I/O operations. This ensures fair comparisons with traditional optimization-based methods, which typically have CPU-only implementations. All experiments were conducted on a Linux workstation equipped with an Intel\textsuperscript{\textregistered} Xeon\textsuperscript{\textregistered} Silver 4410Y processor and 535~GB of RAM.
    We note that, in practical settings where GPUs are widely available, participant methods and baseline models with GPU acceleration can achieve substantially lower runtimes.
    \textcolor{black}{Accordingly, the reported CPU runtimes are intended to provide a relative point of reference rather than to reflect deployment performance, and were not factored into the final ranking.}
\end{itemize}

In line with previous editions of the Learn2Reg Challenge~\citep{hering2023tmi}, which followed the standard set by the Medical Decathlon~\citep{antonelli2022natc}, we rank the methods using the Wilcoxon signed-rank test.
Specifically for each of DSC, HD95, TRE, and NDV, all methods are compared against each other using the paired Wilcoxon signed rank test with a one-sided alternative hypothesis, a per comparison $\alpha = 0.05$, and no adjustment for multiple comparisons.
The ranking for DSC, HD95, TRE, and NDV, is then determined based on the number of comparisons a method ``\textit{won}'', with the number of wins for each method assigned to values between $0.1$ and $1$ with the possibility of multiple methods having the same rank.
Finally, a task rank score is obtained by taking the geometric mean of the individual rank scores, which determines the ACC ranking.
An almost identical procedure is used to rank the methods based on their DSC30 and TRE30, with the modification that an unpaired Wilcoxon rank-sum test (Mann–Whitney U test) is used.
The unpaired test is used, as the 30\textsuperscript{th} percentile of data represented in the DSC30 or TRE30 is different between methods.
The rankings for each of the methods across the challenge test set and zero-shot evaluations are reported in Fig.~\ref{fig:rank_all_heatmaps}.
The summary results for each of the evaluations for the challenge test set are presented in Table~\ref{t:testset}.

\section{LUMIR Challenge Setup}
\label{s:setup}
The LUMIR challenge consisted of three phases (mainly
organized on grand-challenge.org).

\begin{itemize}
\item \textbf{\textit{Phase 1 -- Validation Phase}}:
The challenge began on April 1\textsuperscript{st}, 2024, allowing participants to download the training and validation datasets and train their registration networks using the training images.
Participants then generated displacement fields for the validation cases, which were subsequently submitted and automatically evaluated via grand-challenge.org. 
Participants whose submissions ranked within the top 10 on the leaderboard were invited to provide method descriptions by June 24\textsuperscript{th}, 2024, and present posters at the Workshop on Biomedical Image Registration (WBIR), held during the 27\textsuperscript{th} International Conference on Medical Image Computing and Computer Assisted Intervention~(MICCAI 2024).
\item \textbf{\textit{Phase 2 -- Test phase}}:
After the validation phase, participants had until September 8\textsuperscript{th}, 2024 to submit a Docker container to the organizers.
This container included their registration models in executable form, designed to read input image pairs from a provided file listing all test set pairs, and subsequently generate the corresponding displacement fields.
The organizers evaluated these displacement fields using both label maps and anatomical landmarks.
The performance metrics and final rankings were then computed and presented at the MICCAI conference.
\item \textbf{\textit{Phase 3 -- Further analysis phase}}:
After the challenge concluded at MICCAI 2024, the organizers conducted further analyses of the submitted algorithms by evaluating their zero-shot performance on out-of-domain datasets, whose data distributions differed from those used during training.
The results and findings of this analysis are summarized in this paper. 
\end{itemize}


The LUMIR challenge attracted a total of 39 participants across 31 teams, each submitting at least one set of computed displacement fields during the validation phase.
Collectively, these participants contributed \textcolor{black}{305} valid entries to the challenge leaderboard, \textcolor{black}{see Table~\ref{tab:submission_summary} for a summary of the number of submissions from each team}.
In the subsequent test phase, 12 teams submitted their trained models for evaluation.
Notably, while one team consisted of industry professionals, the remaining teams were affiliated with academic institutions.

\section{Methods}
\label{s:methods}

\begin{table*}[!t]
\caption{Overview of the 21 registration methods evaluated in the LUMIR challenge. The image similarity measures, regularization terms used in the registration formulation, and model design choices for each method are marked with a $\sbullet[1.5]$. For diffeomorphic registration, methods that do not mathematically guarantee diffeomorphisms but adopt alternative formulations that approximate them are denoted by $\hbullet[1.6]$. The average runtime per image pair (measured on CPU) and the public availability of the source code are also provided. The runtime for \twBox{zhuoyuanw210} is omitted, as the implementation requires a GPU and cannot be fairly compared with CPU-only methods. \textcolor{black}{We note that the reported runtimes are intended as points of reference; with GPU acceleration, most participant methods and several baseline models can complete registration in roughly 30 seconds or less, depending on the use of instance-specific optimization.}}\label{tab:models}
\centering
\fontsize{8.5}{10}\selectfont
    \begin{tabular}{ r ? c | c | c | c ? c | c | c | c | c ? c | c | c | c | c | c | c | c | c ? c ? c}
 \hline
  & \multicolumn{4}{c?}{\textbf{Similarity}} & \multicolumn{5}{c?}{\textbf{Regularizer}} & \multicolumn{9}{c?}{\textbf{Model Design}} & \multicolumn{1}{c?}{ } & \multicolumn{1}{c}{ }\\[0.3em]
  & \rotatebox[origin=c]{90}{\textcolor{black}{NCC}} & \rotatebox[origin=c]{90}{\textcolor{black}{LNCC}} & \rotatebox[origin=c]{90}{MIND-SSC} & \rotatebox[origin=c]{90}{Dice} & \rotatebox[origin=c]{90}{Diffusion} & \rotatebox[origin=c]{90}{Inverse Consist.} & \rotatebox[origin=c]{90}{Jacobian Det.} & \rotatebox[origin=c]{90}{Kinetic Energy} & \rotatebox[origin=c]{90}{Gaussian Smooth.} & \rotatebox[origin=c]{90}{\textcolor{black}{Deep Learning}} & \rotatebox[origin=c]{90}{Dual-Stream Enc.} & \rotatebox[origin=c]{90}{Coarse-to-Fine} & \rotatebox[origin=c]{90}{Progressive Reg.} & \rotatebox[origin=c]{90}{Cost Volume} & \rotatebox[origin=c]{90}{Attention} & 
  \rotatebox[origin=c]{90}{IC by Construction} & \rotatebox[origin=c]{90}{Diffeomorphic Reg.} & \rotatebox[origin=c]{90}{Instance-spec. Opt.} & \rotatebox[origin=c]{90}{Run Time (CPU)} & \rotatebox[origin=c]{90}{Open Source}\\
 \hline
 \rowcolor{Gray}
 honkamj~\raisebox{1.5pt}{\fcolorbox{white}{honkamj}{\rule{0pt}{2pt}\rule{2pt}{0pt}}} 
    & & $\sbullet[1.5]$ & & & 
    $\sbullet[1.5]$ & & & & & 
    $\sbullet[1.5]$ & $\sbullet[1.5]$& $\sbullet[1.5]$& $\sbullet[1.5]$& & &$\sbullet[1.5]$&$\sbullet[1.5]$&
    & 51.70s & \href{https://github.com/honkamj/SITReg}{\faGithub}\\

 MadeForLife~\raisebox{1.5pt}{\fcolorbox{white}{MadeForLife}{\rule{0pt}{2pt}\rule{2pt}{0pt}}} 
    & & $\sbullet[1.5]$ & & & 
    $\sbullet[1.5]$ & & & & & 
    $\sbullet[1.5]$ & $\sbullet[1.5]$&$\sbullet[1.5]$ & $\sbullet[1.5]$& &$\sbullet[1.5]$ & &$\sbullet[1.5]$ &
    & 8.79s &\href{https://github.com/TVayne/GroupMorph}{\faGithub}\\

 \rowcolor{Gray}
 LYU1~\raisebox{1.5pt}{\fcolorbox{white}{LYU1}{\rule{0pt}{2pt}\rule{2pt}{0pt}}} 
    & & $\sbullet[1.5]$ & & & 
    $\sbullet[1.5]$ & & & & & 
    $\sbullet[1.5]$ & $\sbullet[1.5]$ & $\sbullet[1.5]$ & & & $\sbullet[1.5]$& & &
    & 15.86s &\\

  next-gen-nn~\raisebox{1.5pt}{\fcolorbox{white}{next-gen-nn}{\rule{0pt}{2pt}\rule{2pt}{0pt}}}
    & & $\sbullet[1.5]$ & & &
    $\sbullet[1.5]$ & & & & & 
    $\sbullet[1.5]$ & & $\sbullet[1.5]$& & & $\sbullet[1.5]$& & &
    & 16.50s & \href{https://github.com/XiangChen1994/EOIR}{\faGithub}\\

 \rowcolor{Gray}
 zhuoyuanw210~\raisebox{1.5pt}{\fcolorbox{white}{zhuoyuanw210}{\rule{0pt}{2pt}\rule{2pt}{0pt}}} 
    & & $\sbullet[1.5]$ & & & 
    $\sbullet[1.5]$& & & & & 
    $\sbullet[1.5]$ & $\sbullet[1.5]$&$\sbullet[1.5]$ & & & $\sbullet[1.5]$& & &
    & - &\\

 DutchMasters~\raisebox{1.5pt}{\fcolorbox{white}{DutchMasters}{\rule{0pt}{2pt}\rule{2pt}{0pt}}} 
    & & $\sbullet[1.5]$ & & &
    & $\sbullet[1.5]$ & & & & 
    $\sbullet[1.5]$ & & $\sbullet[1.5]$& $\sbullet[1.5]$& & & &$\hbullet[1.6]$&
    $\sbullet[1.5]$& 1458.44s &\\

 \rowcolor{Gray}
 lukasf~\raisebox{1.5pt}{\fcolorbox{white}{lukasf}{\rule{0pt}{2pt}\rule{2pt}{0pt}}} 
    & & $\sbullet[1.5]$ & & & 
    $\sbullet[1.5]$ & & & & &
    $\sbullet[1.5]$ & & & $\sbullet[1.5]$& & $\sbullet[1.5]$& & &
    & 4.62s &\\

 Bailiang~\raisebox{1.5pt}{\fcolorbox{white}{bailiang}{\rule{0pt}{2pt}\rule{2pt}{0pt}}} 
    & & $\sbullet[1.5]$ & & &
    $\sbullet[1.5]$ & & & & &
    $\sbullet[1.5]$ & $\sbullet[1.5]$& $\sbullet[1.5]$& $\sbullet[1.5]$& $\sbullet[1.5]$ & & & &
    & 14.93s & \href{https://github.com/BailiangJ/rethink-reg}{\faGithub}\\

 \rowcolor{Gray}
 VROC~\raisebox{1.5pt}{\fcolorbox{white}{VROC}{\rule{0pt}{2pt}\rule{2pt}{0pt}}} 
    & $\sbullet[1.5]$ & & & & 
    & & & &$\sbullet[1.5]$ & 
    & & &$\sbullet[1.5]$ & & & &
    & $\sbullet[1.5]$& 152.70s &\href{https://github.com/IPMI-ICNS-UKE/vroc}{\faGithub}\\

  TimH~\raisebox{1.5pt}{\fcolorbox{white}{TimH}{\rule{0pt}{2pt}\rule{2pt}{0pt}}} 
    & & $\sbullet[1.5]$ & & & 
    & & $\sbullet[1.5]$& & & 
    $\sbullet[1.5]$ & & & & & &$\sbullet[1.5]$ &$\sbullet[1.5]$&
    & 9.41s &\href{https://github.com/timH6502/MultiStepConsistent-LUMIRReg}{\faGithub}\\

 \rowcolor{Gray}
 LoRA-FT~\raisebox{1.5pt}{\fcolorbox{white}{LoRA-FT}{\rule{0pt}{2pt}\rule{2pt}{0pt}}} 
    & & $\sbullet[1.5]$ & & & 
    & $\sbullet[1.5]$ & & & & 
    $\sbullet[1.5]$ & & $\sbullet[1.5]$& $\sbullet[1.5]$& & & &$\hbullet[1.6]$ &
    $\sbullet[1.5]$& 18.20s &\\
 
 \hline
 
 ANTs SyN~\raisebox{1.5pt}{\fcolorbox{black}{ANTsSyN}{\rule{0pt}{2pt}\rule{2pt}{0pt}}} 
    & & $\sbullet[1.5]$ & & & 
    & & & $\sbullet[1.5]$& & 
    & & $\sbullet[1.5]$& $\sbullet[1.5]$& & & $\sbullet[1.5]$&$\sbullet[1.5]$&
    $\sbullet[1.5]$& 988.70s&\href{https://github.com/ANTsX/ANTsPy}{\faGithub}\\

 \rowcolor{Gray}
 deedsBCV~\raisebox{1.5pt}{\fcolorbox{black}{deedsBCV}{\rule{0pt}{2pt}\rule{2pt}{0pt}}} 
    & & & $\sbullet[1.5]$ & & 
    $\sbullet[1.5]$& & & & & 
    & & $\sbullet[1.5]$& $\sbullet[1.5]$& $\sbullet[1.5]$& & & &
    $\sbullet[1.5]$& 43.34s &\href{https://github.com/mattiaspaul/deedsBCV}{\faGithub}\\
 
 FireANTs Greedy~\raisebox{1.5pt}{\fcolorbox{black}{FireANTsGreedy}{\rule{0pt}{2pt}\rule{2pt}{0pt}}} 
    & & $\sbullet[1.5]$ & & & 
    & & & &$\sbullet[1.5]$& 
    & & $\sbullet[1.5]$& $\sbullet[1.5]$& & & &$\sbullet[1.5]$&
    $\sbullet[1.5]$& 61.24s &\href{https://github.com/rohitrango/fireants}{\faGithub}\\
    
 \rowcolor{Gray}
 FireANTs SyN~\raisebox{1.5pt}{\fcolorbox{black}{FireANTsSyN}{\rule{0pt}{2pt}\rule{2pt}{0pt}}} 
    & & $\sbullet[1.5]$ & & & 
    & & & $\sbullet[1.5]$ & & 
    & & $\sbullet[1.5]$& $\sbullet[1.5]$& & & &$\sbullet[1.5]$&
    $\sbullet[1.5]$& 2289.42s &\href{https://github.com/rohitrango/fireants}{\faGithub}\\
 
 SynthMorph~\raisebox{1.5pt}{\fcolorbox{black}{SynthMorph}{\rule{0pt}{2pt}\rule{2pt}{0pt}}} 
    & & & & $\sbullet[1.5]$ & 
    $\sbullet[1.5]$ & & & & & 
    $\sbullet[1.5]$ & & & & & &$\sbullet[1.5]$&
    $\sbullet[1.5]$ & & 28.62s &\href{https://hub.docker.com/r/freesurfer/synthmorph}{\faDocker}\\
    
 \rowcolor{Gray}
 TransMorph~\raisebox{1.5pt}{\fcolorbox{black}{TransMorph}{\rule{0pt}{2pt}\rule{2pt}{0pt}}} 
    & & $\sbullet[1.5]$ & & & 
    $\sbullet[1.5]$ & & & & & 
    $\sbullet[1.5]$ & & & $\sbullet[1.5]$& & $\sbullet[1.5]$& & &
    & 8.00s &\href{https://github.com/junyuchen245/TransMorph_Transformer_for_Medical_Image_Registration}{\faGithub}\\
 
 uniGradICON~\raisebox{1.5pt}{\fcolorbox{black}{uniGradICON}{\rule{0pt}{2pt}\rule{2pt}{0pt}}} 
    & & $\sbullet[1.5]$ & & & 
    & $\sbullet[1.5]$ & & & & 
    $\sbullet[1.5]$ & & $\sbullet[1.5]$& $\sbullet[1.5]$& & & & $\hbullet[1.6]$&
    & 16.69s &\href{https://github.com/uncbiag/uniGradICON}{\faGithub}\\
    
 \rowcolor{Gray}
 uniGradICON w/ ISO~\raisebox{1.5pt}{\fcolorbox{black}{uniGradICONiso}{\rule{0pt}{2pt}\rule{2pt}{0pt}}} 
    & & $\sbullet[1.5]$ & & & 
    & $\sbullet[1.5]$ & & & & 
    $\sbullet[1.5]$ & & $\sbullet[1.5]$& $\sbullet[1.5]$& & & & $\hbullet[1.6]$&
    $\sbullet[1.5]$& 1265.90s& \href{https://github.com/uncbiag/uniGradICON}{\faGithub}\\
 
 VFA~\raisebox{1.5pt}{\fcolorbox{black}{VFA}{\rule{0pt}{2pt}\rule{2pt}{0pt}}} 
    & & $\sbullet[1.5]$ & & & 
    $\sbullet[1.5]$ & & & & & 
    $\sbullet[1.5]$ & $\sbullet[1.5]$&$\sbullet[1.5]$ & & & $\sbullet[1.5]$& &
    & & 13.97s& \href{https://github.com/yihao6/vfa}{\faGithub}\\
 
 \rowcolor{Gray}
 VoxelMorph~\raisebox{1.5pt}{\fcolorbox{black}{VoxelMorph}{\rule{0pt}{2pt}\rule{2pt}{0pt}}} 
    & & $\sbullet[1.5]$ & & & 
    $\sbullet[1.5]$ & & & & & 
    $\sbullet[1.5]$ & & & & & & & &
    & 4.11s & \href{https://github.com/voxelmorph/voxelmorph}{\faGithub}\\
 \hline
\end{tabular}
\end{table*}

\subsection{Submissions}
\label{ss:submissions}
The evaluation phase of the challenge received submissions from 11 teams, of which only one employed a purely optimization-based method; the remaining ten submissions were deep learning-based.
Among these, two incorporated instance-specific optimization, where the deformation fields predicted by the network were further refined through pairwise optimization at inference time.
\textcolor{black}{All submitted methods employed cross-correlation as the similarity metric for registration, with \twBox{VROC} using normalized cross-correlation (NCC) computed over the entire moving and fixed images, whereas all others used local NCC (LNCC) as the similarity metric.}
All participant methods were required to be trained or developed using the LUMIR dataset.
The use of external data was permitted only if explicitly disclosed; based on the submitted method descriptions, none of the participants relied on external datasets.
Some approaches, specifically \twBox{lukasf} and \twBox{LoRA-FT}, were initialized from previously pretrained models but were subsequently retrained or fine-tuned using the LUMIR dataset.

In addition to participant submissions, we evaluated ten widely used baseline methods, including traditional optimization-based approaches such as \tbBox{ANTsSyN}~\citep{avants2008mia} and representative deep learning models such as \tbBox{VoxelMorph}~\citep{balakrishnan2019voxelmorph}.
We include an eleventh baseline method of no deformation, \texttt{ZeroDisplacement}, to establish how the images differ prior to registration.

We summarize all evaluated registration methods in Table~\ref{tab:models}, highlighting their image similarity metrics, deformation regularization techniques, model design choices, runtime, and availability as open-source models.
For comprehensive descriptions of image similarity metrics and deformation regularization methods, readers are referred to a recent survey~\citep{chen2024survey}.
The main architectural design choices listed in the table are defined as follows:

\begin{itemize} 

\item \textbf{\textit{Dual-Stream Encoders}}: Two separate encoders extracting features from moving and fixed images, which may either share weights or be trained independently. 
\item \textbf{\textit{Coarse-to-Fine Pyramid}}: Models (either deep learning- or optimization-based) that generate deformation fields or perform registration in a coarse-to-fine manner. Note that this is distinct from whether the underlying network employs multi-resolution architectures. 
\item \textbf{\textit{Progressive Registration}}: Models that progressively refine deformation fields. For deep learning methods, this involves using multi-stage networks, with each subsequent network refining the deformation produced by the previous stage. For optimization-based methods, progressive refinement is inherently built-in, as each iteration updates and improves the deformation estimate.
\item \textbf{\textit{Cost Volume}}: Computation of a cost volume containing similarity or distance measures (e.g., correlation or Euclidean distance) between intensity or feature values at corresponding voxels or patches in the fixed and moving images.
\item \textbf{\textit{Attention}}: Models incorporating attention mechanisms (e.g., self-attention, cross-attention, or other variants) within their architectures.
\item \textbf{\textit{Inverse Consistency by Construction}}: Models explicitly designed to guarantee inverse consistency, such that the forward deformation $\phi_{A\rightarrow B}$ is the inverse of the backward deformation $\phi_{B\rightarrow A}$. In deep learning-based methods, this is often achieved by exponentiating the difference between intermediate fields computed by an auxiliary network~\citep{greer2023inverse}; in optimization-based approaches, symmetric optimization schemes are used, as in~\citep{avants2008mia}.
\item \textbf{\textit{Diffeomorphic Registration}}: Models that employ formulations (e.g., time-varying velocity fields or stationary velocity fields) to achieve diffeomorphic (smooth, invertible) deformation fields.
\item \textbf{\textit{Instance-Specific Optimization}}: Models that iteratively refine deformation fields for each specific image pair by optimizing the registration objective function individually per instance.

\end{itemize}
Following this overview, we introduce each method individually using a concise three-line summary: the first line includes a colored square (that is used in plots and figures for quick reference) and the name of the method; second is a one line summary of the method; and finally in parentheses is the Participant(s) that contributed the method. Each summary is then followed by a single sentence description of the respective method.
In \ref{a:methods}, we provide more detailed descriptions of each of the submission methods.

\desc{white}{honkamj}{Symmetric, inverse consistent, and topology
preserving by construction}{Joel Honkamaa \& Pekka
Marttinen~\citep{honkamaa2024melba}}

\twBox{honkamj} is a multi-resolution deep learning registration
architecture which is inverse consistent, symmetric, and preserves
topology by construction.

\desc{white}{MadeForLife}{Dual-stream encoders with a specialized decoder for coarse-to-fine pyramid and progressive registration}{Yichao Zhou \& Zuopeng Tan}

\twBox{MadeForLife} uses a combination of
GroupMorph~\citep{tan2024tmi} and
\tbBox{VoxelMorph}~\citep{balakrishnan2019tmi} for their registration
framework.

\desc{white}{LYU1}{Pyramid network with auxiliary decoder}{Hongchao Zhou \& Shunbo Hu}

\twBox{LYU1} proposes a shared encoder, a shared auxiliary decoder,
and a fusion pyramid decoder.

\desc{white}{next-gen-nn}{Robust backbone network}{Yuxi Zhang \& Hang Zhang}

\twBox{next-gen-nn} uses a robust backbone network based on Zhang
\textit{et al.}~\citep{zhang2024miccai} which comprises an encoder
with a series of convolutional layers and a cascaded mechanism
designed to refine the prediction of deformation fields.

\desc{white}{zhuoyuanw210}{Dual stream pyramid encoder and motion decomposition transformer}{Zhuoyuan Wang \& Yi Wang}

This method employs a dual-stream pyramid encoder and a
GPU-accelerated motion decomposition
Transformer~(ModeTv2)~\citep{wang2024modetv2} decoder.

\desc{white}{DutchMasters}{Gradient projection optimization}{Yi Zhang \& Qian Tao~\citep{zhang2024arxiv}}

\twBox{DutchMasters} uses gradient projection techniques to address
the multi-objective optimization~(MOO) challenge that is at the core
of image registration.

\desc{white}{lukasf}{TransMorph refinement}{Lukas F\"{o}rner \& Thomas Wendler}

The authors propose a fine-tuning of a pre-trained
\tbBox{TransMorph}~\citep{chen2022mia} model to improve convergence
stability and deformation smoothness.

\desc{white}{Bailiang}{Dual-stream encoders, coarse-to-fine pyramid, correlation layers, and progressive registration}{Bailiang Jian \& Benedikt Wiestler~\citep{jian2024mamba}}

\twBox{Bailiang} uses a ConvNet architecture with dual-stream encoders
to independently extract image features from moving and fixed images,
followed by a decoder that integrates these features with a
coarse-to-fine optimization pyramid to progressively refine
deformations across scales.

\desc{white}{VROC}{Variational Registration on Crack}{Frederic Madesta \& Thilo Sentker}

\twBox{VROC} uses classic registration components like Demons based forces~\citep{thirion1998mia} and image preprocessing steps to perform
the registration.

\desc{white}{TimH}{ConstrICON framework multistep inverse consistency}{Tim Hable}

\twBox{TimH} uses the  ConstrICON framework~\citep{greer2023miccai} to
build an inverse consistent multistep registration model.

\desc{white}{LoRA-FT}{Low-rank adaptation of foundation models}{Jin Kim \& Matthew S. Brown}

\twBox{LoRA-FT} uses a Low-rank Adaptation~(LoRA)~\citep{hu2021lora}
technique to adapt the \tbBox{uniGradICON}~\citep{tian2024miccai}
foundation models to the challenge datasets.

\subsection{Baseline Methods}
\label{ss:baselines}
For each of the Baseline methods we also provide a three line summary: the first line includes a colored square (that is used in plots and figures for quick reference) and the name of the method; the second is a one line summary of the method; and finally in parentheses is the reference for the method.
Following the one line summaries is a single sentence description of the respective methods.
Table~\ref{tab:models} includes additional details about the methods and complete descriptions of each of the baseline methods is included in \ref{a:methods}.
\textcolor{black}{Among the deep learning-based baseline methods, \tbBox{TransMorph}, \tbBox{VFA}, and \tbBox{VoxelMorph} were trained using the LUMIR training data.
In contrast, the remaining approaches, including \tbBox{SynthMorph} and \tbBox{uniGradICON}, were treated as registration foundation models, and we used the pretrained weights provided by the respective authors.
Note that \tbBox{SynthMorph} was developed specifically for brain image registration, whereas \tbBox{uniGradICON} is a more general foundation model that has demonstrated applicability across multiple anatomical regions.
All optimization-based methods do not require training and were directly applied to the test set, performing instance-specific optimization for each test image pair.}
\textcolor{black}{All baseline deep learning based methods, except the pretrained foundation registration models (i.e., \tbBox{SynthMorph} and \tbBox{uniGradICON}/\tbBox{uniGradICONiso}), were trained on the LUMIR training data, which was brain-specific.}

\desc{black}{ANTsSyN}{Optimization-based, symmetric, time-varying diffeomorphic registration}{\citet{avants2007miccai, avants2008mia, avants2018ants}}

\textcolor{black}{The symmetric image normalization method~(SyN), packaged as part of the Advanced Normalization Tools~(ANTs), is a traditional registration framework that maximizes the cross-correlation~(CC) within the space of diffeomorphic maps.}

\desc{black}{deedsBCV}{Optimization-based, discrete registration}{\citet{heinrich2013mrf}}

\tbBox{deedsBCV} uses Markov random field~(MRF)-based discrete optimization strategies to address large motion of small features, sliding motions between organs, and changes in image contrast due to compression in lung CT registration.

\doubleDesc{black}{FireANTsGreedy}{black}{FireANTsSyN}{Optimization-based, GPU-accelerated, time-varying diffeomorphic registration}{\citet{jena2024fireants}}

\textcolor{black}{\tbBox{FireANTsSyN} and \tbBox{FireANTsGreedy} use multi-scale adaptive Riemannian optimization for diffeomorphic image matching that overcome both the slow convergence of traditional optimization based registration and the poor out-of-distribution generalization typical of deep learning based registration.}

\desc{black}{SynthMorph}{Learning contrast-agnostic registration without acquired images}{\citet{hoffmann2021synthmorph}}

\tbBox{SynthMorph} is a general strategy for learning contrast-agnostic registration, which uses a generative model to provide random label maps during raining.

\desc{black}{TransMorph}{Vision Transformer-based registration model}{\citet{chen2022mia}}

\tbBox{TransMorph} is a deep learning-based image registration framework that incorporates a Transformer-based architecture for both affine and deformable registration.

\doubleDesc{black}{uniGradICON}{black}{uniGradICONiso}{Inverse consistent, learning-based foundation registration model}{\citet{tian2024miccai}}

\tbBox{uniGradICON} uses gradient inverse consistency regularization~\citep{tian2023cvpr} to train a universal foundation model for registration. \tbBox{uniGradICONiso} incorporates instance-specific optimization through iterative fine-tuning of the network parameters for each image pair.

\desc{black}{VFA}{Dual-stream encoders, vector field cross-attention decoder, multi-scale registration model}{\citet{liu2024vector}}

\tbBox{VFA} uses neural networks to extract multiresolution feature maps from the fixed and moving images and then retrieves pixel-level correspondences based on an attention module without the need for learnable parameters.

\desc{black}{VoxelMorph}{U-Net-based registration model}{\citet{balakrishnan2019voxelmorph}}

\textcolor{black}{\tbBox{VoxelMorph} is a learning-based framework for deformable, pairwise medical image registration that uses parameterized CNNs.}

\section{Results}
\label{s:results}

\begin{figure*}[!tb]
    \centering
    \includegraphics[width = 0.85\linewidth]{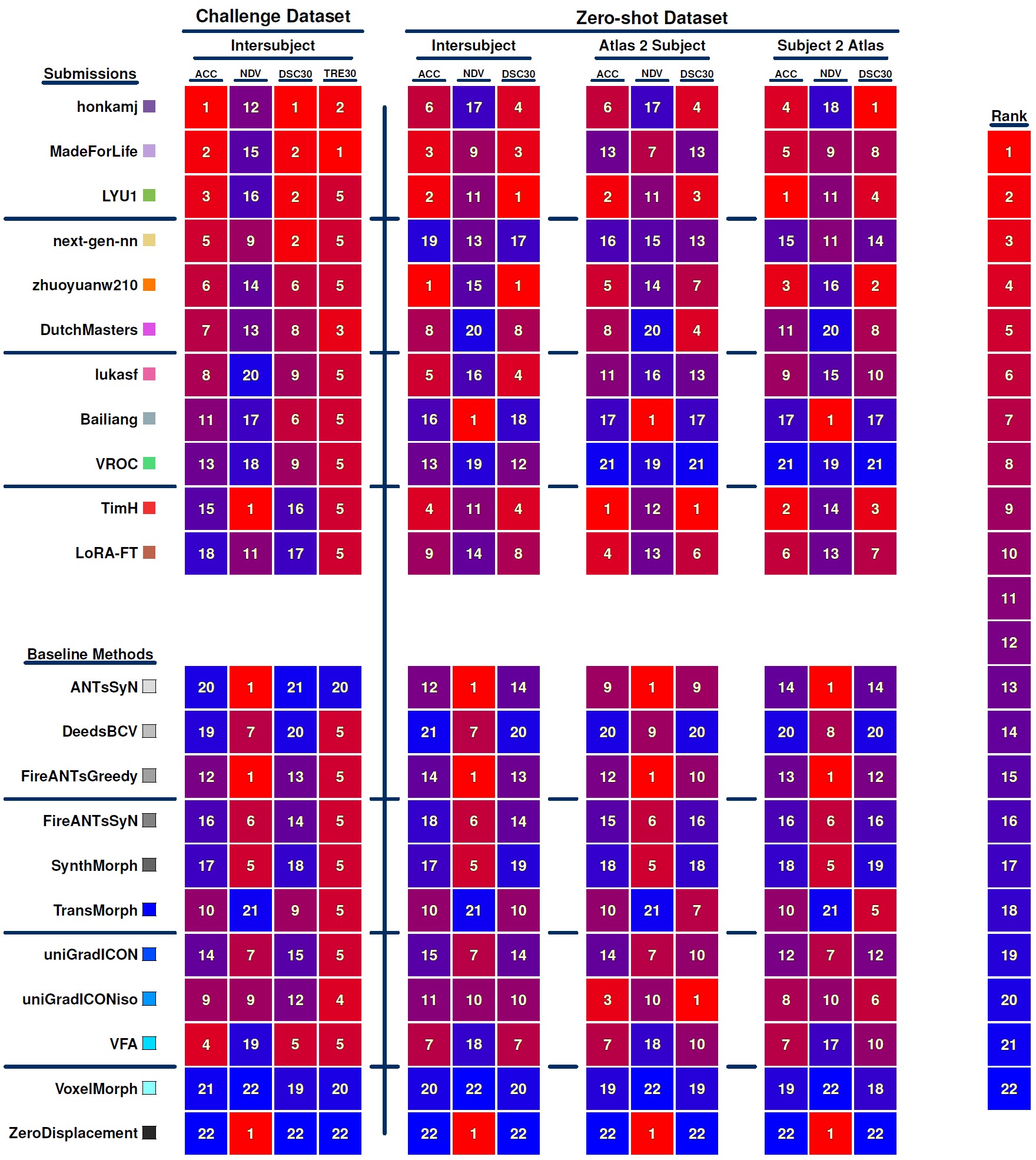}
    \caption{Shown for each subtask is a ranking for registration accuracy~(ACC), 30\textsuperscript{th} percentile of Dice Similarity Coefficient~(DSC30), non-diffeomorphic volume~(NDV), and 30\textsuperscript{th} percentile of total registration error~(TRE30), across all 21 reported methods and the baseline of no registration~(\tbBox{ZeroDisplacement}\,). \textcolor{black}{The reported rankings aggregate all zero-shot sub-tasks within each category (inter-subject, atlas-to-subject, and subject-to-atlas), including unseen MRI contrasts and intra-species macaque registration used to evaluate cross-species generalization of human-trained models. Detailed per-dataset results are reported in the Appendix. For brief descriptions of the evaluated methods see Sec.~\ref{s:methods}, with more detailed descriptions in~\ref{a:methods}. A description of the ranking scheme is given in Sec.~\ref{ss:comparisons}, with a detailed example of the ranking in \ref{a:ranking-example}.}}
    \label{fig:rank_all_heatmaps}
\end{figure*}

\begin{table*}[!tb]
\caption{Shown for the challenge test set are the mean and standard deviation for the DSC, HD95, TRE, NDV, DSC30, and TRE30. Also shown is the ranking based on registration accuracy~(ACC) which for the Testing set is a combination of the rankings of DSC, HD95, and TRE. For a complete description of the reported statistics and the how the ranking scheme was computed see Sec.~\ref{ss:comparisons}, with a detailed example of the ranking in \ref{a:ranking-example}.}
\label{t:testset}
\centering
\resizebox{0.85\paperwidth}{!}{%
\begin{tabular}{r c c c c c c c c}
\toprule
&&\textbf{DSC}$\uparrow$ & \textbf{HD95}$\downarrow$~\textcolor{black}{(mm)} & \textbf{TRE}$\downarrow$~\textcolor{black}{(mm)} & \textbf{Ranking} & \textbf{\%NDV}$\downarrow$ & \textbf{DSC30}$\uparrow$& \textbf{TRE30}$\downarrow$~\textcolor{black}{(mm)} \\

\cmidrule(lr){3-3}
\cmidrule(lr){4-4}\cmidrule(lr){5-5}
\cmidrule(lr){6-6}
\cmidrule(lr){7-7}
\cmidrule(lr){8-8}
\cmidrule(lr){9-9}
\textbf{Method}&& \textbf{Mean(}$\bm{\PM}$\textbf{Std. Dev.)} 
& \textbf{Mean(}$\bm{\PM}$\textbf{Std. Dev.)} 
& \textbf{Mean(}$\bm{\PM}$\textbf{Std. Dev.)} 
& \scriptsize{\textbf{(DSC + HD95 + TRE)}}& \textbf{Mean(}$\bm{\PM}$\textbf{Std. Dev.)} 
& \textbf{Mean(}$\bm{\PM}$\textbf{Std. Dev.)} 
& \textbf{Mean(}$\bm{\PM}$\textbf{Std. Dev.)} 
\\
\midrule 

\rowcolor{white}
\mBox{white}{honkamj}  &
 &0.785($\mypm$0.027)
 &3.035($\mypm$0.413)
 &3.088($\mypm$0.962)
 &\rCirc{1}
 &2.52e-03($\mypm$6.08e-04)
 &0.753($\mypm$0.021)
 &3.389($\mypm$0.997)
 \\

\rowcolor{Gray}
\mBox{white}{MadeForLife}  &
 &0.778($\mypm$0.027)
 &3.285($\mypm$0.436)
 &3.071($\mypm$0.942)
 &\rCirc{2}
 &1.21e-02($\mypm$3.25e-03)
 &0.746($\mypm$0.021)
 &3.362($\mypm$0.980)
 \\

\rowcolor{white}
\mBox{white}{LYU1}  &
 &0.778($\mypm$0.027)
 &3.246($\mypm$0.427)
 &3.132($\mypm$0.953)
 &\rCirc{3}
 &1.50e-02($\mypm$3.70e-03)
 &0.745($\mypm$0.020)
 &3.427($\mypm$0.994)
 \\

\rowcolor{Gray}
\mBox{white}{next-gen-nn}  &
 &0.777($\mypm$0.027)
 &3.278($\mypm$0.430)
 &3.125($\mypm$0.938)
 &\rCirc{5}
 &1.39e-04($\mypm$1.29e-04)
 &0.745($\mypm$0.021)
 &3.430($\mypm$0.955)
 \\

\rowcolor{white}
\mBox{white}{zhuoyuanw210}  &
 &0.773($\mypm$0.027)
 &3.233($\mypm$0.434)
 &3.143($\mypm$0.935)
 &\rCirc{6}
 &4.47e-03($\mypm$1.37e-03)
 &0.740($\mypm$0.022)
 &3.434($\mypm$0.973)
 \\

\rowcolor{Gray}
\mBox{white}{DutchMasters}  &
 &0.770($\mypm$0.033)
 &3.256($\mypm$0.456)
 &3.114($\mypm$0.967)
 &\rCirc{7}
 &3.02e-03($\mypm$1.59e-03)
 &0.731($\mypm$0.027)
 &3.409($\mypm$1.009)
 \\

\rowcolor{white}
\mBox{white}{lukasf}  &
 &0.764($\mypm$0.030)
 &3.422($\mypm$0.443)
 &3.144($\mypm$0.950)
 &\rCirc{8}
 &2.76e-01($\mypm$7.98e-02)
 &0.728($\mypm$0.024)
 &3.436($\mypm$0.991)
 \\

\rowcolor{Gray}
\mBox{white}{Bailiang}  &
 &0.773($\mypm$0.028)
 &3.329($\mypm$0.434)
 &3.156($\mypm$0.931)
 &\rCirc{11}
 &2.22e-02($\mypm$1.16e-02)
 &0.740($\mypm$0.023)
 &3.439($\mypm$0.975)
 \\

\rowcolor{white}
\mBox{white}{VROC}  &
 &0.760($\mypm$0.029)
 &3.630($\mypm$0.499)
 &3.225($\mypm$0.980)
 &\rCirc{13}
 &4.75e-02($\mypm$3.06e-02)
 &0.726($\mypm$0.024)
 &3.532($\mypm$1.014)
 \\

\rowcolor{Gray}
\mBox{white}{TimH}  &
 &0.730($\mypm$0.027)
 &3.569($\mypm$0.410)
 &3.193($\mypm$0.939)
 &\rCirc{15}
 &0.00e+00($\mypm$0.00e+00)
 &0.698($\mypm$0.019)
 &3.492($\mypm$0.967)
 \\

\rowcolor{white}
\mBox{white}{LoRA-FT}  &
 &0.735($\mypm$0.037)
 &3.733($\mypm$0.546)
 &3.235($\mypm$0.933)
 &\rCirc{18}
 &3.27e-03($\mypm$5.57e-03)
 &0.692($\mypm$0.026)
 &3.526($\mypm$0.970)
 \\
\midrule

\rowcolor{Gray}
\mBox{black}{ANTsSyN}  &
 &0.703($\mypm$0.059)
 &3.688($\mypm$0.577)
 &3.484($\mypm$1.062)
 &\rCirc{20}
 &0.00e+00($\mypm$0.00e+00)
 &0.633($\mypm$0.055)
 &3.845($\mypm$1.068)
 \\

\rowcolor{white}
\mBox{black}{DeedsBCV}  &
 &0.696($\mypm$0.026)
 &3.945($\mypm$0.410)
 &3.104($\mypm$0.981)
 &\rCirc{19}
 &1.68e-04($\mypm$3.72e-04)
 &0.665($\mypm$0.019)
 &3.424($\mypm$1.000)
 \\

\rowcolor{Gray}
\mBox{black}{FireANTsGreedy}  &
 &0.752($\mypm$0.030)
 &3.671($\mypm$0.491)
 &3.152($\mypm$0.958)
 &\rCirc{12}
 &0.00e+00($\mypm$0.00e+00)
 &0.717($\mypm$0.022)
 &3.466($\mypm$0.975)
 \\

\rowcolor{white}
\mBox{black}{FireANTsSyN}  &
 &0.749($\mypm$0.030)
 &3.729($\mypm$0.515)
 &3.174($\mypm$0.952)
 &\rCirc{16}
 &3.40e-05($\mypm$2.60e-05)
 &0.714($\mypm$0.021)
 &3.476($\mypm$0.979)
 \\

\rowcolor{Gray}
\mBox{black}{SynthMorph}  &
 &0.722($\mypm$0.028)
 &3.614($\mypm$0.463)
 &3.228($\mypm$0.938)
 &\rCirc{17}
 &1.12e-05($\mypm$1.31e-05)
 &0.688($\mypm$0.019)
 &3.514($\mypm$0.983)
 \\

\rowcolor{white}
\mBox{black}{TransMorph}  &
 &0.762($\mypm$0.030)
 &3.462($\mypm$0.457)
 &3.142($\mypm$0.951)
 &\rCirc{10}
 &3.62e-01($\mypm$8.70e-02)
 &0.726($\mypm$0.023)
 &3.439($\mypm$0.988)
 \\

\rowcolor{Gray}
\mBox{black}{uniGradICON}  &
 &0.742($\mypm$0.036)
 &3.575($\mypm$0.526)
 &3.240($\mypm$0.937)
 &\rCirc{14}
 &8.40e-05($\mypm$1.23e-04)
 &0.699($\mypm$0.028)
 &3.522($\mypm$0.983)
 \\

\rowcolor{white}
\mBox{black}{uniGradICONiso}  &
 &0.760($\mypm$0.032)
 &3.401($\mypm$0.479)
 &3.135($\mypm$0.961)
 &\rCirc{9}
 &1.65e-04($\mypm$2.21e-04)
 &0.721($\mypm$0.025)
 &3.421($\mypm$1.012)
 \\

\rowcolor{Gray}
\mBox{black}{VFA}  &
 &0.777($\mypm$0.027)
 &3.150($\mypm$0.422)
 &3.138($\mypm$0.944)
 &\rCirc{4}
 &7.04e-02($\mypm$4.67e-02)
 &0.744($\mypm$0.020)
 &3.428($\mypm$0.984)
 \\

\rowcolor{white}
\mBox{black}{VoxelMorph}  &
 &0.714($\mypm$0.036)
 &4.072($\mypm$0.578)
 &3.528($\mypm$0.968)
 &\rCirc{21}
 &1.22e+00($\mypm$2.69e-01)
 &0.671($\mypm$0.023)
 &3.851($\mypm$0.982)
 \\

\rowcolor{Gray}
\mBox{black}{ZeroDisplacement}  &
 &0.555($\mypm$0.038)
 &4.915($\mypm$0.556)
 &4.384($\mypm$1.028)
 &\rCirc{22}
 &0.00e+00($\mypm$0.00e+00)
 &0.510($\mypm$0.024)
 &4.772($\mypm$0.984)
 \\

\bottomrule
\end{tabular}}
\end{table*}

In the following, we summarize the results from both the challenge test set {(Sec.~\ref{subsec:test_set}) and the zero-shot out-of-domain datasets (Sec.~\ref{subsec:zero_shot})}.
Figures~\ref{fig:result_challenge_samples} and~\ref{fig:result_zeroshot_samples} present representative qualitative results from the challenge test set and the zero-shot evaluation tasks~(respectively), showcasing the outputs of the top three performing methods for each main task.
A brief description of all the evaluated methods is provided in Sec.~\ref{s:methods} with a detailed presentation of the methods in \ref{a:methods}, and a summary of the properties of each method is in Table~\ref{tab:models}.
\textcolor{black}{We note that across many of the challenge and zero-shot evaluations, the strongest methods were often closely clustered across the reported metrics.
Accordingly, the rankings are most informative when interpreted together with the underlying metric values, which provide additional context on the magnitude of the observed differences.}

\subsection{Challenge Test Set}
\label{subsec:test_set}
The LUMIR challenge test set comprises images drawn from the same distribution as the training data, which consists exclusively of T1-w brain MRIs from healthy controls.
The test set includes 590 images, of which 460 are annotated with anatomical label maps and 130 with landmark points.
Using these annotations, we evaluated the registration performance based on segmentation overlap and landmark alignment.
%
%
Details of the dataset are provided in Sec.~\ref{sec:challenge_set}, and a detailed description of the evaluation metrics is presented in Sec.~\ref{ss:comparisons}.

The test set results are summarized in Table~\ref{t:testset}, with a more detailed ranking presented in Fig.~\ref{fig:rank_all_heatmaps}. The top ten methods demonstrated comparable registration accuracy, with DSC scores ranging from 0.76 to 0.79 and TRE between 3.15~\textcolor{black}{mm} and 3.00~\textcolor{black}{mm}. Even for more challenging cases, reflected by DSC30 and TRE30, the performance differences among the top ten methods remained similar.

Overall, deep learning-based methods outperformed traditional optimization-based approaches, with all top ten methods being DL-based. Among these, eight were developed by participants, and two were baseline models.
The top three methods in overall registration accuracy were \twBox{honkamj} (1\textsuperscript{st}), \twBox{MadeForLife} (2\textsuperscript{nd}), and \twBox{LYU1} (3\textsuperscript{rd}).
For DSC30, the top performers were \twBox{honkamj} (1\textsuperscript{st}), followed by \twBox{MadeForLife}, \twBox{LYU1}, and \twBox{next-gen-nn}, which were all tied for 2\textsuperscript{nd}.
Of particular note, \twBox{next-gen-nn} ranked 5\textsuperscript{th} overall, however, it performed better on the more difficult cases.
For TRE30, the top three methods were \twBox{MadeForLife} (1\textsuperscript{st}), \twBox{honkamj} (2\textsuperscript{nd}), and \twBox{DutchMasters} (3\textsuperscript{rd}), reflecting a slight change in the classification between segmentation-based and landmark-based evaluations.
The highest-ranked optimization-based method was \tbBox{FireANTsGreedy}, which placed 12\textsuperscript{th}, followed by the participant-developed optimization method \tbBox{VROC} in 13\textsuperscript{th} place.

In terms of deformation regularity, the widely used deep learning baseline \tbBox{VoxelMorph} produced approximately 1\% non-diffeomorphic volumes within the brain region. All other top-performing methods generated smoother fields, with the top three methods achieving an NDV well below 1\%, approaching diffeomorphic quality.
Notably, all optimization-based methods excelled in generating smooth deformation fields.

\subsection{Zero-shot Out-of-domain Evaluation Tasks}
\label{subsec:zero_shot}

\begin{table}[!ht]
\centering
\caption{We present a medal table for the methods that covers all the subtasks presented in this paper. Medals are assigned for 1\textsuperscript{st}, 2\textsuperscript{nd}, and 3\textsuperscript{rd} place based on their performance in the various subtasks. We include only the top five medal receiving methods.}
\label{t:medals}
\resizebox{0.48\textwidth}{!}{\begin{tabular}{r c cc cc cc c}
\toprule
\raisebox{0.5em}{\textbf{Method}} && \gold && \silver && \bronze &&
\textbf{Total}\\
\cmidrule(lr){1-1}
\cmidrule(lr){3-9}
\rowcolor{Gray}
\mBox{white}{honkamj} && 9 && 4 && 4 && 17\\
\mBox{white}{MadeForLife} && 4 && 6 && 4 && 14\\
\rowcolor{Gray}
\mBox{white}{DutchMasters} && 4 && 3 && 0 && 7\\
\mBox{white}{next-gen-nn} && 1 && 2 && 2 && 5\\
\rowcolor{Gray}
\mBox{black}{VFA} && 0 && 3 && 2 && 5\\
%
%
%
%
\bottomrule
\end{tabular}}
\end{table}

We evaluated the zero-shot performance on a diverse set of out-of-domain datasets:
(1)~T1-w MRI from a different acquisition site (NIMH), (2)~pediatric T1-w MRI scans from ADHD patients, (3)~T1-w MRI scans with neurodegenerative pathology from ADNI, (4)~intra-contrast registration for unseen MRI contrasts (T2-w, T2*-w, FLAIR), (5)~ultra-high-field T1-w MRI at 9.4T, and (6)~macaque brain MRI.
Details of the dataset \textcolor{black}{and the atlas} are provided in Sec.~\ref{sec:zeroshot_set}
Across these datasets, we assessed inter-subject, atlas-to-subject, and subject-to-atlas registration tasks.

Performance across subtasks is summarized in Table~\ref{t:medals}, ranking methods by the number of gold, silver, and bronze medals earned for top-three placement in each evaluation setting. Consistently strong performers across zero-shot tasks include \twBox{honkamj}, \twBox{MadeForLife}, and \tbBox{DutchMasters}.
Despite the overall success of deep learning-based methods, several optimization-based methods, such as \tbBox{FireANTsGreedy}, performed competitively in specific domains, notably in the macaque dataset where anatomical priors diverge significantly from the training data.
Contrast-specific generalization remained an open challenge, as all methods exhibited a noticeable drop in performance when evaluated on unseen contrasts, especially T2*-w and FLAIR images.
\textcolor{black}{This is discussed in further detail in \ref{sec:ood_contrasts}.}

\textcolor{black}{To facilitate a more detailed analysis, we provide a task-wise breakdown and separate discussions for each zero-shot setting in the Appendix. These evaluations include out-of-domain T1-weighted data acquired from sources distinct from the training set~(\ref{sec:ood_t1w}) and MRI acquisition field strengths unseen during training~(\ref{sec:highfield_mr}); the application of models trained on LUMIR healthy controls to pathological brains with neurodegenerative disease~(\ref{sec:adni}) and across different age groups~(\ref{sec:adhd}); inter-contrast registration involving out-of-domain MRI contrasts such as T2-, T2*-weighted, and FLAIR images~(\ref{sec:ood_contrasts}); and evaluations under different anatomical priors using macaque brain data~(\ref{sec:macaque}). Several of these datasets, particularly those involving out-of-domain MRI contrasts, exhibit heterogeneous image quality, including thick-slice acquisitions that were resampled to an isotropic voxel size of 1~mm. Detailed task definitions, evaluation metrics, and per-method results are provided in \ref{sec:detailed_results}.}

\section{Discussion}
\label{s:discussion}
\subsection{Deep Learning Models Excel in Zero-Shot Registration.} 
Deep learning models tend to degrade when evaluated on data from a distribution different from the one they were trained on~\citep{torralba2011unbiased}.
Addressing such domain shift remains an active research area, prompting the development of specialized techniques for domain generalization~\citep{zhou2022domain} and domain adaptation~\citep{wang2018deep}.
In the LUMIR challenge, models were trained exclusively on inter-subject registration tasks using T1-w brain MRIs of healthy controls.
\textcolor{black}{The zero-shot evaluations, in contrast, exposed these models to distributional shifts including different MRI contrasts (T2-, T2*-weighted, and FLAIR), atlas-to-subject registration, images from pathological populations, and scans from non-human species.}

\textcolor{black}{Under these evaluation settings, several learning-based registration methods exhibited strong performance relative to traditional optimization-based approaches.
In particular, methods such as \twBox{honkamj}, \twBox{MadeForLife}, and \twBox{next-gen-nn} consistently ranked among the top-performing approaches across zero-shot tasks.
As summarized in Fig.~\ref{fig:rank_all_heatmaps}, learning-based methods occupied the majority of the top-five rankings and outperformed established optimization-based baselines, including \tbBox{ANTsSyN}, \tbBox{DeedsBCV}, and the recently proposed \texttt{FireANTs} variants.
While performance degradation was observed for more challenging contrasts such as T2* and FLAIR, the relative ranking of the top-performing deep learning methods remained stable.}

\textcolor{black}{We emphasize that these observations should be interpreted as empirical trends rather than as evidence of universal robustness.
In particular, all zero-shot datasets were processed using the same preprocessing pipeline as the training data, including pre-affine alignment, resampling to a common 1$\times$1$\times$1~mm resolution, and cropping to a uniform image size.
Deep learning-based registration methods are sensitive to preprocessing discrepancies, and their performance is not expected to generalize out of the box when preprocessing pipelines differ substantially.
For example, changes in image resolution alter the effective spatial context observed by convolutional kernels or Transformer patches, leading to differences in foreground-to-background ratios and anatomical scale relative to the learned receptive fields.
Such mismatches can substantially degrade performance, even when the underlying anatomy remains similar.
Accordingly, the generalization observed in this study should be understood as conditional on consistent preprocessing rather than as an inherent property of the learned models.}

\textcolor{black}{Within this context, we offer two interpretations that may help explain the observed behavior.
First, as the training data aggregates images from over 60 imaging centers it thus encompasses substantial variability in anatomy and acquisition.
While this factor was not isolated through controlled experiments, the scale and diversity of the training set may contribute to the improved generalization compared to prior studies relying on more homogeneous data.
Second, we hypothesize that the formulation of image registration itself may play a role.
Registration emphasizes learning relative spatial correspondences between image pairs, rather than modeling the semantic content of individual images.
This focus on geometric alignment may reduce sensitivity to appearance variations across contrasts, a behavior also reported in prior work such as \texttt{SynthMorph}~\citep{hoffmann2021synthmorph}.}
\textcolor{black}{However deep learning methods, despite their observed robustness, have clear limitations which we describe in Sec.~\ref{sec:DL_limitations}.}

Finally, despite excellent performance on zero-shot tasks, all tasks evaluated in the challenge remained within a single MRI contrast type.
We anticipate that performance would degrade substantially on inter-contrast registration tasks (e.g., T1-w to T2-w MRI).
Despite the comprehensive nature of our Zero-shot Dataset, there are still unanswered questions.
For example, none of our presented data sources represent severe pathologies such as hemispherectomy or neuroblastoma.
The pathological cases included in this study primarily involve neurodegenerative changes that preserve overall brain topology.
Scenarios involving substantial topological alterations, such as large tumors or edema, would likely violate the assumptions underlying deformable registration and are beyond the scope of the present evaluation.
\textcolor{black}{For example, given two subjects with tumors in different brain regions, a registration algorithm should not (incorrectly) align these tumors.
Likewise, when registering images between healthy controls and a subject with a tumor, there is no clear or universally accepted definition of what constitutes successful registration.}

%
%
\subsection{\textcolor{black}{Limitations of Deep Learning-based Methods.}}
\label{sec:DL_limitations}
\textcolor{black}{While learning-based methods demonstrate strong performance on unseen contrasts, several limitations remain.
In this study, we show that models trained solely on large-scale T1-w images can generalize robustly to unseen contrasts and datasets.
However, this evaluation relies on the assumption that zero-shot datasets undergo the same preprocessing pipeline as the training data, including voxel-size standardization via resampling and cropping to a uniform image size.
Under different conditions commonly encountered in practice, such as substantially different spatial resolutions, sampling schemes, image corruptions, or input image sizes arising from non-standardized acquisition protocols, learning-based methods may struggle to operate out of the box.
This limitation is particularly evident when such variations are not represented during training, as recently discussed in~\citep{jena2024deep, jena2025lumirage}.
In this sense, learning-based methods are not yet sufficiently robust to generalize to all unseen tasks through training on T1-w images alone.}
\textcolor{black}{Additionally, the deep-learning based approaches do have the significant drawback of requiring high-end GPUs for their training.
This is mitigated in many cases as these methods provide publicly available models, however it is not clear at this time how well these pre-trained models behave when applied to a never before seen organ or other more complex registration task (ie. inter-modal registration).}

%
%

\subsection{\textcolor{black}{Optimization-based Methods: Strengths and Limitations.}}
Optimization-based refers to those methods that rely on pairwise optimization without any prior training stage, and they have historically seen widespread use in neuroimaging studies.
Common pipelines for statistical parametric mapping~\citep{penny2011statistical}, voxel-based morphometry~\citep{ashburner2000voxel}, brain segmentation~\citep{fischl2012freesurfer}, and computational anatomy~\citep{younes2009evolutions} heavily depend on optimization-based methods.
In the LUMIR challenge, the optimization-based methods evaluated included \twBox{VROC}, \tbBox{ANTsSyN}, \tbBox{DeedsBCV}, and \texttt{FireANTs}~\pBox{black}{FireANTsGreedy}/\pBox{black}{FireANTsSyN}.
Despite their historical significance and conceptual appeal, these methods consistently underperformed compared to learning-based approaches.

On the challenge test set, \tbBox{ANTsSyN}, a widely used optimization-based method, ranked 20\textsuperscript{th} out of 21 methods, outperforming only \tbBox{VoxelMorph}.
This is consistent with the original VoxelMorph publication, which reported comparable or slightly inferior performance relative to \tbBox{ANTsSyN}.
\tbBox{DeedsBCV} followed closely, ranking 19\textsuperscript{th}.
\texttt{FireANTs}~\pBox{black}{FireANTsGreedy}/\pBox{black}{FireANTsSyN}, a recent method introduced as a strong optimization-based alternative to deep learning models, achieved rankings of 12\textsuperscript{th} for its Greedy variant~(\tbBox{FireANTsGreedy}) and 16\textsuperscript{th} for its SyN variant~(\tbBox{FireANTsSyN}).
Another optimization-based method, \twBox{VROC}, specifically tuned for the LUMIR test set by challenge participants, ranked 13\textsuperscript{th} overall.

On the zero-shot datasets, optimization-based methods demonstrated certain advantages.
Because these methods optimize deformation fields independently for each image pair, they are inherently less sensitive to domain shifts.
As such, \tbBox{ANTsSyN} and \texttt{FireANTs}~\pBox{black}{FireANTsGreedy}/\pBox{black}{FireANTsSyN} achieved higher accuracy and modest improvements in their overall rankings (see Fig.\ref{fig:rank_all_heatmaps}).
Nevertheless, they consistently remained outside the top ten, with the best-performing method on each zero-shot task being deep learning-based.

%
Since the introduction of \tbBox{VoxelMorph} in 2019~\citep{balakrishnan2019voxelmorph}---when deep learning-based registration was only comparable to \texttt{ANTsSyN}---the field has advanced markedly.
Optimization-based approaches still hold value, particularly for their well-defined properties such as symmetric deformations and explicit modeling of time-varying velocity fields.
However, the deep learning methods that perform the best in the LUMIR challenge now substantially outperform the traditional techniques on a wide range of tasks.
%

\textcolor{black}{That said, as discussed in Sec.~\ref{sec:DL_limitations}, the zero shot datasets evaluated in this study were processed using the same preprocessing pipeline as the LUMIR challenge training data.
Prior work has shown that learning-based registration methods are sensitive to preprocessing discrepancies and can underperform when applied to data processed with different pipelines~\citep{jena2025lumirage}.
As a result, these methods are unlikely to generalize out-of-the-box to arbitrary zero-shot scenarios.
In contrast, optimization-based approaches can be more adaptable to unseen cases, as their performance can often be recovered through careful tuning of hyperparameters or loss functions, highlighting a complementary role for these methods in practical deployment.}

%
\subsection{Identifying Effective Model Designs.} 
In this section, we explore common architectural patterns among the top-performing models, based on observed trends, as direct ablation studies are not feasible.
Notably, the top five models on the challenge test set---\twBox{honkamj}, \twBox{MadeForLife}, \twBox{LYU1}, \tbBox{VFA}, and \twBox{next-gen-nn}---all adopted a dual-stream encoder design.
In this setup, separate encoders are used to extract features from the moving and fixed images independently, rather than jointly feeding both images into a shared encoder.
This design encourages each encoder to focus on extracting features specific to its input, without early fusion of information across the image pair.
The responsibility of identifying differences between the two images is thus shifted to the decoder, which estimates the deformation field based on the contrast between the separately encoded feature maps.
The results suggest a strong correlation between this architectural design and effective registration performance.
\textcolor{black}{Another common feature among these models is a coarse-to-fine strategy for deformation estimation, where the deformation fields are estimated at multiple resolutions and then composed to form the final output.
This hierarchical approach is beneficial for capturing both global and local deformations effectively.}

Additionally, progressive registration appears to play a key role in promoting generalization across domains.
Models employing this design---where the model iteratively estimates full-resolution deformation fields, gradually refining the alignment by progressively warping the moving image toward the fixed image over multiple steps---tended to maintain consistent performance between the challenge and zero-shot tasks.
In contrast, models that did not adopt progressive registration, such as \twBox{LYU1} and \twBox{next-gen-nn}, exhibited notable performance drops, often falling out of the top ten on zero-shot tasks.

Notably, \twBox{DutchMasters} and \twBox{Bailiang}, both employing a progressive registration design, ranked 7\textsuperscript{th} and 11\textsuperscript{th}, respectively, on the challenge test set and showed substantial gains on zero-shot tasks.
\twBox{DutchMasters}, in particular, placed among the top three in atlas-based registration.
These results highlight progressive registration as a promising strategy for developing robust and generalizable registration models, which warrants further investigation through controlled ablation studies.

\textcolor{black}{In this challenge, several deep learning-based registration foundation models were evaluated as baseline methods, including \tbBox{uniGradICON}/\tbBox{uniGradICONiso} and \tbBox{SynthMorph}, the latter of which was developed specifically for brain registration.
We note that these models were used as pretrained baselines and were not trained on the LUMIR dataset.
On the challenge test set, they did not outperform the strongest challenge-specific methods and generally ranked in the lower half of the evaluated methods.
Compared with the top-performing method, \twBox{honkamj}, the mean DSC was lower by 0.063 for \tbBox{SynthMorph} and by 0.043 for \tbBox{uniGradICON}. Even with instance-specific optimization enabled, \tbBox{uniGradICONiso} ranked 9\textsuperscript{th} overall.
On the zero-shot evaluations, these models largely retained similar relative ranks, with the notable exception of \tbBox{uniGradICONiso}, which achieved third place overall on the atlas-to-subject tasks across the zero-shot datasets.
However, the strongest challenge-specific deep learning methods also generalized well to the zero-shot settings and continued to outperform the evaluated foundation models in most cases.
Taken together, these results suggest that, in this benchmark, challenge-specific models remained the most competitive overall, while pretrained foundation models are an attractive alternative in settings where developing or retraining a task-specific model may not be practical.
Additionally, as \tbBox{uniGradICON} was trained beyond brain registration, it may offer broader utility across anatomical regions.}



%
\subsection{\textcolor{black}{Impact of Instance-specific Optimization on Registration Performance}}
%
Learning-based methods can further refine the deformation field at test time by iteratively optimizing the registration loss function for each specific image pair.
This process, known as instance-specific optimization~(ISO), treats the DNN as an implicit optimizer corresponding to an unrolled iterative optimization procedure~\citep{gregor2010learning}.
Several previous studies have shown that ISO can significantly improve the performance of learning-based registration models~\citep{balakrishnan2019voxelmorph, tian2024miccai}.

The top-performing methods on both the ``in-domain'' challenge test set and the ``out-of-domain'' zero-shot tasks did not employ ISO.
Among ISO-based approaches, the best-performing method was \twBox{DutchMasters}, achieving the rank of 7\textsuperscript{th} on the challenge test set.
For the zero-shot tasks, \twBox{DutchMasters} exhibited enhanced robustness, even reaching top performance in several instances.
Such improvements could be attributed to ISO bridging performance gaps caused by data distribution shifts.

Although direct ablation studies were not feasible for participant-submitted methods, comparison between \texttt{uniGradICON}~\pBox{black}{uniGradICON} and \texttt{uniGradICONiso}~\pBox{black}{uniGradICONiso} highlights the benefit of the ISO component.
Incorporating ISO improved the rankings by five places (14\textsuperscript{th} $\rightarrow$ 9\textsuperscript{th}) for the challenge test set, alongside noticeable gains for the zero-shot set.
However, a clear disadvantage of ISO is its substantial computational burden, resulting in significantly increased runtimes.
For instance, the runtime for \twBox{honkamj} was 56.87s per image, compared to 1604.29s for \twBox{DutchMasters} using a CPU, although employing GPUs is expected to significantly alleviate this runtime overhead.
Overall, these findings emphasize that while ISO can enhance robustness and model performance, optimal DNN design that integrates inductive bias relevant to image registration continues to play a critical role in achieving the best performance.

\begin{table}[!tb]
\caption{Mean absolute error~(MAE) quantifying the deviation of the composed forward and inverse deformation fields from the identity, computed as $\vert\phi_{A\rightarrow B} \circ \phi_{B\rightarrow A} -Id\vert$.
The forward and inverse fields are obtained from the ``Subject to Atlas'' and ``Atlas to Subject'' tasks, respectively.
\textcolor{black}{The dagger ($\dagger$) next to \tbBox{ANTsSyN} and \tbBox{FireANTsSyN} indicates that both the forward and the inverse deformation fields were generated from a single symmetric run on the ``Subject to Atlas'' task, whereas entries without the dagger correspond to MAE values computed from deformation fields obtained in separate runs for the two tasks.}}
%
%
\label{tab:inverse_consistency}
\centering
\begin{tabular}{rc}
\toprule
 & \textbf{MAE$\downarrow$ (mm)}\\
\cmidrule(lr){2-2}
\textbf{Method}& \textbf{Mean(}$\mathbf{\pm}$\textbf{Std. Dev.)} \\
\cmidrule{1-2}

\rowcolor{Gray}
\mBox{white}{honkamj} & 0.049($\pm$0.004) \\

\rowcolor{white}
\mBox{white}{MadeForLife} & 0.567($\pm$0.031) \\

\rowcolor{Gray}
\mBox{white}{LYU1} & 0.651($\pm$0.047)\\

\rowcolor{white}
\mBox{white}{next-gen-nn} & 0.890($\pm$0.072)\\

\rowcolor{Gray}
\mBox{white}{zhuoyuanw210} & 0.628($\pm$0.067)\\

\rowcolor{white}
\mBox{white}{DutchMasters} & 0.341($\pm$0.065)\\

\rowcolor{Gray}
\mBox{white}{lukasf} & 0.725($\pm$0.061) \\

\rowcolor{white}
\mBox{white}{Bailiang} &  1.063($\pm$0.059)\\

\rowcolor{Gray}
\mBox{white}{VROC} & 0.725($\pm$0.061)\\

\rowcolor{white}
\mBox{white}{TimH} & 0.094($\pm$0.024)\\

\rowcolor{Gray}
\mBox{white}{LoRA-FT} & 0.476($\pm$0.049) \\

\cmidrule{1-2}

\rowcolor{white}
\mBox{black}{ANTsSyN} & 0.190($\pm$0.051)\\

\rowcolor{Gray}
\mBox{black}{ANTsSyN$^\dagger$} & 0.057($\pm$0.007)\\

\rowcolor{white}
\mBox{black}{DeedsBCV} & 0.163($\pm$0.022)\\

\rowcolor{Gray}
\mBox{black}{FireANTsGreedy} & 0.793($\pm$0.075)\\

\rowcolor{white}
\mBox{black}{FireANTsSyN} & 0.608($\pm$0.057)\\

\rowcolor{Gray}
\mBox{black}{FireANTsSyN$^\dagger$} & 0.098($\pm$0.003)\\

\rowcolor{white}
\mBox{black}{SynthMorph} & 0.052($\pm$0.005)\\

\rowcolor{Gray}
\mBox{black}{TransMorph} & 0.811($\pm$0.064)\\

\rowcolor{white}
\mBox{black}{uniGradICON} & 0.516($\pm$0.041)\\

\rowcolor{Gray}
\mBox{black}{uniGradICONiso} & 0.331($\pm$0.055)\\

\rowcolor{white}
\mBox{black}{VFA} & 0.679($\pm$0.034)\\

\rowcolor{Gray}
\mBox{black}{VoxelMorph} & 0.602($\pm$0.047)\\

\rowcolor{white}
\mBox{black}{ZeroDisplacement} & -\\

\bottomrule
\end{tabular}
\end{table}

%
%
\subsection{Inverse Consistency.}
\textcolor{black}{Let the deformation field that maps image $A$ to image $B$ be denoted as $\phi_{A\rightarrow B}$.} Inverse consistency~(IC) requires that the inverse transformation $\phi_{B\rightarrow A}$ to satisfy that $\phi_{A\rightarrow B} \circ \phi_{B\rightarrow A}^{-1} = \phi_{B\rightarrow A}^{-1} \circ \phi_{A\rightarrow B} = id$, where id is the identity transformation.
The rationale for the importance of IC is the preservation of anatomical landmarks when moving between the images of two patients; several other benefits and implications of IC that are discussed in greater detail in~\citep{chen2024survey, greer2023inverse}.
Without explicit architectural designs or dedicated loss terms enforcing IC, neural networks typically generate different deformation fields when the order of the input images is swapped, even when the generated deformation fields are diffeomorphic.

\textcolor{black}{To study IC, we computed the mean absolute error~(MAE) between the composed deformation fields~(forwards and backwards) and the identity map.
Specifically, we evaluated the performance of each method in paired subtasks, ``atlas-to-subject'' and ``subject-to-atlas'', across the zero-shot datasets.
These tasks differ only in the order of input images; thus, an inverse-consistent method should produce deformation fields whose composition approximates the identity transformation.
A lower MAE indicates stronger IC, with a value of zero corresponding to perfect consistency.
Table~\ref{tab:inverse_consistency} summarizes these results, with scores averaged across five datasets: ADHD, ADNI1.5T, ADNI3T, NIMH, and UltraCortex.}

In the LUMIR challenge, four evaluated methods explicitly guarantee IC by design: \twBox{honkamj}, \twBox{TimH}, \tbBox{ANTsSyN}, and \tbBox{FireANTsSyN} (see Table~\ref{tab:models}).
Specifically, \twBox{honkamj} and \twBox{TimH} achieve IC through an auxiliary network structure that computes deformation fields by exponentiating the difference of intermediate fields generated with reversed input orders.
In contrast, the optimization-based methods \tbBox{ANTsSyN} and \tbBox{FireANTsSyN} ensure IC by simultaneously optimizing forward and inverse deformation maps in a symmetric manner.
\textcolor{black}{For these optimization-based methods, we report MAE values computed from the forward and inverse maps obtained from the ``subject-to-atlas'' task alone, since both maps can be generated within a single run of the algorithms.
We additionally report MAE values computed from separate runs, where the forward map was obtained from the ``subject-to-atlas'' task and the inverse map from the ``atlas-to-subject'' task with swapped input order.
Because optimization-based methods are sensitive to initialization, IC is not necessarily preserved across separate runs.}

\textcolor{black}{None of the methods that theoretically guarantee IC achieved perfect IC, as indicated by non-zero MAE values. Minor deviations are observed, likely due to interpolation errors introduced during the resampling step. Nevertheless, these methods consistently exhibited the lowest MAE values, with \twBox{honkamj}, \twBox{TimH}, \tbBox{ANTsSyN}, and \tbBox{FireANTsSyN} achieving mean MAEs of 0.049, 0.094, 0.057, and 0.098~\textcolor{black}{mm}, respectively.}

In addition, four methods encourage IC through the GradICON regularizer~\citep{tian2023cvpr}: \twBox{LoRA-FT}, \twBox{DutchMasters}, \tbBox{uniGradICON}, and \tbBox{uniGradICONiso}.
However, as noted in~\citep{tian2023cvpr}, GradICON represents a relaxed version of IC regularization, penalizing derivatives of deformation rather than strictly enforcing inverse consistency.
\textcolor{black}{Nonetheless, as observed in Table~\ref{tab:inverse_consistency}, their MAE values typically fall within the 0.3--0.5 mm range, which is the lowest among methods that do not explicitly enforce IC, while other methods without such regularization generally exhibit MAE values above 0.5 mm.}

\textcolor{black}{Interestingly, the best-performing method in terms of IC is \tbBox{SynthMorph}, achieving an MAE of 0.052.
This improvement is consistent with updates in the latest \texttt{SynthMorph} model, which explicitly enhanced inverse consistency, as noted in its Docker release documentation~\href{https://hub.docker.com/r/freesurfer/synthmorph}{\faDocker}.}

\begin{table}[H]
\caption{The Pearson correlation coefficient~(Correlation) and slope of a linear fit between the mean Dice Similarity Coefficient and the mean Total Registration Error on the Challenge Test dataset.}
\label{tab:correlation}
\centering
\begin{tabular}{rcc}
\toprule
\textbf{Method} & \textbf{Correlation} & \textbf{Slope}\\
\cmidrule{1-3}

\rowcolor{Gray}
\mBox{white}{honkamj} & -0.072 & -2.180\\

\rowcolor{white}
\mBox{white}{MadeForLife} & -0.078 & -2.274\\

\rowcolor{Gray}
\mBox{white}{LYU1} & -0.073 & -2.236\\

\rowcolor{white}
\mBox{white}{next-gen-nn} & -0.089 & -2.568\\

\rowcolor{Gray}
\mBox{white}{zhuoyuanw210} & -0.094 & -2.856\\

\rowcolor{white}
\mBox{white}{DutchMasters} & -0.054 & -1.361\\

\rowcolor{Gray}
\mBox{white}{lukasf} & -0.085 & -2.312\\

\rowcolor{white}
\mBox{white}{Bailiang} & -0.080 & -2.252\\

\rowcolor{Gray}
\mBox{white}{VROC} & -0.151 & -4.287\\

\rowcolor{white}
\mBox{white}{TimH} & -0.125 & -4.085\\

\rowcolor{Gray}
\mBox{white}{LoRA-FT} & -0.132 & -3.113\\

\cmidrule{1-3}

\rowcolor{white}
\mBox{black}{ANTsSyN} & -0.347 & -6.815\\

\rowcolor{Gray}
\mBox{black}{DeedsBCV} & -0.177 & -5.994\\

\rowcolor{white}
\mBox{black}{FireANTsGreedy} & -0.135 & -4.016\\

\rowcolor{Gray}
\mBox{black}{FireANTsSyN} & -0.101 & -2.876\\

\rowcolor{white}
\mBox{black}{SynthMorph} & -0.167 & -5.428\\

\rowcolor{Gray}
\mBox{black}{TransMorph} & -0.081 & -2.291\\

\rowcolor{white}
\mBox{black}{uniGradICON} & -0.126 & -3.061\\

\rowcolor{Gray}
\mBox{black}{uniGradICONiso} & -0.107 & -2.847\\

\rowcolor{white}
\mBox{black}{VFA} & -0.089 & -2.721\\

\rowcolor{Gray}
\mBox{black}{VoxelMorph} & -0.208 & -5.246\\

\rowcolor{white}
\mBox{black}{ZeroDisplacement} & -0.377 & -9.926\\

\bottomrule
\end{tabular}

\end{table}

\subsection{\textcolor{black}{Correlation between DSC and TRE}}
\label{sec:dsc_tre}
Torsten Rohlfing's 2012 paper~\citep{rohlfing2012tmi} makes a compelling argument for not relying on tissue label overlap to evaluate registration accuracy.
Given the broad range of registration methods available to us as organizers and the presence of both labels and landmark data in our Test Set, we have the opportunity to return to Rohlfing's premise and see if we can offer any further insight.
We choose to see if there is any correlation between DSC and TRE on the challenge test set.
We computed the Pearson correlation coefficient between the mean DSC and the mean TRE across our challenge test set, as the slope of the linear fit.
The correlation and slope are reported in Table~\ref{tab:correlation}.

The slopes, in Table~\ref{tab:correlation}, are all negative, which is expected---as DSC improves, it goes up and TRE goes down, and vice versa for worsening results.
However, the correlations are poor, indicating that the relationship between DSC and TRE is not particularly strong.
This is not particularly surprising; first the landmarks are sparse, and the labeled structures run the gambit of small to large.
Second, the landmarks are not in any reasonable correspondence with the available labels.
As such, this simple analysis has informed us about how to better evaluate the relationship between DSC and TRE, if any such relationship exists.
Moving forward with future challenges, we anticipate using several dedicated landmarks that are readily focused on specific structures, with the landmarks capturing the important boundary positions of the structures under investigation.

\subsection{Deformation Regularity.}
Deformation regularization plays a crucial role in ensuring that estimated transformations are anatomically plausible.
Without appropriate constraints, deformable registration algorithms may produce biologically unrealistic mappings, such as foldings, even when achieving high image similarity.
While the assumption of smooth, invertible deformations may not hold in cases of severe brain pathology, it is well-justified in the context of this challenge, which focuses primarily on healthy subjects and individuals with neurodegenerative conditions such as Alzheimer’s disease and mild cognitive impairment.
In these populations, overall brain topology is largely preserved, and the transformations used to register them are expected to be diffeomorphic.

To quantify deformation regularity, we \textcolor{black}{compute the percentage of non-diffeomorphic volume (\%NDV)~\citep{liu2024ijcv} for each estimated transformation, defined as the percentage of folded volume within the total brain volume.} 
\textcolor{black}{A zero \%NDV confirms the absence of folding, but does not imply that the registration is accurate.
Conversely, a non-zero \%NDV provides direct evidence of anatomically implausible behavior in the affected region(s).
However, the practical interpretation of \%NDV depends not only on whether it is non-zero, but also on the extent of the folded volume.
Therefore, \%NDV should not be interpreted as a purely binary measure in which any non-zero value is equally problematic.
Although non-zero folded volume is undesirable in principle, prior literature does not establish a universal threshold for when such behavior becomes practically unacceptable; rather, acceptable tolerance depends on the intended application and the extent of the affected volume. For example, in radiotherapy applications~\citep{nenoff2023review}, suggested tolerances for deformation smoothness at locations with negative Jacobian have been discussed on the order of 0.2\% to 2.0\%, highlighting the application-dependent nature of such criteria.}

\textcolor{black}{Across the challenge test set and the zero-shot datasets, the observed \%NDV values fall into several practical regimes.
First, some methods consistently produced exactly diffeomorphic transformations in practice, including the deep learning-based \twBox{TimH} and the optimization based \tbBox{ANTsSyN}, both of which achieved \%NDV=0 across all evaluated cases.
Second, several methods yielded extremely small non-zero \%NDV values, including \tbBox{FireANTsSyN}, \tbBox{SynthMorph}, \tbBox{uniGradICON}, \tbBox{DeedsBCV}, and \twBox{next-gen-nn}; these values are effectively negligible in practice and indicate only minimal local folding.
Third, a group of methods, including \twBox{honkamj}, \twBox{DutchMasters}, \twBox{zhuoyuanw210}, \twBox{MadeForLife}, \twBox{LYU1}, and \twBox{Bailiang}, exhibited small but measurable \%NDV values, indicating that folding was present but remained spatially limited.
In contrast, methods such as \twBox{lukasf}, \twBox{VROC}, \tbBox{VFA}, \tbBox{TransMorph}, and especially \tbBox{VoxelMorph}, produced substantially larger \%NDV values, corresponding to more extensive non-diffeomorphic volumes.
Accordingly, the NDV rankings should be interpreted with care, especially among methods whose \%NDV values are all very small and close to zero, as differences in numerical rank in this regime may correspond to only minor practical differences in the extent of folded volume.}

\textcolor{black}{To aid interpretation, we relate representative \%NDV values to absolute folded volume.
For a typical skull-stripped brain volume of 1,886.54~cc in 1$\times$1$\times$1~mm$^3$ MNI space, a \%NDV of 0.01\% corresponds to approximately 0.19~cc of non-diffeomorphic volume, whereas a \%NDV of 1\% corresponds to approximately 18.86~cc.
Thus, although several methods yielded non-zero \%NDV values, the practical extent of folding varies substantially across methods.
Very small non-zero values indicate limited local folding that is negligible relative to the full brain volume, whereas larger values reflect increasingly extensive non-diffeomorphic regions that are more likely to compromise transformation plausibility.}

\textcolor{black}{Across the test and zero-shot datasets, \%NDV $\geq 1\%$ was exceedingly rare.
The only method reaching this regime was \tbBox{VoxelMorph}.
This suggests that, for nearly all evaluated methods, folding was either absent or spatially limited, even when explicit diffeomorphic guarantees were not built into the formulation.
Additionally, methods that incorporate inverse-consistency constraints, as discussed in the following paragraph, generally performed favorably in terms of deformation regularity.
While deformation regularization has traditionally been viewed as being in tension with registration accuracy, the top-performing methods on each dataset were not those with the highest \%NDV.
This indicates that, with thoughtful architectural or algorithmic design, it is possible to simultaneously achieve strong registration accuracy and maintain highly regular deformations.}

\subsection{Use of automated label maps}
An additional criticism of \textcolor{black}{the benchmark design} is our use of automatically generated anatomical label maps to enable large-scale benchmarking of all the submitted results.
We believe that this approach makes systematic and reproducible evaluation across thousands of scans feasible, however {we acknowledge that it also introduces the possibility that segmentation inaccuracies may influence Dice- and HD95-based scores.}
As \citet{rohlfing2012tmi} notes, tissue overlap measures are an imperfect surrogate for registration validation.
To mitigate these limitations, we used fine-grained parcellation of the brain into 133 cortical and subcortical structures to reduces the gap between Dice and registration performance.
Even if an automatic tool introduces errors, consistent errors across images can still support valid registration assessment, since overlapping mislabeled regions continue to reflect alignment accuracy.
SLANT has been shown to be robust in this regard.
We further complemented segmentation-based metrics with landmark-based TRE, providing an orthogonal evaluation that is independent of the segmentation quality.

\subsection{\textcolor{black}{Interpreting Rankings and Practical Significance}}
\textcolor{black}{We pool DSC, TRE, and HD95 to evaluate the overall robustness of each method. The inclusion of HD95 is particularly important because it captures extreme spatial outliers and penalizes localized boundary errors that may not be reflected by a high overall DSC. Although the combined ranking provides a useful summary of general-purpose performance, many of the top-performing methods achieved very similar metric values. Therefore, small differences in rank should be interpreted together with the underlying metric magnitudes and should not necessarily be taken to imply large practical differences. Method selection should ultimately be guided by the intended downstream clinical application. For instance, applications such as radiation oncology or surgical planning may place greater emphasis on boundary accuracy and spatial correspondence, making HD95 and TRE especially relevant, whereas general volumetric estimation may place greater emphasis on DSC. It is also important to distinguish statistical significance in the rankings from practical significance in clinical settings. A statistically significant difference in TRE may have limited practical relevance if both errors are substantially smaller than the image voxel size. In contrast, differences in HD95 or DSC30 may be practically meaningful when they separate clinically acceptable performance from substantial failure on challenging cases.}

\section{Conclusion}
\label{s:conclusion}
The LUMIR challenge has highlighted the transformative impact of deep learning on brain MRI image registration, by presenting both promising insights and remaining challenges. 
The LUMIR challenge is characterized by two major focuses: large-scale dataset and unsupervised nature.
By incorporating datasets with substantial distributional variations relative to the training data, the challenge establishes a rigorous benchmark for assessing model robustness and adaptability. 
Additionally, by deliberately excluding label maps from training, the challenge encourages the development of models that inherently learn spatial correspondences and adapt to diverse data conditions without relying on predefined labels, further promoting the creation of generalizable and scalable registration solutions. 
This framework also fosters innovation in the development of foundational models that can operate effectively in environments where annotated data is scarce.

From our comprehensive experiments and comparisons, we found that deep learning-based registration methods significantly outperform traditional optimization-based approaches in inter-subject and atlas-based brain MRI registration.
Sec.~\ref{s:discussion} provided additional insights into the promising advancements and remaining challenges of various deformable registration approaches. 
Our evaluation of zero-shot registration revealed that deep learning models consistently outperform traditional optimization-based techniques, even in scenarios with substantial domain shifts. 
\textcolor{black}{Furthermore, our results suggest an empirical trend in which architectural choices, such as dual-stream encoders, are associated with more effective registration performance; however, without controlled ablation studies it is not possible to establish whether this architectural design is the causal factor, or whether other co-occurring design decisions (e.g., training strategies, loss functions, data augmentation) play an equally or more important role.
Achieving perfect inverse consistency also remains an open challenge: even methods that encourage inverse consistency by construction exhibited small but non-zero deviations. 
Finally, while instance-specific optimization can enhance performance in certain settings, the top-ranked single-step approaches in this benchmark achieved competitive accuracy with substantially lower runtime, suggesting---though not conclusively proving---that well-designed single-step approaches can deliver a favorable trade-off between accuracy and computational cost.}
Collectively, these insights from the LUMIR challenge advance our understanding of medical image registration and guide future research directions in model development and optimization.

In conclusion, the LUMIR challenge embodies the ongoing advances in medical image registration.
As the field progresses, we believe LUMIR will play an important role in shaping the future of medical image registration, ultimately unlocking new possibilities for clinical and research advancements.

\renewcommand\thefigure{A\arabic{figure}}
\setcounter{figure}{0}
\renewcommand\thetable{A\arabic{table}}
\setcounter{table}{0}

\appendix
\section{Methods}
\label{a:methods}
\subsection{Submissions}
\textcolor{black}{The challenge submissions are listed in the order in which they ranked, see Table~\ref{t:testset} for details.}

\desc{white}{honkamj}{Symmetric, inverse consistent, and topology preserving by construction}{Joel Honkamaa \& Pekka Marttinen~\citep{honkamaa2024melba}}

\twBox{honkamj} is a novel multi-resolution deep learning registration architecture which is inverse consistent, symmetric, and preserves topology by construction.
These properties are fulfilled for the whole multi-resolution pipeline and not just separately at each resolution.

\twBox{honkamj} is designed to be cycle-consistent.
\textcolor{black}{This means that the deformation map between two images $A$ and $B$ is denoted by $\phi(A, B)$, and we want the composition of $\phi(A, B)$ and $\phi(B, A)$ to be the identity.
The authors achieve this through an auxiliary network $u$ that predicts deformations and defines $\phi$ as $\phi(A, B) = \phi_u(A, B) \circ \phi_u(B, A)^{-1}$.
By construction, $\phi(A, B) = \phi(B, A)^{-1}$, up to numerical precision.}
This approach of using an intermediate network and creating the desired deformation field through an inverse consistent construction has been previously proposed by Estienne~\textit{et al.}~\citep{estienne2021arxiv}.
Inverse consistency is not guaranteed by Estienne~\textit{et al.} as they enforce it via a loss term, whereas the formulation of \twBox{honkamj} is symmetric by construction.
To implement this inverse consistent approach effectively, the authors encode features of the inputs separately before feeding them to the deformation extraction network $\phi$.

The approach is implemented in a multi-scale manner; from the lowest resolution, features are extracted and used to create the coarsest deformation field.
Then the approach recursively builds deformations between the inputs at each level using the extracted features.
To ensure symmetry, two deformations are built: one deforming the first image half-way towards the second image, and the other deforming the second image half-way towards the first image.
The final deformation is the composition of these two deformations at the highest resolution level.
For more details, see Honkamaa and Marttinen~\citep{honkamaa2024melba}; we note that the model trained for the challenge used the group consistency loss over image triplets~\citep{gu2020miccai}.

\desc{white}{MadeForLife}{Dual-stream encoders with a specialized decoder for coarse-to-fine pyramid and progressive registration}{Yichao Zhou \& Zuopeng Tan}

The authors use the GroupMorph~\citep{tan2024tmi} structure for their approach.
Using a dual encoder with shared weights to extract multilevel features for both the fixed and moving images.
The encoder adopts a structure similar to that of \tbBox{VoxelMorph}~\citep{balakrishnan2019tmi}.
The features extracted by the dual encoder are fed into grouping update modules~(GUMs) for the production of deformation fields and contextual features, where the contextual features provide explicit prior information and intergroup correspondence for the generation of the deformation field.
The GUMs consist of two parts: the grouping module~(GM), which calculates feature correlation and generates deformation fields, and the contextual fusion module~(CFM), which fuses contextual features to facilitate inter-group communication for the grouping field estimator.

In summary, GMs use a different number of convolutional layers to compute multiple deformation subfields per level, with these subfields having different receptive fields.
The intralevel multiscale fields, along with the interlevel multiscale fields, form the final deformation fields.
Compared to the manner of predicting one field per level, the finer grained decomposing can further promote model optimization, thereby obtaining an optimal alignment.
The CFMs facilitate collaboration among subfields within each group, the CFMs fuse the contextual features from adjacent groups, thus their intergroup communication is strengthened.
This enables every group to have access to explicit prior information from the other groups while estimating the deformation fields.

\desc{white}{LYU1}{Pyramid network with auxiliary decoder}{Hongchao Zhou \& Shunbo Hu}

\twBox{LYU1} proposes a shared encoder, a shared auxiliary decoder, and a fusion pyramid decoder.
The fusion pyramid decoder includes five scales.
Except for the smallest scale, each scale is constructed from a multi-scale feature fusion block~(MSFB).
The pyramid decoder first merges the smallest-scale low-level features and predicts a coarse deformation field.
This deformation field is then upsampled and applied to the moving feature map at the current scale.
The moving feature map, the fixed feature map, and the high-level moving and fixed features from the shared auxiliary decoder at the same scale are input into the MSFB.
This process removes redundant information and produces fused features, predicting the deformation field at the current scale.
This process is repeated at each scale, with the final refined deformation field obtained by iteratively combining the coarse deformation fields from each scale.

\desc{white}{next-gen-nn}{Robust backbone network}{Yuxi Zhang \& Hang Zhang}

\twBox{next-gen-nn} uses a robust backbone network based on Zhang \textit{et al.}~\citep{zhang2024miccai}.
Their network architecture comprises an encoder with a series of convolutional layers and a cascaded mechanism designed to refine the prediction of deformation fields.
Specifically, \twBox{next-gen-nn} uses modifications to the base network, including a co-attention mechanism~\citep{chen2021arxiv}, large kernel convolutions~\citep{jia2022mlmi}, and an increased number of convolutional channels in the prediction stage of the deformation field.
Additionally, \twBox{next-gen-nn} uses an image-guided bilateral filtering technique~\citep{wagner2022sr}, to refine the predicted registration results.

\desc{white}{zhuoyuanw210}{Dual stream pyramid encoder and motion decomposition transformer}{Zhuoyuan Wang \& Yi Wang}

This method employs a dual-stream pyramid encoder and a GPU-accelerated motion decomposition Transformer~(ModeTv2)~\citep{wang2024modetv2} decoder.
The dual-stream pyramid encoder extracts features from both the moving image and the fixed image, generating corresponding feature hierarchies for each image.
The feature hierarchies are used from coarse to fine to generate deformations between the moving and fixed image at each scale level.

\desc{white}{DutchMasters}{Gradient projection optimization}{Yi Zhang \& Qian Tao~\citep{zhang2024arxiv}}

\twBox{DutchMasters} introduces a novel instance optimization~(IO) algorithm that addresses the multi-objective optimization~(MOO) challenge that is inherent to image registration.
The approach applies gradient projection techniques, inspired by advances in MOO~\citep{yu2020gradient}, to enhance
the effectiveness of IO.
IO effectively combines the generalization capabilities of deep learning with the fine-tuning advantages of instance-specific optimization.
The authors framework uses gradient projection to mitigate conflicting updates in the MOO.
The technique projects conflicting gradients into a common space, better aligning the dual objectives and enhancing optimization stability.
The authors consider the components of the loss being optimized (i.e. similarity and regularization) as components in a MOO.
The multiple losses can sometimes be in conflict with each other, affecting the optimization which is reflected in the direction and magnitude of the loss gradients.
\twBox{DutchMasters} address this by using the gradient projection technique from multi-task learning~\citep{yu2020gradient}.
The conflicting direction of the loss gradients is corrected by randomly projecting one gradient to the normal space of the other.
Conflicting gradients are identified by evaluating the cosine similarity between two gradients.
Unlike mini-batch calculation~\citep{yu2020gradient}, the authors focus on instance optimization, thus reducing the scope from multi-task learning to multi-objective optimization.
If the gradients are not conflicting, the update will remain unchanged compared to gradient-based methods.
If, however, the gradients are in conflict, one of the gradients will be randomly projected to the normal space of the other, and that gradient will be used for the update.
For complete details, see Zhang~\emph{et al.}~\citep{zhang2024arxiv}.

\desc{white}{lukasf}{TransMorph refinement}{Lukas Förner \& Thomas Wendler}

The authors propose a fine-tuning of a pre-trained \tbBox{TransMorph}~\citep{chen2022mia} model to improve convergence stability and deformation smoothness.
Convergence stability is achieved through the Fisher Adam~(FAdam)~\citep{hwang2024arxiv} optimizer, and the deformation smoothness comes from the addition of gradient correlation~\citep{hiasa2018sashimi} in the similarity measure, which improves anatomical alignment.
%
%

The authors use the FAdam optimizer to improve optimization during training.
FAdam is a variant of the Adam optimizer that incorporates principles from natural gradient descent and Riemannian geometry to improve convergence and stability.
The key modifications in FAdam include adjustments to momentum, bias correction, adaptive gradient scaling, and a refined weight decay that respects the geometry of the parameter space.
\twBox{lukasf} also incorporates the gradient correlation~\citep{hiasa2018sashimi} in their loss computation along with the local normalized cross-correlation, as part of the similarity loss between the fixed and moving images.

\desc{white}{Bailiang}{Dual-stream encoders, coarse-to-fine pyramid, correlation layers, and progressive registration}{Bailiang Jian \& Benedikt Wiestler~\citep{jian2024mamba}}

\twBox{Bailiang} introduces a ConvNet architecture designed with registration-specific components.
It employs dual-stream encoders to independently extract image features from moving and fixed images, followed by a decoder that integrates these features with a coarse-to-fine optimization pyramid to progressively refine deformations across scales.
The network also employs progressive registration to iteratively compose deformation fields to enable large deformations, while correlation layers establish voxel-wise correspondences.

\desc{white}{VROC}{Variational Registration on Crack}{Frederic Madesta \& Thilo Sentker}

The Variational Registration on Crack~(\twBox{VROC}) toolbox employs a classic registration approach using task-specific hyperparameters and image preprocessing steps to improve performance.
For the LUMIR Challenge the pre-processing includes Gaussian smoothing with a $\sigma$ of $1$ to reduce the noise.
Intensity-based label maps are generated using a multi-Otsu thresholding with seven classes.
A registration mask is created by taking the union of the fixed and moving images where pixel values are greater than zero, ensuring that only relevant regions are considered during the registration process.
Deformable registration is performed for the first time using NCC forces applied to the fixed and moving label maps based on thresholds.
This is followed by fine-tuning the initial registration result with Demons forces~\citep{thirion1998mia} (on the original intensity-based images) to achieve precise alignment and handle more complex deformations.

\desc{white}{TimH}{ConstrICON framework multistep inverse consistency}{Tim Hable}

This approach addresses the challenge of unsupervised registration of anatomical structures in brain MRI by using the ConstrICON framework~\citep{greer2023miccai}, which is used to build an inverse consistent multistep registration model.
The approach employs residual U-Nets and applies additional smoothing to the stationary velocity fields to reduce the inverse consistency error.

The approach leverages the \textit{TwoStepConsistent}~(\texttt{TSC}) operator from the ConstrICON framework.
In contrast to other approaches, that utilize penalty terms during training to approximate inverse
consistency, ConstrICON achieves this property intrinsically through its architectural design and by parameterizing the output transformations of the neural networks by a Lie group.
Parameterizing the output of a neural network by a Lie group is assumed to be sufficient, since many types of transformations that are used in medical image registration, such as affine or diffeomorphic transforms, are Lie groups.
The \texttt{TSC} operator preserves inverse consistency by construction---see Greer \textit{et al.}~\citep{greer2023miccai} for complete details---and can be recursively applied, thereby extending the procedure from a two-step registration to an $N$-step approach that remains inverse consistent.
The proposed model uses three residual U-Nets, denoted as $\tilde{\Xi}$, $\tilde{\Phi}$, and $\Psi$, each producing an intermediate transformation to align the moving image $m$ to the fixed image $f$. These networks are integrated into a three-step architecture of the form \texttt{TSC}\{$\Xi$, \texttt{TSC}\{$\Phi,\Psi$\}\}. Here, \(\Xi\) and \(\Phi\) are defined by composing the outputs of \(\tilde{\Xi}\) and \(\tilde{\Phi}\) with themselves, e.g., $\Phi(m,f)=\tilde{\Phi}(m,f)\circ\tilde{\Phi}(m,f)$.
This ensures that \(\tilde{\Xi}\) and \(\tilde{\Phi}\) each learn only the square root of the corresponding transformation required by the \texttt{TSC} formulation. In contrast, \(\Psi\) directly predicts a full intermediate transformation.
The stationary velocity field is smoothed via average pooling (kernel size 3, applied twice), which empirically reduces inverse consistency error. Each registration network also includes selective Jacobian determinant regularization~\citep{mok2020cvpr} to mitigate folding.
The final output of the combined registration networks is regularized using bending energy loss~\citep{johnson2001landmark} and selective Jacobian determinant regularization loss, with \textcolor{black}{LNCC} as the similarity metric.

\desc{white}{LoRA-FT}{Low-rank adaptation of foundation models}{Jin Kim \& Matthew S. Brown}

Adapting \tbBox{uniGradICON}~\citep{tian2024miccai} foundation models to new domains or datasets typically requires extensive fine-tuning of millions of parameters, which is computationally expensive and can lead to catastrophic forgetting.
\twBox{LoRA-FT} addresses this limitation by using Low-rank Adaptation~(LoRA)~\citep{hu2021lora} to achieve parameter-efficient adaptation.
Rather than fine-tuning all network parameters, \twBox{LoRA-FT} decomposes weight updates into low-rank matrices, reducing the number of trainable parameters by orders of magnitude while maintaining registration performance.
The approach enables rapid adaptation of the foundation model to any dataset without compromising its general registration capabilities or requiring direct supervision of anatomical label maps.

\subsection{Baseline Methods}
\label{as:baselines}
For each of the Baseline methods we also provide a three line summary: the first line includes a colored square (that is used in plots and figures for quick reference) and the name of the method; second is a one line summary of the method; and finally in parentheses is the reference for the method. 
Following the one line summaries is a brief overview of the respective methods and Table~\ref{tab:models} includes additional details about the methods.
\textcolor{black}{All baseline deep learning based methods, except the pretrained foundation registration models (i.e., \tbBox{SynthMorph} and \tbBox{uniGradICON}/\tbBox{uniGradICONiso}), were trained on the LUMIR training data, which was brain-specific.}

\desc{black}{ANTsSyN}{Optimization-based, symmetric, time-varying diffeomorphic registration}
{\citet{avants2007miccai, avants2008mia, avants2018ants}}

Originally proposed circa 2007~\citep{avants2007miccai, avants2008mia}, the symmetric image normalization method~(SyN), now packaged as part of the Advanced Normalization Tools~(ANTs), is a traditional registration framework that maximizes the cross-correlation~(CC) within the space of diffeomorphic maps using Euler-Lagrange equations during optimization.

\textcolor{black}{\tbBox{ANTsSyN} uses a symmetric, diffeomorphic formulation in which the local normalized CC similarity metric is optimized jointly in the forward and inverse directions. The symmetric energy leads to Euler–Lagrange update equations that ensure inverse‐consistent registration.}
It works by iteratively applying transformations to align the images in a symmetric manner, thus optimizing the transformation in both directions simultaneously.
The symmetry used in \tbBox{ANTsSyN} ensures that the transformation is optimal in both directions, which removes bias from the directionality of the transformation.
ANTsSyN coming from the pre-deep learning era can be computationally expensive, which is often reported as a criticism despite its continued good results on a wide range of imaging data and organs~\citep{klein2009ni, murphy2011tmi, tustison2011jmri}.

\desc{black}{deedsBCV}{Optimization-based, discrete registration}
{\citet{heinrich2013mrf}}

\tbBox{deedsBCV} was designed to address three outstanding issues of lung CT registration: large motion of small features, sliding motions between organs, and changes in image contrast due to compression.
The approach takes advantage of Markov random field~(MRF)-based discrete optimization strategies while also keeping computational complexity tractable without reducing the registration accuracy.

\tbBox{deedsBCV} made three distinct contributions, first the use of an image-derived minimum spanning tree as a simplified graph structure to handle the complexity of sliding motion and allowed for the estimation of the global optimum very efficiently.
Secondly, a stochastic sampling approach for the similarity cost between images is introduced within a symmetric, diffeomorphic B-spline transformation model with diffusion regularization.
This helps reduce the complexity of the computation by orders of magnitude and enables the minimization of much larger label spaces.
Thirdly, the addition of proxy-labels, dubbed hyper-labels, are introduced which capture the local intensity variations arising within this task.
\tbBox{deedsBCV} was originally designed for lung CT registration; however, it has been shown to be quite adapt at a broad range of tasks~\citep{hering2023tmi}.

\doubleDesc{black}{FireANTsGreedy}{black}{FireANTsSyN}{Optimization-based, GPU-accelerated, time-varying diffeomorphic registration}
{\citet{jena2024fireants}}

Traditional registration algorithms such as ANTsSyN~\citep{avants2007miccai, avants2008mia} have endured over the years due to their precision, reliability, and robustness in a wide spectrum of modalities and acquisition settings~\textcolor{black}{\citep{klein2009ni, murphy2011tmi, tustison2011jmri}}.
However, these algorithms converge slowly, are prohibitively expensive to run, and their usage requires a steep learning curve, limiting their scalability to larger clinical and scientific studies.
The authors of the \texttt{FireANTs} approach have developed a multiscale adaptive Riemannian optimization algorithm for diffeomorphic image registration.
Their framework leverages mathematical correspondences to avoid expensive operations like the Riemannian metric tensor, and parallel transport of the optimization state, which are needed for implementing first-order adaptive algorithms.
\texttt{FireANTs} can compose transformations, avoiding resampling artifacts across transformations.
All of this with the aim of improving scalability in image registration.

\desc{black}{SynthMorph}{Learning contrast-agnostic registration without acquired images}
{\citet{hoffmann2021synthmorph}}

\tbBox{SynthMorph} is a general strategy for learning contrast-agnostic registration.
\tbBox{SynthMorph} enables the registration of real images within and between contrasts, learning only from synthetic data.
During training, \tbBox{SynthMorph} synthesizes images from label maps and trains the registration framework to align these synthetic images.
The approach uses a generative model to provide random label maps of variable geometric shapes.
Conditioned on these generated label maps, they synthesize images with arbitrary contrasts, deformations, and artifacts.
\tbBox{SynthMorph} can then train with a contrast-agnostic loss that measures label overlap, rather than an image-based loss.

\desc{black}{TransMorph}{Vision Transformer-based registration model}
{\citet{chen2022mia}}

\tbBox{TransMorph} is a deep learning-based image registration framework that incorporates a Transformer-based architecture for both affine and deformable registration.
Its encoder consists of a three-stage Swin Transformer~\citep{liu2021swin}, which extracts hierarchical features from the input images. These features are then passed to a decoder that progressively refines the deformation field, composing each new estimate with the previously generated field~\citep{chen2022unsupervised}.
\tbBox{TransMorph} is trained using an unsupervised learning paradigm similar to that of \tbBox{VoxelMorph}, with both fully unsupervised and semi-supervised variants (leveraging anatomical label maps during training) described in the literature.

\doubleDesc{black}{uniGradICON}{black}{uniGradICONiso}{Inverse consistent, learning-based foundation registration model}
{\citet{tian2024miccai}}

The authors propose to train a universal foundation model for registration.
A hurdle to developing a foundation model for registration is the obstacle of training a universal registration network over many different datasets to obtain such a model.
It is incredibly difficult to pick a set of registration hyperparameters (for regularizers and similarity measures) that provide good performance across a wide-range of registration tasks.
The issue of different registration tasks requiring different hyperparameters is obviously in direct conflict with a universal registration network being trained with a fixed set of hyperparameters.
However, this is made possible by replacing conventional regularizers with gradient inverse consistency~(\textit{GradICON}) regularization~\citep{tian2023cvpr}.
\textit{GradICON} acts as a weak regularizer that only encourages invertibility of the transform.
This weak constraint, allows the network to discover what transformations are supported by the data and
thereby facilitates training with the same hyperparameters across different datasets.
The uniGradICONiso version incorporates instance-specific optimization, iteratively fine-tuning the network parameters for 50 iterations for each image pair.

\desc{black}{VFA}{Dual-stream encoders, vector field cross-attention decoder, multi-scale registration model}
{\citet{liu2024vector}}

Vector field attention~(\tbBox{VFA}) is an extension of the previously published \textit{im2grid}~\citep{liu2022mmmi}.
VFA uses neural networks to extract multiresolution feature maps from the fixed and moving images and then retrieves pixel-level correspondences based on
feature similarity.
Retrieval of the correspondences is achieved with a novel attention module without the need for learnable parameters.
VFA is trained end-to-end in either a supervised or unsupervised manner.

\desc{black}{VoxelMorph}{U-Net-based registration model}
{\citet{balakrishnan2019voxelmorph}}

\tbBox{VoxelMorph} is a fast learning-based framework for deformable, pairwise medical image registration.
In contrast to traditional registration methods, \tbBox{VoxelMorph} is formulated as a function that maps an input image pair to a deformation field that aligns the input images.
The underlying network is parameterized by CNNs. Both supervised and unsupervised versions have been reported in the literature.

\section{Results}
\subsection{Detailed Results on Zero-shot Evaluation}
\label{sec:detailed_results}
\subsubsection{Diverse T1-w MRI Data Sources}
\label{sec:ood_t1w}
%
The first and smallest step we can take in establishing if these methods are approaching foundation model ability is to see how they behave on a T1-w dataset that is distinct from the available T1-w training data made available for the Challenge.
In essence, this is the first (and easiest) obstacle that these methods have to overcome to be considered foundational.
To assess this, we compare the results on the Challenge test set with T1-w inter-subject results from the NIMH dataset.
These results are shown in Tables~\ref{t:testset} and~\ref{tab:nimh_is_t1w}, specifically by looking at the comparable columns of DSC, HD95, and also DSC30.
The top four methods for the Challenge test set are \twBox{honkamj}~(1\textsuperscript{st}), \twBox{MadeForLife}~(2\textsuperscript{nd}, \twBox{next-gen-nn}~(3\textsuperscript{rd}), and \tbBox{VFA}~(4\textsuperscript{th}).
These same four methods are also the top four methods on the T1-w NIMH inter-subject data with the order changing slightly: \twBox{honkamj}~(1\textsuperscript{st}), \tbBox{VFA}~(2\textsuperscript{nd}), \twBox{MadeForLife}~(3\textsuperscript{rd}), and \twBox{next-gen-nn}~(4\textsuperscript{th}).
For all four methods, their DSC values are almost identical between the two datasets, the biggest change being \twBox{MadeForLife} with a difference of $0.003$ between the mean DSCs for the two cohorts.
For HD95, \twBox{honkamj}, \twBox{MadeForLife}, \twBox{next-gen-nn}, and \tbBox{VFA}, all experienced an almost identical drop in performance of approximately $0.3$~mm.
Across nearly all the methods, there is stability in the DSC and HD95 between the two cohorts of T1-w images.
With respect to the DSC, the biggest changes between the two cohorts occur with \tbBox{ANTsSyN} and \tbBox{VoxelMorph}, with changes of approximately $0.04$ and $0.01$, respectively.
With three methods seeing a change of approximately $0.007$ (\twBox{LoRA-FT}, \twBox{DutchMasters}, and \tbBox{uniGradICON}), with other methods seeing a change of $0.003$ or less.
For HD95 the change for all methods, except \tbBox{ANTsSyN}, is very uniform with an approximate change of $0.3$~mm; with \tbBox{ANTsSyN} having a change of $0.5$~mm.
However, the most interesting observation from these results is that while the DSC results saw a drop-off in performance, the HD95 saw a uniform improvement, which is a most surprising finding.

\subsubsection{Pathological Brain: ADHD}
\label{sec:adhd}
Next, we evaluate the methods on the same T1-w contrast in which they were trained but in distinct cohorts, with attention deficit hyperactivity disorder~(ADHD) in this section and subjects from the Alzheimer's Disease Neuroimaging Initiative~(ADNI) in the next.
Although ADHD typically does not cause distinct anatomical changes that can be clearly observed at the individual level in structural MRI scans, we included this dataset because it focuses on a specific age group, namely children aged 8 to 12, whose brain anatomy can differ substantially from that of adults.
Although the training dataset does include subjects within this age range, as shown in Sec.~\ref{sec:challenge_set}, they represent a relatively small portion of the overall data. 
This setup allows us to assess whether the registration models can perform well on this specific subset, despite the imbalance in age representation during training.

The detailed scores for inter-subject, atlas-to-subject, and subject-to-atlas registration are shown in Tables~\ref{t:ADHD_is}, \ref{t:ADHD_a2s}, and~\ref{t:ADHD_s2a}, respectively.
For inter-subject registration, the top three performing methods are \twBox{honkamj}, \tbBox{VFA}, and \twBox{MadeForLife}, all of which are deep learning-based. Compared to the challenge test set, we observed a slight degradation in DSC and HD95 performance, with DSC decreasing by less than 0.01 and HD95 by less than 0.05~\textcolor{black}{mm}. 
This suggests potential influences from dataset shift, as well as the focus on a younger age group.
Interestingly, \tbBox{ANTsSyN} performed notably better on this dataset than on the challenge test set, showing an improvement of 0.031~(4\%) in DSC and 0.205~\textcolor{black}{mm}~(6\%) in HD95.
Despite this improvement, it still did not outperform deep learning-based methods and was ranked 15\textsuperscript{th} overall.

For atlas-based registration tasks, the top three performing methods are \twBox{honkamj}, \twBox{DutchMasters}, and \twBox{zhuoyuanw210} for atlas-to-subject registration, and \twBox{Bailiang}, \twBox{next-gen-nn}, and \twBox{zhuoyuanw210} for the subject-to-atlas registration task.
Interestingly, we observe that all methods perform better, or at least equally well (with equal scores likely due to the inverse consistency constraint enforced by some methods), on the subject-to-atlas task than the atlas-to-subject task.
However, the specific reason for this performance asymmetry remains unclear.

\subsubsection{Pathological Brain: ADNI}
\label{sec:adni}
Subjects with mild cognitive impairment and Alzheimer’s disease often present with significant brain atrophy, ventricular enlargement, and other structural changes compared to healthy controls.
These anatomical changes introduce a significant domain shift relative to structurally normal training data, posing substantial challenges for deep learning-based registration models. Evaluating zero-shot performance on these pathological cases provides valuable insight into how effectively deep learning-based registration models generalize beyond their training domain.
Our evaluation includes both 1.5T and 3T MRI scans, allowing for an assessment of model robustness across different field strengths.
The detailed scores for inter-subject, atlas-to-subject, and subject-to-atlas registration are reported in Tables~\ref{t:ADNI15_is},~\ref{t:ADNI15_a2s},~\ref{t:ADNI15_s2a},~\ref{t:ADNI3_is},~\ref{t:ADNI3_a2s}, and~\ref{t:ADNI3_s2a}.

For all three registration tasks, the top-performing methods were based on deep learning.
For inter-subject registration, the top-performing methods for both 1.5T and 3T scans were \twBox{honkamj}, \twBox{MadeForLife}, and \twBox{next-gen-nn}.
For atlas-to-subject registration, the top-performing methods across both field strengths were \twBox{DutchMasters}, \twBox{honkamj}, and \tbBox{VFA}.
For subject-to-atlas registration, the top three methods for 1.5T were \twBox{DutchMasters}, \twBox{MadeForLife}, and \twBox{honkamj}, while for 3T they were \twBox{DutchMasters}, \twBox{honkamj}, and \twBox{Bailiang}.
Despite the anatomical variability associated with neurodegeneration, the top-performing deep learning-based methods consistently outperformed traditional optimization-based approaches across all registration tasks.
The similar performance observed between the 1.5T and 3T results further suggest that incorporating training data from both field strengths enhances model generalizability across different acquisition protocols.

\subsubsection{Out-of-Domain MRI Contrasts}
\label{sec:ood_contrasts}
%
%
%
Out-of-domain contrast tests include intra-contrast registration tasks within one of T2-weighted~(T2-w), T2*-weighted~(T2*-w), and FLAIR, all of which come from the NIMH dataset; this is an out-of-domain task as the methods have been trained on only T1-weighted~(T1-w) intra-contrast registration.
We want to see if the methods generalize well to unseen contrasts and maintain the same level of accuracy as the in-domain contrast.
Using DSC and HD95 as our proxies for accuracy, we see an almost uniform drop in the performance characteristics of the various methods between the intra-contrast registration on the T1-w data and the intra-contrast registration with the other contrasts.
We see this based on comparing the intra-contrast results on the T1-w data~(Table~\ref{tab:nimh_is_t1w}) with any of the intra-contrast results for the other three contrasts~(Table~\ref{t:nimh_is_t2*w} for T2*-w, Table~\ref{t:nimh_is_t2w} for T2-w, or Table~\ref{t:nimh_is_flair} for FLAIR).
The only exception is \tbBox{DeedsBCV}, which had slightly better mean DSC and HD95 for intra-contrast registration with the T2-w images than it had for intra-contrast registration with the T1-w images.

A key observation from examination of Tables~\ref{tab:nimh_is_t1w} through~\ref{t:nimh_is_flair}, is that while all methods see a degradation in performance (except \tbBox{DeedsBCV} in one sub-task), the impact varies considerably; this is readily discernible from the change in the ordering of the methods in these tables.
Consider the top three performing methods with respect to the T1-w inter-subject task, they are \twBox{honkamj}, \tbBox{VoxelMorph}, and \twBox{MadeForLife}.
However, the same methods have dramatically different rankings (and as noted above poorer performance) on the inter-subject task for the T2*-w images.
Specifically, \twBox{honkamj} drops to 7\textsuperscript{th}, \tbBox{VoxelMorph} falls dramatically to 20\textsuperscript{th}, while \twBox{MadeForLife} manages to move up to rank 2\textsuperscript{nd} despite its own drop in performance; this is due to the much greater performance drop of the other methods.
Similar changes in the rankings are obvious when considering any methods across the NIMH collection of sub-tasks.
It might be argued that the issue is not with the methods, but rather with the choice of statistics; however, this is not borne out by a review of the statistics used or by replacing those statistics with more robust counterparts.
Firstly, the issue of the inconsistency in the rankings exists in HD95 which itself is already a robust statistic on the Haudorff distance.
Secondly, if we look at the DSC30 which is a robust statistic on Dice score, we again see drops in performance and large changes in the rankings between the T1-weighted contrast and any of the other contrasts.
For example, the DSC30 on the intra-contrast sub-task on the NIMH data has the top three methods being \twBox{honkamj}, \twBox{MadeForLife}, and \twBox{next-gen-nn}.
While for the T2*-w registration sub-task, with respect to DSC30, \twBox{honkamj} drops to 7\textsuperscript{th}, \twBox{MadeForLife} moves up to 1\textsuperscript{st}, and \twBox{next-gen-nn} drops to 9\textsuperscript{th}.
In fact, for each of the inter-subject sub-tasks on the NIMH data, the ranking of the methods for both DSC and DSC30 is almost identical; which is not the case when looking at rankings between the different sub-tasks.
These drops in performance and chaotic changes in the ranking when applying the methods to the NIMH zero-shot tasks highlight that there is still work to be done to establish foundational models for registration that are robust to imaging contrast.

\subsubsection{Ultra-High-Field MRI}
\label{sec:highfield_mr}
Recent advances in high-field and ultra-high-field MRI technologies are enabling neuroimaging at substantially improved spatial resolution and tissue contrast compared to conventional 1.5T or 3T MRI scanners.
These improvements allow for better separation of fine anatomical structures, but they also introduce domain shifts in image appearance that can challenge registration algorithms.
In neuroimaging studies using high- or ultra-high-field MRIs, accurate registration is critical, and it is desirable for top-performing deep learning-based registration models to generalize to these new \textcolor{black}{acquisition settings} without requiring retraining.

To assess the robustness of the submitted models, we evaluated their zero-shot performance on the UltraCortex 9.4T dataset across three registration tasks: inter-subject, atlas-to-subject, and subject-to-atlas registration.
The results of each task are summarized in Tables~\ref{t:ultracortex_is},~\ref{t:ultracortex_a2s}, and~\ref{t:ultracortex_s2a}, respectively.

Inter-subject registration is inherently the most challenging due to anatomical variability. 
As expected, overall Dice scores were lower across the three subtasks. 
In the inter-subject registration task, \twBox{honkamj} achieved the top performance, leading both in terms of DSC and HD95, earning 1\textsuperscript{st} place overall. 
Close competitors included \twBox{MadeForLife} and \tbBox{VFA}, which also demonstrated high DSC scores and reasonably low HD95 values, indicating strong spatial alignment. 
In particular, \tbBox{VFA} achieved these results with a relatively high NDV, suggesting some trade-off between anatomical plausibility and spatial accuracy.

In the atlas-to-subject registration task, \twBox{honkamj} again led with the highest accuracy ranking, while maintaining anatomical plausibility with a low NDV. 
\twBox{MadeForLife} and \tbBox{VFA} closely followed in performance, showing similarly high DSC scores and low HD95 values, affirming their robustness across tasks.
Notably, \tbBox{VFA}'s relatively high NDV shows again its tendency towards less regularized deformation.

The subject-to-atlas registration task shows slightly better performance across methods compared to atlas-to-subject registration, indicating that the direction of registration affects accuracy. 
\twBox{MadeForLife} claimed the highest accuracy ranking, closely followed by \twBox{next-gen-nn} and \tbBox{VFA}. 
These top-performing methods consistently achieved high overlap and low boundary errors, while generally keeping NDV within acceptable ranges. 
Notably, \twBox{next-gen-nn} had one of the lowest NDVs, which emphasizes its effectiveness in preserving diffeomorphic properties while delivering high accuracy.

Overall, consistent top performers across all tasks include \twBox{honkamj}, \twBox{MadeForLife}, and \tbBox{VFA}, which show robust registration accuracy and good diffeomorphic behavior.
Registration accuracy declines from subject-to-atlas to atlas-to-subject, emphasizing the importance of evaluating across multiple directions.
Deep learning methods outperform traditional methods in alignment metrics, but often at the cost of increased non-diffeomorphic deformations, unless specifically regularized.

\subsubsection{Non-Human Species}
\label{sec:macaque}
%
The macaque brain inter-subject registration task involves anatomical structures that differ significantly from those in the human brain dataset.
Evaluating on this dataset introduces a domain shift in structural anatomy and imaging content, beyond the intensity differences assessed in the out-of-domain contrast experiments.
However, due to the differences in anatomical structures, it is not feasible to evaluate the alignment of the same types of anatomical labels as those used in the human brain datasets.
Additionally, the annotations in the macaque dataset are relatively coarse, making direct performance comparisons with human brain registration results inappropriate.

Given these limitations, we discuss the results of this task separately and focus on comparing the methods directly. As shown in Table~\ref{t:macaque}, without registration (i.e., \tbBox{ZeroDisplacement}), the DSC is already 0.665.
Most methods perform reasonably close to each other, with DSC values ranging from 0.75 to 0.78 and HD95 from 4.6~\textcolor{black}{mm} to 4.3~\textcolor{black}{mm}.
An exception is \tbBox{VFA}, which, despite consistently ranking in the top ten on human brain tasks, was the only method to fall below 0.7 in DSC and above 5~\textcolor{black}{mm} in HD95, ranking lowest among all methods on this task.

The three best performing methods on the macaque dataset are \twBox{MadeForLife}, \tbBox{FireANTsGreedy}, and \twBox{honkamj}.
Notably, this is the only task in which the optimization-based method \tbBox{FireANTsGreedy} ranked among the top three, outperforming several deep learning-based methods that had surpassed it in the other tasks.

\subsection{\textcolor{black}{Explaining the Ranking}}
\label{a:ranking-example}
\textcolor{black}{We provide a simplified example to explain how the rankings are determined.
Consider the case in which we have three algorithms A, B, \& C and we are evaluating these algorithms with respect to two measurements I \& II.
Using the Wilcoxon signed rank test~(WSRT) we compare algorithm A against B with respect to measure I.
Algorithm A has a better measure I score than B and it is significant based on the WSRT, thus A \textit{wins} the comparison with B.
For algorithm A and C with respect to I, we have A has a better I score than C which is significant and thus A wins the comparison with C.
For algorithm B and C, B has a better I score than C which is significant and thus B wins the comparison with C.
Thus for measure I, A has two wins and B has one win, thus the ranking would be A, B, C.
This would mean for measure I, that A gets assigned the value 0.1 (for winning) and C get assigned the value 1.0 (for coming last).
B would get a value of 0.55 for coming second in the comparison (in the instance of more methods between the first and last algorithms, the range [0.1, 1.0] is evenly divided up and values are assigned in order from 0.1 to 1.0).
This is summarized in the two columns under Measure~I in Table~\ref{t:ranking-example}.}

\begin{table*}[!tb]
\centering
\caption{\textcolor{black}{A worked example of how the ranking is computed: In this instance for two measures across three algorithms. See Sec.~\ref{a:ranking-example} for complete details.}}
\label{t:ranking-example}
\begin{tabular}{lc ccc c ccc cc cc}
\toprule
&& \multicolumn{3}{c}{\textbf{Measure I}}
&& \multicolumn{3}{c}{\textbf{Measure II}}\\
\cmidrule(lr){3-5}
\cmidrule(lr){7-9}
\textbf{Algorithm} && \textbf{Wins} && \textbf{Score} && \textbf{Wins} && \textbf{Score}
&& \textbf{Geometric Mean}
&& \textbf{Final Ranking}\\
\cmidrule(lr){1-1}
\cmidrule(lr){3-5}
\cmidrule(lr){7-9}
\cmidrule(lr){11-13}
A && 2 && 0.1\phantom{0} && 2 && 0.1 && 0.1\phantom{00} && 1$^{\text{\tiny{st}}}$\\
B && 1 && 0.55 && 0 && 1.0 && 0.742 && 2$^{\text{\tiny{nd}}}$\\
C && 0 && 1.0\phantom{0} && 0 && 1.0 && 1.0\phantom{00}  && 3$^{\text{\tiny{rd}}}$\\
\bottomrule
\end{tabular}
\end{table*}

\textcolor{black}{For measure II, algorithm A is again significantly better than both B and C, and is denoted as having two wins in the appropriate column of Table~\ref{t:ranking-example}.
Whereas neither algorithm B or C is significantly better than the other resulting in neither of them having a win.
Thus the ranking with respect to measure~II is A first, with B and C tied in last place.
The corresponding scores are 0.1 for A and 1.0 for both B and C.
It is then straightforward to compute the corresponding geometric means and determine the final ranking of the three algorithms over the two measures.
This procedure was used to compute all of our reported rankings.}

\subsection{Detailed Quantitative Results for Each Zero-Shot Subtask}
\subsubsection{NIMH - T1w}
\begin{table}[H]
\caption{NIMH T1w Atlas to Subject.}
\label{t:nimh_a2s_t1w}
\centering
\resizebox{0.9\columnwidth}{!}{%
\begin{tabular}{r c c c c c c c c}
\toprule
&&\textbf{DSC}$\uparrow$ & \textbf{HD95}$\downarrow$~\textcolor{black}{(mm)} & \textbf{Ranking} & \textbf{\%NDV}$\downarrow$ & \textbf{DSC30}$\uparrow$\\

\cmidrule(lr){3-3}
\cmidrule(lr){4-4}\cmidrule(lr){5-5}
\cmidrule(lr){6-6}
\cmidrule(lr){7-7}
\textbf{Method}&& \textbf{Mean(}$\bm{\PM}$\textbf{Std. Dev.)} 
& \textbf{Mean(}$\bm{\PM}$\textbf{Std. Dev.)} 
& \scriptsize{\textbf{(DSC + HD95)}}
& \textbf{Mean(}$\bm{\PM}$\textbf{Std. Dev.)} 
& \textbf{Mean(}$\bm{\PM}$\textbf{Std. Dev.)} 
\\
\midrule 

\rowcolor{white}
\mBox{white}{honkamj}  &
 &0.801($\mypm$0.012)
 &2.365($\mypm$0.237)
 &\rCirc{1}
 &8.11e-04($\mypm$2.08e-04)
 &0.786($\mypm$0.008)
 \\

\rowcolor{Gray}
\mBox{white}{MadeForLife}  &
 &0.796($\mypm$0.013)
 &2.515($\mypm$0.247)
 &\rCirc{3}
 &3.56e-03($\mypm$1.27e-03)
 &0.779($\mypm$0.009)
 \\

\rowcolor{white}
\mBox{white}{LYU1}  &
 &0.786($\mypm$0.014)
 &2.565($\mypm$0.233)
 &\rCirc{9}
 &1.26e-02($\mypm$1.57e-03)
 &0.767($\mypm$0.009)
 \\

\rowcolor{Gray}
\mBox{white}{next-gen-nn}  &
 &0.791($\mypm$0.013)
 &2.591($\mypm$0.237)
 &\rCirc{8}
 &1.31e-04($\mypm$1.61e-04)
 &0.775($\mypm$0.010)
 \\

\rowcolor{white}
\mBox{white}{zhuoyuanw210}  &
 &0.793($\mypm$0.009)
 &2.463($\mypm$0.221)
 &\rCirc{4}
 &2.98e-03($\mypm$5.19e-04)
 &0.781($\mypm$0.007)
 \\

\rowcolor{Gray}
\mBox{white}{DutchMasters}  &
 &0.795($\mypm$0.008)
 &2.450($\mypm$0.208)
 &\rCirc{2}
 &1.53e-03($\mypm$6.56e-04)
 &0.786($\mypm$0.006)
 \\
\bottomrule
\end{tabular}}

\end{table}

\begin{table}[H]
\caption{NIMH T1w Subject to Atlas.}
\label{t:nimh_s2a_t1w}
\centering
\resizebox{0.9\columnwidth}{!}{%
\begin{tabular}{r c c c c c c c c}
\toprule
&&\textbf{DSC}$\uparrow$ & \textbf{HD95}$\downarrow$~\textcolor{black}{(mm)} & \textbf{Ranking} & \textbf{\%NDV}$\downarrow$ & \textbf{DSC30}$\uparrow$\\

\cmidrule(lr){3-3}
\cmidrule(lr){4-4}\cmidrule(lr){5-5}
\cmidrule(lr){6-6}
\cmidrule(lr){7-7}
\textbf{Method}&& \textbf{Mean(}$\bm{\PM}$\textbf{Std. Dev.)} 
& \textbf{Mean(}$\bm{\PM}$\textbf{Std. Dev.)} 
& \scriptsize{\textbf{(DSC + HD95)}}
& \textbf{Mean(}$\bm{\PM}$\textbf{Std. Dev.)} 
& \textbf{Mean(}$\bm{\PM}$\textbf{Std. Dev.)} 
\\
\midrule 

\rowcolor{white}
\mBox{white}{honkamj}  &
 &0.806($\mypm$0.012)
 &2.424($\mypm$0.226)
 &\rCirc{2}
 &2.05e-03($\mypm$4.09e-04)
 &0.791($\mypm$0.008)
 \\

\rowcolor{Gray}
\mBox{white}{MadeForLife}  &
 &0.809($\mypm$0.011)
 &2.482($\mypm$0.219)
 &\rCirc{6}
 &5.96e-03($\mypm$1.66e-03)
 &0.795($\mypm$0.008)
 \\

\rowcolor{white}
\mBox{white}{LYU1}  &
 &0.806($\mypm$0.012)
 &2.476($\mypm$0.221)
 &\rCirc{7}
 &5.72e-03($\mypm$8.02e-04)
 &0.790($\mypm$0.009)
 \\

\rowcolor{Gray}
\mBox{white}{next-gen-nn}  &
 &0.811($\mypm$0.009)
 &2.448($\mypm$0.189)
 &\rCirc{1}
 &1.80e-04($\mypm$1.59e-04)
 &0.800($\mypm$0.006)
 \\

\rowcolor{white}
\mBox{white}{zhuoyuanw210}  &
 &0.809($\mypm$0.011)
 &2.440($\mypm$0.208)
 &\rCirc{3}
 &2.06e-03($\mypm$5.49e-04)
 &0.795($\mypm$0.008)
 \\

\rowcolor{Gray}
\mBox{white}{DutchMasters}  &
 &0.801($\mypm$0.008)
 &2.444($\mypm$0.204)
 &\rCirc{8}
 &3.47e-03($\mypm$1.31e-03)
 &0.791($\mypm$0.006)
 \\
\bottomrule
\end{tabular}}

\end{table}

\begin{table}[H]
\caption{NIMH T1w Inter-subject.}
\label{tab:nimh_is_t1w}
\centering
\resizebox{0.9\columnwidth}{!}{%
\begin{tabular}{r c c c c c c c c}
\toprule
&&\textbf{DSC}$\uparrow$ & \textbf{HD95}$\downarrow$~\textcolor{black}{(mm)} & \textbf{Ranking} & \textbf{\%NDV}$\downarrow$ & \textbf{DSC30}$\uparrow$\\

\cmidrule(lr){3-3}
\cmidrule(lr){4-4}\cmidrule(lr){5-5}
\cmidrule(lr){6-6}
\cmidrule(lr){7-7}
\textbf{Method}&& \textbf{Mean(}$\bm{\PM}$\textbf{Std. Dev.)} 
& \textbf{Mean(}$\bm{\PM}$\textbf{Std. Dev.)} 
& \scriptsize{\textbf{(DSC + HD95)}}
& \textbf{Mean(}$\bm{\PM}$\textbf{Std. Dev.)} 
& \textbf{Mean(}$\bm{\PM}$\textbf{Std. Dev.)} 
\\
\midrule 

\rowcolor{white}
\mBox{white}{honkamj}  &
 &0.783($\mypm$0.012)
 &2.713($\mypm$0.239)
 &\rCirc{1}
 &2.91e-03($\mypm$4.82e-04)
 &0.769($\mypm$0.008)
 \\

\rowcolor{Gray}
\mBox{white}{MadeForLife}  &
 &0.781($\mypm$0.013)
 &2.935($\mypm$0.301)
 &\rCirc{2}
 &1.34e-02($\mypm$2.68e-03)
 &0.766($\mypm$0.010)
 \\

\rowcolor{white}
\mBox{white}{LYU1}  &
 &0.777($\mypm$0.013)
 &2.953($\mypm$0.292)
 &\rCirc{5}
 &1.39e-02($\mypm$1.31e-03)
 &0.761($\mypm$0.009)
 \\

\rowcolor{Gray}
\mBox{white}{next-gen-nn}  &
 &0.779($\mypm$0.012)
 &2.927($\mypm$0.263)
 &\rCirc{4}
 &2.75e-04($\mypm$1.53e-04)
 &0.765($\mypm$0.009)
 \\

\rowcolor{white}
\mBox{white}{zhuoyuanw210}  &
 &0.773($\mypm$0.013)
 &2.940($\mypm$0.266)
 &\rCirc{8}
 &6.09e-03($\mypm$1.02e-03)
 &0.758($\mypm$0.010)
 \\

\rowcolor{Gray}
\mBox{white}{DutchMasters}  &
 &0.776($\mypm$0.012)
 &2.901($\mypm$0.246)
 &\rCirc{6}
 &3.33e-03($\mypm$1.21e-03)
 &0.762($\mypm$0.008)
 \\
\bottomrule
\end{tabular}}

\end{table}

\subsubsection{ADHD}
\begin{table}[H]
\caption{ADHD Inter-subject.}
\label{t:ADHD_is}
\centering
\resizebox{0.9\columnwidth}{!}{%
\begin{tabular}{r c c c c c c c c}
\toprule
&&\textbf{DSC}$\uparrow$ & \textbf{HD95}$\downarrow$~\textcolor{black}{(mm)} & \textbf{Ranking} & \textbf{\%NDV}$\downarrow$ & \textbf{DSC30}$\uparrow$\\

\cmidrule(lr){3-3}
\cmidrule(lr){4-4}\cmidrule(lr){5-5}
\cmidrule(lr){6-6}
\cmidrule(lr){7-7}
\textbf{Method}&& \textbf{Mean(}$\bm{\PM}$\textbf{Std. Dev.)} 
& \textbf{Mean(}$\bm{\PM}$\textbf{Std. Dev.)} 
& \scriptsize{\textbf{(DSC + HD95)}}
& \textbf{Mean(}$\bm{\PM}$\textbf{Std. Dev.)} 
& \textbf{Mean(}$\bm{\PM}$\textbf{Std. Dev.)} 
\\
\midrule 

\rowcolor{white}
\mBox{white}{honkamj}  &
 &0.778($\mypm$0.017)
 &3.048($\mypm$0.451)
 &\rCirc{1}
 &2.26e-03($\mypm$3.58e-04)
 &0.755($\mypm$0.005)
 \\

\rowcolor{Gray}
\mBox{white}{MadeForLife}  &
 &0.773($\mypm$0.017)
 &3.281($\mypm$0.475)
 &\rCirc{3}
 &7.11e-03($\mypm$2.99e-03)
 &0.751($\mypm$0.007)
 \\

\rowcolor{white}
\mBox{white}{LYU1}  &
 &0.770($\mypm$0.016)
 &3.286($\mypm$0.437)
 &\rCirc{4}
 &1.23e-02($\mypm$1.47e-03)
 &0.750($\mypm$0.006)
 \\

\rowcolor{Gray}
\mBox{white}{next-gen-nn}  &
 &0.769($\mypm$0.018)
 &3.330($\mypm$0.479)
 &\rCirc{7}
 &2.73e-04($\mypm$2.16e-04)
 &0.747($\mypm$0.006)
 \\

\rowcolor{white}
\mBox{white}{zhuoyuanw210}  &
 &0.768($\mypm$0.015)
 &3.230($\mypm$0.425)
 &\rCirc{5}
 &2.98e-03($\mypm$6.74e-04)
 &0.749($\mypm$0.006)
 \\

\rowcolor{Gray}
\mBox{white}{DutchMasters}  &
 &0.767($\mypm$0.019)
 &3.253($\mypm$0.461)
 &\rCirc{6}
 &1.93e-03($\mypm$8.37e-04)
 &0.743($\mypm$0.008)
 \\
\bottomrule
\end{tabular}}

\end{table}

\begin{table}[H]
\caption{ADHD Atlas to Subject.}
\label{t:ADHD_a2s}
\centering
\resizebox{0.9\columnwidth}{!}{%
\begin{tabular}{r c c c c c c c c}
\toprule
&&\textbf{DSC}$\uparrow$ & \textbf{HD95}$\downarrow$~\textcolor{black}{(mm)} & \textbf{Ranking} & \textbf{\%NDV}$\downarrow$ & \textbf{DSC30}$\uparrow$\\

\cmidrule(lr){3-3}
\cmidrule(lr){4-4}\cmidrule(lr){5-5}
\cmidrule(lr){6-6}
\cmidrule(lr){7-7}
\textbf{Method}&& \textbf{Mean(}$\bm{\PM}$\textbf{Std. Dev.)} 
& \textbf{Mean(}$\bm{\PM}$\textbf{Std. Dev.)} 
& \scriptsize{\textbf{(DSC + HD95)}}
& \textbf{Mean(}$\bm{\PM}$\textbf{Std. Dev.)} 
& \textbf{Mean(}$\bm{\PM}$\textbf{Std. Dev.)} 
\\
\midrule 

\rowcolor{white}
\mBox{white}{honkamj}  &
 &0.764($\mypm$0.012)
 &3.132($\mypm$0.342)
 &\rCirc{1}
 &7.74e-04($\mypm$1.95e-04)
 &0.751($\mypm$0.007)
 \\

\rowcolor{Gray}
\mBox{white}{MadeForLife}  &
 &0.755($\mypm$0.013)
 &3.330($\mypm$0.356)
 &\rCirc{5}
 &2.87e-03($\mypm$1.11e-03)
 &0.740($\mypm$0.006)
 \\

\rowcolor{white}
\mBox{white}{LYU1}  &
 &0.748($\mypm$0.013)
 &3.335($\mypm$0.369)
 &\rCirc{7}
 &1.30e-02($\mypm$2.38e-03)
 &0.732($\mypm$0.007)
 \\

\rowcolor{Gray}
\mBox{white}{next-gen-nn}  &
 &0.750($\mypm$0.013)
 &3.410($\mypm$0.369)
 &\rCirc{9}
 &1.72e-04($\mypm$1.78e-04)
 &0.734($\mypm$0.007)
 \\

\rowcolor{white}
\mBox{white}{zhuoyuanw210}  &
 &0.759($\mypm$0.011)
 &3.246($\mypm$0.354)
 &\rCirc{3}
 &2.24e-03($\mypm$3.50e-04)
 &0.745($\mypm$0.006)
 \\

\rowcolor{Gray}
\mBox{white}{DutchMasters}  &
 &0.763($\mypm$0.012)
 &3.174($\mypm$0.377)
 &\rCirc{2}
 &1.34e-03($\mypm$6.08e-04)
 &0.748($\mypm$0.008)
 \\
\bottomrule
\end{tabular}}

\end{table}

\begin{table}[H]
\caption{ADHD Subject to Atlas.}
\label{t:ADHD_s2a}
\centering
\resizebox{0.9\columnwidth}{!}{%
\begin{tabular}{r c c c c c c c c}
\toprule
&&\textbf{DSC}$\uparrow$ & \textbf{HD95}$\downarrow$~\textcolor{black}{(mm)} & \textbf{Ranking} & \textbf{\%NDV}$\downarrow$ & \textbf{DSC30}$\uparrow$\\

\cmidrule(lr){3-3}
\cmidrule(lr){4-4}\cmidrule(lr){5-5}
\cmidrule(lr){6-6}
\cmidrule(lr){7-7}
\textbf{Method}&& \textbf{Mean(}$\bm{\PM}$\textbf{Std. Dev.)} 
& \textbf{Mean(}$\bm{\PM}$\textbf{Std. Dev.)} 
& \scriptsize{\textbf{(DSC + HD95)}}
& \textbf{Mean(}$\bm{\PM}$\textbf{Std. Dev.)} 
& \textbf{Mean(}$\bm{\PM}$\textbf{Std. Dev.)} 
\\
\midrule 

\rowcolor{white}
\mBox{white}{honkamj}  &
 &0.762($\mypm$0.012)
 &3.119($\mypm$0.365)
 &\rCirc{5}
 &1.72e-03($\mypm$2.73e-04)
 &0.749($\mypm$0.006)
 \\

\rowcolor{Gray}
\mBox{white}{MadeForLife}  &
 &0.767($\mypm$0.012)
 &3.188($\mypm$0.390)
 &\rCirc{6}
 &4.06e-03($\mypm$2.09e-03)
 &0.753($\mypm$0.007)
 \\

\rowcolor{white}
\mBox{white}{LYU1}  &
 &0.763($\mypm$0.013)
 &3.197($\mypm$0.382)
 &\rCirc{7}
 &5.03e-03($\mypm$7.45e-04)
 &0.749($\mypm$0.008)
 \\

\rowcolor{Gray}
\mBox{white}{next-gen-nn}  &
 &0.772($\mypm$0.011)
 &3.130($\mypm$0.383)
 &\rCirc{2}
 &1.73e-04($\mypm$1.64e-04)
 &0.758($\mypm$0.006)
 \\

\rowcolor{white}
\mBox{white}{zhuoyuanw210}  &
 &0.766($\mypm$0.012)
 &3.125($\mypm$0.371)
 &\rCirc{3}
 &1.30e-03($\mypm$4.02e-04)
 &0.752($\mypm$0.007)
 \\

\rowcolor{Gray}
\mBox{white}{DutchMasters}  &
 &0.760($\mypm$0.012)
 &3.165($\mypm$0.383)
 &\rCirc{9}
 &1.64e-03($\mypm$1.02e-03)
 &0.746($\mypm$0.008)
 \\
\bottomrule
\end{tabular}}

\end{table}

\subsubsection{ADNI 1.5T}
\begin{table}[H]
\caption{ADNI 1.5T Inter-subject.}
\label{t:ADNI15_is}
\centering
\resizebox{0.9\columnwidth}{!}{%
\begin{tabular}{r c c c c c c c c}
\toprule
&&\textbf{DSC}$\uparrow$ & \textbf{HD95}$\downarrow$~\textcolor{black}{(mm)} & \textbf{Ranking} & \textbf{\%NDV}$\downarrow$ & \textbf{DSC30}$\uparrow$\\

\cmidrule(lr){3-3}
\cmidrule(lr){4-4}\cmidrule(lr){5-5}
\cmidrule(lr){6-6}
\cmidrule(lr){7-7}
\textbf{Method}&& \textbf{Mean(}$\bm{\PM}$\textbf{Std. Dev.)} 
& \textbf{Mean(}$\bm{\PM}$\textbf{Std. Dev.)} 
& \scriptsize{\textbf{(DSC + HD95)}}
& \textbf{Mean(}$\bm{\PM}$\textbf{Std. Dev.)} 
& \textbf{Mean(}$\bm{\PM}$\textbf{Std. Dev.)} 
\\
\midrule 

\rowcolor{white}
\mBox{white}{honkamj}  &
 &0.774($\mypm$0.011)
 &2.759($\mypm$0.227)
 &\rCirc{1}
 &3.24e-03($\mypm$6.01e-04)
 &0.761($\mypm$0.008)
 \\

\rowcolor{Gray}
\mBox{white}{MadeForLife}  &
 &0.772($\mypm$0.012)
 &2.985($\mypm$0.271)
 &\rCirc{2}
 &1.81e-02($\mypm$3.97e-03)
 &0.757($\mypm$0.009)
 \\

\rowcolor{white}
\mBox{white}{LYU1}  &
 &0.766($\mypm$0.013)
 &3.032($\mypm$0.286)
 &\rCirc{6}
 &1.38e-02($\mypm$1.70e-03)
 &0.750($\mypm$0.009)
 \\

\rowcolor{Gray}
\mBox{white}{next-gen-nn}  &
 &0.769($\mypm$0.013)
 &2.984($\mypm$0.255)
 &\rCirc{4}
 &4.09e-04($\mypm$2.90e-04)
 &0.754($\mypm$0.009)
 \\

\rowcolor{white}
\mBox{white}{zhuoyuanw210}  &
 &0.760($\mypm$0.013)
 &3.036($\mypm$0.273)
 &\rCirc{8}
 &8.84e-03($\mypm$1.84e-03)
 &0.745($\mypm$0.009)
 \\

\rowcolor{Gray}
\mBox{white}{DutchMasters}  &
 &0.768($\mypm$0.012)
 &2.947($\mypm$0.249)
 &\rCirc{3}
 &4.20e-03($\mypm$1.62e-03)
 &0.754($\mypm$0.008)
 \\
\bottomrule
\end{tabular}}

\end{table}

\begin{table}[H]
\caption{ADNI 1.5T Atlas to Subject.}
\label{t:ADNI15_a2s}
\centering
\resizebox{0.9\columnwidth}{!}{%
\begin{tabular}{r c c c c c c c c}
\toprule
&&\textbf{DSC}$\uparrow$ & \textbf{HD95}$\downarrow$~\textcolor{black}{(mm)} & \textbf{Ranking} & \textbf{\%NDV}$\downarrow$ & \textbf{DSC30}$\uparrow$\\

\cmidrule(lr){3-3}
\cmidrule(lr){4-4}\cmidrule(lr){5-5}
\cmidrule(lr){6-6}
\cmidrule(lr){7-7}
\textbf{Method}&& \textbf{Mean(}$\bm{\PM}$\textbf{Std. Dev.)} 
& \textbf{Mean(}$\bm{\PM}$\textbf{Std. Dev.)} 
& \scriptsize{\textbf{(DSC + HD95)}}
& \textbf{Mean(}$\bm{\PM}$\textbf{Std. Dev.)} 
& \textbf{Mean(}$\bm{\PM}$\textbf{Std. Dev.)} 
\\
\midrule 

\rowcolor{white}
\mBox{white}{honkamj}  &
 &0.765($\mypm$0.020)
 &2.594($\mypm$0.284)
 &\rCirc{3}
 &9.02e-04($\mypm$2.38e-04)
 &0.741($\mypm$0.013)
 \\

\rowcolor{Gray}
\mBox{white}{MadeForLife}  &
 &0.756($\mypm$0.020)
 &2.740($\mypm$0.291)
 &\rCirc{7}
 &4.04e-03($\mypm$1.64e-03)
 &0.731($\mypm$0.012)
 \\

\rowcolor{white}
\mBox{white}{LYU1}  &
 &0.740($\mypm$0.022)
 &2.792($\mypm$0.313)
 &\rCirc{13}
 &1.07e-02($\mypm$1.20e-03)
 &0.713($\mypm$0.014)
 \\

\rowcolor{Gray}
\mBox{white}{next-gen-nn}  &
 &0.743($\mypm$0.025)
 &2.879($\mypm$0.333)
 &\rCirc{14}
 &1.21e-04($\mypm$1.40e-04)
 &0.713($\mypm$0.015)
 \\

\rowcolor{white}
\mBox{white}{zhuoyuanw210}  &
 &0.761($\mypm$0.021)
 &2.748($\mypm$0.365)
 &\rCirc{4}
 &3.18e-03($\mypm$6.37e-04)
 &0.735($\mypm$0.014)
 \\

\rowcolor{Gray}
\mBox{white}{DutchMasters}  &
 &0.774($\mypm$0.015)
 &2.588($\mypm$0.260)
 &\rCirc{1}
 &2.15e-03($\mypm$9.33e-04)
 &0.756($\mypm$0.009)
 \\
\bottomrule
\end{tabular}}

\end{table}

\begin{table}[H]
\caption{ADNI 1.5T Subject to Atlas.}
\label{t:ADNI15_s2a}
\centering
\resizebox{0.9\columnwidth}{!}{%
\begin{tabular}{r c c c c c c c c}
\toprule
&&\textbf{DSC}$\uparrow$ & \textbf{HD95}$\downarrow$~\textcolor{black}{(mm)} & \textbf{Ranking} & \textbf{\%NDV}$\downarrow$ & \textbf{DSC30}$\uparrow$\\

\cmidrule(lr){3-3}
\cmidrule(lr){4-4}\cmidrule(lr){5-5}
\cmidrule(lr){6-6}
\cmidrule(lr){7-7}
\textbf{Method}&& \textbf{Mean(}$\bm{\PM}$\textbf{Std. Dev.)} 
& \textbf{Mean(}$\bm{\PM}$\textbf{Std. Dev.)} 
& \scriptsize{\textbf{(DSC + HD95)}}
& \textbf{Mean(}$\bm{\PM}$\textbf{Std. Dev.)} 
& \textbf{Mean(}$\bm{\PM}$\textbf{Std. Dev.)} 
\\
\midrule 

\rowcolor{white}
\mBox{white}{honkamj}  &
 &0.791($\mypm$0.015)
 &2.636($\mypm$0.255)
 &\rCirc{3}
 &2.31e-03($\mypm$5.43e-04)
 &0.773($\mypm$0.009)
 \\

\rowcolor{Gray}
\mBox{white}{MadeForLife}  &
 &0.797($\mypm$0.014)
 &2.662($\mypm$0.258)
 &\rCirc{2}
 &7.56e-03($\mypm$3.31e-03)
 &0.780($\mypm$0.009)
 \\

\rowcolor{white}
\mBox{white}{LYU1}  &
 &0.784($\mypm$0.018)
 &2.747($\mypm$0.289)
 &\rCirc{10}
 &6.95e-03($\mypm$1.18e-03)
 &0.762($\mypm$0.011)
 \\

\rowcolor{Gray}
\mBox{white}{next-gen-nn}  &
 &0.787($\mypm$0.018)
 &2.704($\mypm$0.285)
 &\rCirc{7}
 &2.20e-04($\mypm$2.09e-04)
 &0.765($\mypm$0.011)
 \\

\rowcolor{white}
\mBox{white}{zhuoyuanw210}  &
 &0.789($\mypm$0.017)
 &2.739($\mypm$0.299)
 &\rCirc{8}
 &3.52e-03($\mypm$1.63e-03)
 &0.769($\mypm$0.011)
 \\

\rowcolor{Gray}
\mBox{white}{DutchMasters}  &
 &0.793($\mypm$0.012)
 &2.600($\mypm$0.248)
 &\rCirc{1}
 &6.03e-03($\mypm$2.76e-03)
 &0.779($\mypm$0.007)
 \\
\bottomrule
\end{tabular}}

\end{table}

\subsubsection{ADNI 3T}
\begin{table}[H]
\caption{ADNI 3T Inter-subject.}
\label{t:ADNI3_is}
\centering
\resizebox{0.9\columnwidth}{!}{%
\begin{tabular}{r c c c c c c c c}
\toprule
&&\textbf{DSC}$\uparrow$ & \textbf{HD95}$\downarrow$~\textcolor{black}{(mm)} & \textbf{Ranking} & \textbf{\%NDV}$\downarrow$ & \textbf{DSC30}$\uparrow$\\

\cmidrule(lr){3-3}
\cmidrule(lr){4-4}\cmidrule(lr){5-5}
\cmidrule(lr){6-6}
\cmidrule(lr){7-7}
\textbf{Method}&& \textbf{Mean(}$\bm{\PM}$\textbf{Std. Dev.)} 
& \textbf{Mean(}$\bm{\PM}$\textbf{Std. Dev.)} 
& \scriptsize{\textbf{(DSC + HD95)}}
& \textbf{Mean(}$\bm{\PM}$\textbf{Std. Dev.)} 
& \textbf{Mean(}$\bm{\PM}$\textbf{Std. Dev.)} 
\\
\midrule 

\rowcolor{white}
\mBox{white}{honkamj}  &
 &0.777($\mypm$0.012)
 &2.821($\mypm$0.258)
 &\rCirc{1}
 &3.16e-03($\mypm$6.65e-04)
 &0.762($\mypm$0.010)
 \\

\rowcolor{Gray}
\mBox{white}{MadeForLife}  &
 &0.774($\mypm$0.015)
 &3.052($\mypm$0.400)
 &\rCirc{2}
 &2.03e-02($\mypm$4.82e-03)
 &0.756($\mypm$0.015)
 \\

\rowcolor{white}
\mBox{white}{LYU1}  &
 &0.769($\mypm$0.017)
 &3.087($\mypm$0.406)
 &\rCirc{5}
 &1.48e-02($\mypm$1.93e-03)
 &0.749($\mypm$0.017)
 \\

\rowcolor{Gray}
\mBox{white}{next-gen-nn}  &
 &0.772($\mypm$0.015)
 &3.051($\mypm$0.310)
 &\rCirc{3}
 &4.25e-04($\mypm$2.97e-04)
 &0.754($\mypm$0.012)
 \\

\rowcolor{white}
\mBox{white}{zhuoyuanw210}  &
 &0.762($\mypm$0.015)
 &3.112($\mypm$0.342)
 &\rCirc{8}
 &9.34e-03($\mypm$1.89e-03)
 &0.743($\mypm$0.012)
 \\

\rowcolor{Gray}
\mBox{white}{DutchMasters}  &
 &0.768($\mypm$0.013)
 &3.046($\mypm$0.293)
 &\rCirc{6}
 &4.81e-03($\mypm$2.07e-03)
 &0.752($\mypm$0.008)
 \\
\bottomrule
\end{tabular}}

\end{table}

\begin{table}[H]
\caption{ADNI 3T Atlas to Subject.}
\label{t:ADNI3_a2s}
\centering
\resizebox{0.9\columnwidth}{!}{%
\begin{tabular}{r c c c c c c c c}
\toprule
&&\textbf{DSC}$\uparrow$ & \textbf{HD95}$\downarrow$~\textcolor{black}{(mm)} & \textbf{Ranking} & \textbf{\%NDV}$\downarrow$ & \textbf{DSC30}$\uparrow$\\

\cmidrule(lr){3-3}
\cmidrule(lr){4-4}\cmidrule(lr){5-5}
\cmidrule(lr){6-6}
\cmidrule(lr){7-7}
\textbf{Method}&& \textbf{Mean(}$\bm{\PM}$\textbf{Std. Dev.)} 
& \textbf{Mean(}$\bm{\PM}$\textbf{Std. Dev.)} 
& \scriptsize{\textbf{(DSC + HD95)}}
& \textbf{Mean(}$\bm{\PM}$\textbf{Std. Dev.)} 
& \textbf{Mean(}$\bm{\PM}$\textbf{Std. Dev.)} 
\\
\midrule 

\rowcolor{white}
\mBox{white}{honkamj}  &
 &0.770($\mypm$0.020)
 &2.593($\mypm$0.307)
 &\rCirc{2}
 &8.91e-04($\mypm$2.47e-04)
 &0.745($\mypm$0.012)
 \\

\rowcolor{Gray}
\mBox{white}{MadeForLife}  &
 &0.762($\mypm$0.021)
 &2.746($\mypm$0.320)
 &\rCirc{6}
 &4.51e-03($\mypm$1.80e-03)
 &0.735($\mypm$0.012)
 \\

\rowcolor{white}
\mBox{white}{LYU1}  &
 &0.746($\mypm$0.024)
 &2.814($\mypm$0.355)
 &\rCirc{14}
 &1.15e-02($\mypm$1.32e-03)
 &0.716($\mypm$0.015)
 \\

\rowcolor{Gray}
\mBox{white}{next-gen-nn}  &
 &0.752($\mypm$0.026)
 &2.860($\mypm$0.376)
 &\rCirc{13}
 &1.00e-04($\mypm$9.81e-05)
 &0.719($\mypm$0.016)
 \\

\rowcolor{white}
\mBox{white}{zhuoyuanw210}  &
 &0.764($\mypm$0.022)
 &2.738($\mypm$0.415)
 &\rCirc{4}
 &3.58e-03($\mypm$6.19e-04)
 &0.737($\mypm$0.017)
 \\

\rowcolor{Gray}
\mBox{white}{DutchMasters}  &
 &0.774($\mypm$0.016)
 &2.608($\mypm$0.289)
 &\rCirc{1}
 &2.55e-03($\mypm$1.47e-03)
 &0.753($\mypm$0.009)
 \\
\bottomrule
\end{tabular}}

\end{table}

\begin{table}[H]
\caption{ADNI 3T Subject to Atlas.}
\label{t:ADNI3_s2a}
\centering
\resizebox{0.9\columnwidth}{!}{%
\begin{tabular}{r c c c c c c c c}
\toprule
&&\textbf{DSC}$\uparrow$ & \textbf{HD95}$\downarrow$~\textcolor{black}{(mm)} & \textbf{Ranking} & \textbf{\%NDV}$\downarrow$ & \textbf{DSC30}$\uparrow$\\

\cmidrule(lr){3-3}
\cmidrule(lr){4-4}\cmidrule(lr){5-5}
\cmidrule(lr){6-6}
\cmidrule(lr){7-7}
\textbf{Method}&& \textbf{Mean(}$\bm{\PM}$\textbf{Std. Dev.)} 
& \textbf{Mean(}$\bm{\PM}$\textbf{Std. Dev.)} 
& \scriptsize{\textbf{(DSC + HD95)}}
& \textbf{Mean(}$\bm{\PM}$\textbf{Std. Dev.)} 
& \textbf{Mean(}$\bm{\PM}$\textbf{Std. Dev.)} 
\\
\midrule 

\rowcolor{white}
\mBox{white}{honkamj}  &
 &0.791($\mypm$0.014)
 &2.636($\mypm$0.252)
 &\rCirc{2}
 &2.27e-03($\mypm$5.46e-04)
 &0.774($\mypm$0.008)
 \\

\rowcolor{Gray}
\mBox{white}{MadeForLife}  &
 &0.796($\mypm$0.013)
 &2.660($\mypm$0.256)
 &\rCirc{4}
 &8.57e-03($\mypm$3.37e-03)
 &0.779($\mypm$0.008)
 \\

\rowcolor{white}
\mBox{white}{LYU1}  &
 &0.784($\mypm$0.018)
 &2.743($\mypm$0.291)
 &\rCirc{10}
 &7.13e-03($\mypm$1.50e-03)
 &0.761($\mypm$0.012)
 \\

\rowcolor{Gray}
\mBox{white}{next-gen-nn}  &
 &0.790($\mypm$0.018)
 &2.668($\mypm$0.278)
 &\rCirc{5}
 &2.30e-04($\mypm$2.10e-04)
 &0.768($\mypm$0.011)
 \\

\rowcolor{white}
\mBox{white}{zhuoyuanw210}  &
 &0.791($\mypm$0.017)
 &2.730($\mypm$0.310)
 &\rCirc{8}
 &3.65e-03($\mypm$2.11e-03)
 &0.769($\mypm$0.011)
 \\

\rowcolor{Gray}
\mBox{white}{DutchMasters}  &
 &0.792($\mypm$0.012)
 &2.589($\mypm$0.248)
 &\rCirc{1}
 &6.21e-03($\mypm$3.57e-03)
 &0.776($\mypm$0.008)
 \\
\bottomrule
\end{tabular}}

\end{table}

\begin{table}[H]
\caption{NIMH T2*w Inter-subject.}
\label{t:nimh_is_t2*w}
\centering
\resizebox{0.9\columnwidth}{!}{%
\begin{tabular}{r c c c c c c c c}
\toprule
&&\textbf{DSC}$\uparrow$ & \textbf{HD95}$\downarrow$~\textcolor{black}{(mm)} & \textbf{Ranking} & \textbf{\%NDV}$\downarrow$ & \textbf{DSC30}$\uparrow$\\

\cmidrule(lr){3-3}
\cmidrule(lr){4-4}\cmidrule(lr){5-5}
\cmidrule(lr){6-6}
\cmidrule(lr){7-7}
\textbf{Method}&& \textbf{Mean(}$\bm{\PM}$\textbf{Std. Dev.)} 
& \textbf{Mean(}$\bm{\PM}$\textbf{Std. Dev.)} 
& \scriptsize{\textbf{(DSC + HD95)}}
& \textbf{Mean(}$\bm{\PM}$\textbf{Std. Dev.)} 
& \textbf{Mean(}$\bm{\PM}$\textbf{Std. Dev.)} 
\\
\midrule 

\rowcolor{white}
\mBox{white}{honkamj}  &
 &0.651($\mypm$0.032)
 &3.873($\mypm$0.438)
 &\rCirc{7}
 &1.99e-03($\mypm$4.62e-04)
 &0.618($\mypm$0.037)
 \\

\rowcolor{Gray}
\mBox{white}{MadeForLife}  &
 &0.664($\mypm$0.032)
 &3.763($\mypm$0.411)
 &\rCirc{2}
 &2.33e-03($\mypm$1.70e-03)
 &0.632($\mypm$0.039)
 \\

\rowcolor{white}
\mBox{white}{LYU1}  &
 &0.653($\mypm$0.031)
 &3.909($\mypm$0.430)
 &\rCirc{6}
 &6.44e-03($\mypm$9.37e-04)
 &0.619($\mypm$0.035)
 \\

\rowcolor{Gray}
\mBox{white}{next-gen-nn}  &
 &0.649($\mypm$0.034)
 &4.034($\mypm$0.468)
 &\rCirc{9}
 &1.55e-04($\mypm$6.18e-04)
 &0.614($\mypm$0.039)
 \\

\rowcolor{white}
\mBox{white}{zhuoyuanw210}  &
 &0.642($\mypm$0.034)
 &4.044($\mypm$0.507)
 &\rCirc{14}
 &1.51e-03($\mypm$8.24e-04)
 &0.606($\mypm$0.038)
 \\

\rowcolor{Gray}
\mBox{white}{DutchMasters}  &
 &0.640($\mypm$0.035)
 &3.995($\mypm$0.489)
 &\rCirc{12}
 &8.49e-04($\mypm$3.32e-03)
 &0.605($\mypm$0.044)
 \\
\bottomrule
\end{tabular}}

\end{table}

\begin{table}[H]
\caption{NIMH T2w Inter-subject.}
\label{t:nimh_is_t2w}
\centering
\resizebox{0.9\columnwidth}{!}{%
\begin{tabular}{r c c c c c c c c}
\toprule
&&\textbf{DSC}$\uparrow$ & \textbf{HD95}$\downarrow$~\textcolor{black}{(mm)} & \textbf{Ranking} & \textbf{\%NDV}$\downarrow$ & \textbf{DSC30}$\uparrow$\\

\cmidrule(lr){3-3}
\cmidrule(lr){4-4}\cmidrule(lr){5-5}
\cmidrule(lr){6-6}
\cmidrule(lr){7-7}
\textbf{Method}&& \textbf{Mean(}$\bm{\PM}$\textbf{Std. Dev.)} 
& \textbf{Mean(}$\bm{\PM}$\textbf{Std. Dev.)} 
& \scriptsize{\textbf{(DSC + HD95)}}
& \textbf{Mean(}$\bm{\PM}$\textbf{Std. Dev.)} 
& \textbf{Mean(}$\bm{\PM}$\textbf{Std. Dev.)} 
\\
\midrule 

\rowcolor{white}
\mBox{white}{honkamj}  &
 &0.729($\mypm$0.017)
 &3.326($\mypm$0.329)
 &\rCirc{3}
 &2.82e-03($\mypm$5.08e-04)
 &0.709($\mypm$0.013)
 \\

\rowcolor{Gray}
\mBox{white}{MadeForLife}  &
 &0.735($\mypm$0.017)
 &3.282($\mypm$0.346)
 &\rCirc{1}
 &6.63e-03($\mypm$1.17e-03)
 &0.715($\mypm$0.013)
 \\

\rowcolor{white}
\mBox{white}{LYU1}  &
 &0.719($\mypm$0.021)
 &3.523($\mypm$0.395)
 &\rCirc{11}
 &1.54e-02($\mypm$2.03e-03)
 &0.695($\mypm$0.018)
 \\

\rowcolor{Gray}
\mBox{white}{next-gen-nn}  &
 &0.730($\mypm$0.018)
 &3.414($\mypm$0.378)
 &\rCirc{6}
 &1.48e-05($\mypm$2.60e-05)
 &0.708($\mypm$0.014)
 \\

\rowcolor{white}
\mBox{white}{zhuoyuanw210}  &
 &0.715($\mypm$0.019)
 &3.496($\mypm$0.381)
 &\rCirc{12}
 &3.88e-03($\mypm$5.82e-04)
 &0.693($\mypm$0.015)
 \\

\rowcolor{Gray}
\mBox{white}{DutchMasters}  &
 &0.732($\mypm$0.017)
 &3.331($\mypm$0.338)
 &\rCirc{2}
 &1.60e-03($\mypm$6.93e-04)
 &0.711($\mypm$0.013)
 \\
\bottomrule
\end{tabular}}

\end{table}

\begin{table}[H]
\caption{NIMH FLAIR Inter-subject.}
\label{t:nimh_is_flair}
\centering
\resizebox{0.9\columnwidth}{!}{%
\begin{tabular}{r c c c c c c c c}
\toprule
&&\textbf{DSC}$\uparrow$ & \textbf{HD95}$\downarrow$~\textcolor{black}{(mm)} & \textbf{Ranking} & \textbf{\%NDV}$\downarrow$ & \textbf{DSC30}$\uparrow$\\

\cmidrule(lr){3-3}
\cmidrule(lr){4-4}\cmidrule(lr){5-5}
\cmidrule(lr){6-6}
\cmidrule(lr){7-7}
\textbf{Method}&& \textbf{Mean(}$\bm{\PM}$\textbf{Std. Dev.)} 
& \textbf{Mean(}$\bm{\PM}$\textbf{Std. Dev.)} 
& \scriptsize{\textbf{(DSC + HD95)}}
& \textbf{Mean(}$\bm{\PM}$\textbf{Std. Dev.)} 
& \textbf{Mean(}$\bm{\PM}$\textbf{Std. Dev.)} 
\\
\midrule 

\rowcolor{white}
\mBox{white}{honkamj}  &
 &0.662($\mypm$0.050)
 &3.722($\mypm$0.984)
 &\rCirc{3}
 &2.22e-03($\mypm$7.96e-04)
 &0.620($\mypm$0.068)
 \\

\rowcolor{Gray}
\mBox{white}{MadeForLife}  &
 &0.673($\mypm$0.033)
 &3.614($\mypm$0.392)
 &\rCirc{1}
 &4.83e-03($\mypm$4.57e-03)
 &0.637($\mypm$0.029)
 \\

\rowcolor{white}
\mBox{white}{LYU1}  &
 &0.658($\mypm$0.034)
 &3.822($\mypm$0.409)
 &\rCirc{11}
 &8.87e-03($\mypm$2.31e-03)
 &0.621($\mypm$0.035)
 \\

\rowcolor{Gray}
\mBox{white}{next-gen-nn}  &
 &0.659($\mypm$0.047)
 &3.830($\mypm$0.751)
 &\rCirc{9}
 &7.34e-05($\mypm$1.13e-04)
 &0.617($\mypm$0.061)
 \\

\rowcolor{white}
\mBox{white}{zhuoyuanw210}  &
 &0.656($\mypm$0.051)
 &3.858($\mypm$1.043)
 &\rCirc{10}
 &2.78e-03($\mypm$1.33e-03)
 &0.613($\mypm$0.071)
 \\

\rowcolor{Gray}
\mBox{white}{DutchMasters}  &
 &0.655($\mypm$0.057)
 &3.897($\mypm$2.151)
 &\rCirc{7}
 &2.02e-03($\mypm$2.01e-03)
 &0.614($\mypm$0.087)
 \\
\bottomrule
\end{tabular}}

\end{table}

\subsubsection{UltraCortex - 9.4T}
\begin{table}[H]
\caption{UltraCortex - 9.4T Inter-subject.}
\label{t:ultracortex_is}
\centering
\resizebox{0.9\columnwidth}{!}{%
\begin{tabular}{r c c c c c c c c}
\toprule
&&\textbf{DSC}$\uparrow$ & \textbf{HD95}$\downarrow$~\textcolor{black}{(mm)} & \textbf{Ranking} & \textbf{\%NDV}$\downarrow$ & \textbf{DSC30}$\uparrow$\\

\cmidrule(lr){3-3}
\cmidrule(lr){4-4}\cmidrule(lr){5-5}
\cmidrule(lr){6-6}
\cmidrule(lr){7-7}
\textbf{Method}&& \textbf{Mean(}$\bm{\PM}$\textbf{Std. Dev.)} 
& \textbf{Mean(}$\bm{\PM}$\textbf{Std. Dev.)} 
& \scriptsize{\textbf{(DSC + HD95)}}
& \textbf{Mean(}$\bm{\PM}$\textbf{Std. Dev.)} 
& \textbf{Mean(}$\bm{\PM}$\textbf{Std. Dev.)} 
\\
\midrule 

\rowcolor{white}
\mBox{white}{honkamj}  &
 &0.745($\mypm$0.040)
 &3.288($\mypm$0.649)
 &\rCirc{1}
 &2.87e-03($\mypm$5.25e-04)
 &0.688($\mypm$0.024)
 \\

\rowcolor{Gray}
\mBox{white}{MadeForLife}  &
 &0.746($\mypm$0.040)
 &3.479($\mypm$0.659)
 &\rCirc{2}
 &1.70e-02($\mypm$5.09e-03)
 &0.691($\mypm$0.023)
 \\

\rowcolor{white}
\mBox{white}{LYU1}  &
 &0.739($\mypm$0.041)
 &3.510($\mypm$0.658)
 &\rCirc{5}
 &2.23e-02($\mypm$3.37e-03)
 &0.682($\mypm$0.024)
 \\

\rowcolor{Gray}
\mBox{white}{next-gen-nn}  &
 &0.741($\mypm$0.044)
 &3.516($\mypm$0.692)
 &\rCirc{4}
 &4.17e-04($\mypm$3.30e-04)
 &0.680($\mypm$0.026)
 \\

\rowcolor{white}
\mBox{white}{zhuoyuanw210}  &
 &0.730($\mypm$0.041)
 &3.556($\mypm$0.655)
 &\rCirc{8}
 &9.40e-03($\mypm$2.01e-03)
 &0.674($\mypm$0.026)
 \\

\rowcolor{Gray}
\mBox{white}{DutchMasters}  &
 &0.733($\mypm$0.044)
 &3.532($\mypm$0.660)
 &\rCirc{6}
 &6.64e-03($\mypm$2.44e-03)
 &0.672($\mypm$0.025)
 \\
\bottomrule
\end{tabular}}

\end{table}

\begin{table}[H]
\caption{UltraCortex - 9.4T Atlas to Subject.}
\label{t:ultracortex_a2s}
\centering
\resizebox{0.9\columnwidth}{!}{%
\begin{tabular}{r c c c c c c c c}
\toprule
&&\textbf{DSC}$\uparrow$ & \textbf{HD95}$\downarrow$~\textcolor{black}{(mm)} & \textbf{Ranking} & \textbf{\%NDV}$\downarrow$ & \textbf{DSC30}$\uparrow$\\

\cmidrule(lr){3-3}
\cmidrule(lr){4-4}\cmidrule(lr){5-5}
\cmidrule(lr){6-6}
\cmidrule(lr){7-7}
\textbf{Method}&& \textbf{Mean(}$\bm{\PM}$\textbf{Std. Dev.)} 
& \textbf{Mean(}$\bm{\PM}$\textbf{Std. Dev.)} 
& \scriptsize{\textbf{(DSC + HD95)}}
& \textbf{Mean(}$\bm{\PM}$\textbf{Std. Dev.)} 
& \textbf{Mean(}$\bm{\PM}$\textbf{Std. Dev.)} 
\\
\midrule 

\rowcolor{white}
\mBox{white}{honkamj}  &
 &0.767($\mypm$0.029)
 &2.767($\mypm$0.590)
 &\rCirc{1}
 &7.54e-04($\mypm$2.12e-04)
 &0.734($\mypm$0.034)
 \\

\rowcolor{Gray}
\mBox{white}{MadeForLife}  &
 &0.765($\mypm$0.029)
 &2.904($\mypm$0.606)
 &\rCirc{2}
 &3.67e-03($\mypm$1.06e-03)
 &0.732($\mypm$0.034)
 \\

\rowcolor{white}
\mBox{white}{LYU1}  &
 &0.755($\mypm$0.028)
 &2.940($\mypm$0.586)
 &\rCirc{6}
 &1.82e-02($\mypm$2.60e-03)
 &0.722($\mypm$0.031)
 \\

\rowcolor{Gray}
\mBox{white}{next-gen-nn}  &
 &0.762($\mypm$0.033)
 &2.968($\mypm$0.625)
 &\rCirc{5}
 &1.52e-04($\mypm$1.71e-04)
 &0.723($\mypm$0.039)
 \\

\rowcolor{white}
\mBox{white}{zhuoyuanw210}  &
 &0.755($\mypm$0.027)
 &2.860($\mypm$0.547)
 &\rCirc{4}
 &5.56e-03($\mypm$1.12e-03)
 &0.722($\mypm$0.028)
 \\

\rowcolor{Gray}
\mBox{white}{DutchMasters}  &
 &0.736($\mypm$0.027)
 &3.015($\mypm$0.527)
 &\rCirc{13}
 &7.30e-03($\mypm$5.66e-03)
 &0.703($\mypm$0.017)
 \\
\bottomrule
\end{tabular}}

\end{table}

\begin{table}[H]
\caption{UltraCortex - 9.4T Subject to Atlas.}
\label{t:ultracortex_s2a}
\centering
\resizebox{0.9\columnwidth}{!}{%
\begin{tabular}{r c c c c c c c c}
\toprule
&&\textbf{DSC}$\uparrow$ & \textbf{HD95}$\downarrow$~\textcolor{black}{(mm)} & \textbf{Ranking} & \textbf{\%NDV}$\downarrow$ & \textbf{DSC30}$\uparrow$\\

\cmidrule(lr){3-3}
\cmidrule(lr){4-4}\cmidrule(lr){5-5}
\cmidrule(lr){6-6}
\cmidrule(lr){7-7}
\textbf{Method}&& \textbf{Mean(}$\bm{\PM}$\textbf{Std. Dev.)} 
& \textbf{Mean(}$\bm{\PM}$\textbf{Std. Dev.)} 
& \scriptsize{\textbf{(DSC + HD95)}}
& \textbf{Mean(}$\bm{\PM}$\textbf{Std. Dev.)} 
& \textbf{Mean(}$\bm{\PM}$\textbf{Std. Dev.)} 
\\
\midrule 

\rowcolor{white}
\mBox{white}{honkamj}  &
 &0.778($\mypm$0.031)
 &2.794($\mypm$0.616)
 &\rCirc{6}
 &2.35e-03($\mypm$4.15e-04)
 &0.743($\mypm$0.038)
 \\

\rowcolor{Gray}
\mBox{white}{MadeForLife}  &
 &0.787($\mypm$0.030)
 &2.832($\mypm$0.627)
 &\rCirc{1}
 &7.03e-03($\mypm$1.82e-03)
 &0.753($\mypm$0.036)
 \\

\rowcolor{white}
\mBox{white}{LYU1}  &
 &0.781($\mypm$0.031)
 &2.846($\mypm$0.653)
 &\rCirc{5}
 &8.19e-03($\mypm$1.08e-03)
 &0.746($\mypm$0.037)
 \\

\rowcolor{Gray}
\mBox{white}{next-gen-nn}  &
 &0.784($\mypm$0.035)
 &2.844($\mypm$0.691)
 &\rCirc{2}
 &2.65e-04($\mypm$1.91e-04)
 &0.745($\mypm$0.043)
 \\

\rowcolor{white}
\mBox{white}{zhuoyuanw210}  &
 &0.781($\mypm$0.031)
 &2.855($\mypm$0.668)
 &\rCirc{4}
 &2.35e-03($\mypm$7.61e-04)
 &0.745($\mypm$0.037)
 \\

\rowcolor{Gray}
\mBox{white}{DutchMasters}  &
 &0.760($\mypm$0.028)
 &2.919($\mypm$0.688)
 &\rCirc{9}
 &1.14e-02($\mypm$6.84e-03)
 &0.728($\mypm$0.031)
 \\
\bottomrule
\end{tabular}}

\end{table}

\subsubsection{Macaque (Non-human Brain)}
\begin{table}[H]
\caption{Macaque (Non-human Brain) Inter-subject.}
\label{t:macaque}
\centering
\resizebox{0.9\columnwidth}{!}{%
\begin{tabular}{r c c c c c c c c}
\toprule
&&\textbf{DSC}$\uparrow$ & \textbf{HD95}$\downarrow$~\textcolor{black}{(mm)} & \textbf{Ranking} & \textbf{\%NDV}$\downarrow$ & \textbf{DSC30}$\uparrow$\\

\cmidrule(lr){3-3}
\cmidrule(lr){4-4}\cmidrule(lr){5-5}
\cmidrule(lr){6-6}
\cmidrule(lr){7-7}
\textbf{Method}&& \textbf{Mean(}$\bm{\PM}$\textbf{Std. Dev.)} 
& \textbf{Mean(}$\bm{\PM}$\textbf{Std. Dev.)} 
& \scriptsize{\textbf{(DSC + HD95)}}
& \textbf{Mean(}$\bm{\PM}$\textbf{Std. Dev.)} 
& \textbf{Mean(}$\bm{\PM}$\textbf{Std. Dev.)} 
\\
\midrule 

\rowcolor{white}
\mBox{white}{honkamj}  &
 &0.773($\mypm$0.074)
 &4.349($\mypm$3.223)
 &\rCirc{3}
 &3.78e-04($\mypm$2.48e-04)
 &0.674($\mypm$0.032)
 \\

\rowcolor{Gray}
\mBox{white}{MadeForLife}  &
 &0.777($\mypm$0.075)
 &4.318($\mypm$3.233)
 &\rCirc{1}
 &1.22e-03($\mypm$1.04e-03)
 &0.677($\mypm$0.031)
 \\

\rowcolor{white}
\mBox{white}{LYU1}  &
 &0.767($\mypm$0.074)
 &4.361($\mypm$3.128)
 &\rCirc{6}
 &2.72e-02($\mypm$9.19e-03)
 &0.670($\mypm$0.034)
 \\

\rowcolor{Gray}
\mBox{white}{next-gen-nn}  &
 &0.769($\mypm$0.076)
 &4.450($\mypm$3.269)
 &\rCirc{9}
 &2.18e-06($\mypm$1.36e-05)
 &0.669($\mypm$0.032)
 \\

\rowcolor{white}
\mBox{white}{zhuoyuanw210}  &
 &0.760($\mypm$0.073)
 &4.497($\mypm$3.188)
 &\rCirc{15}
 &3.70e-04($\mypm$1.93e-04)
 &0.662($\mypm$0.028)
 \\

\rowcolor{Gray}
\mBox{white}{DutchMasters}  &
 &0.765($\mypm$0.072)
 &4.430($\mypm$3.212)
 &\rCirc{11}
 &1.34e-04($\mypm$2.77e-04)
 &0.670($\mypm$0.029)
 \\
\bottomrule
\end{tabular}}

\end{table}

\section{\textcolor{black}{Statistical significance matrices for all evaluated tasks}}
\textcolor{black}{We present the significance matrices for all evaluated tasks below. Overall differences among methods were first assessed using the Friedman test. When statistical significance was detected, pairwise comparisons were performed using Wilcoxon signed-rank post-hoc tests with Bonferroni correction, which is the most conservative $p$-value correction method.}

\clearpage
\begin{figure*}[!tbh]
    \centering
    \includegraphics[width = 0.9\linewidth]{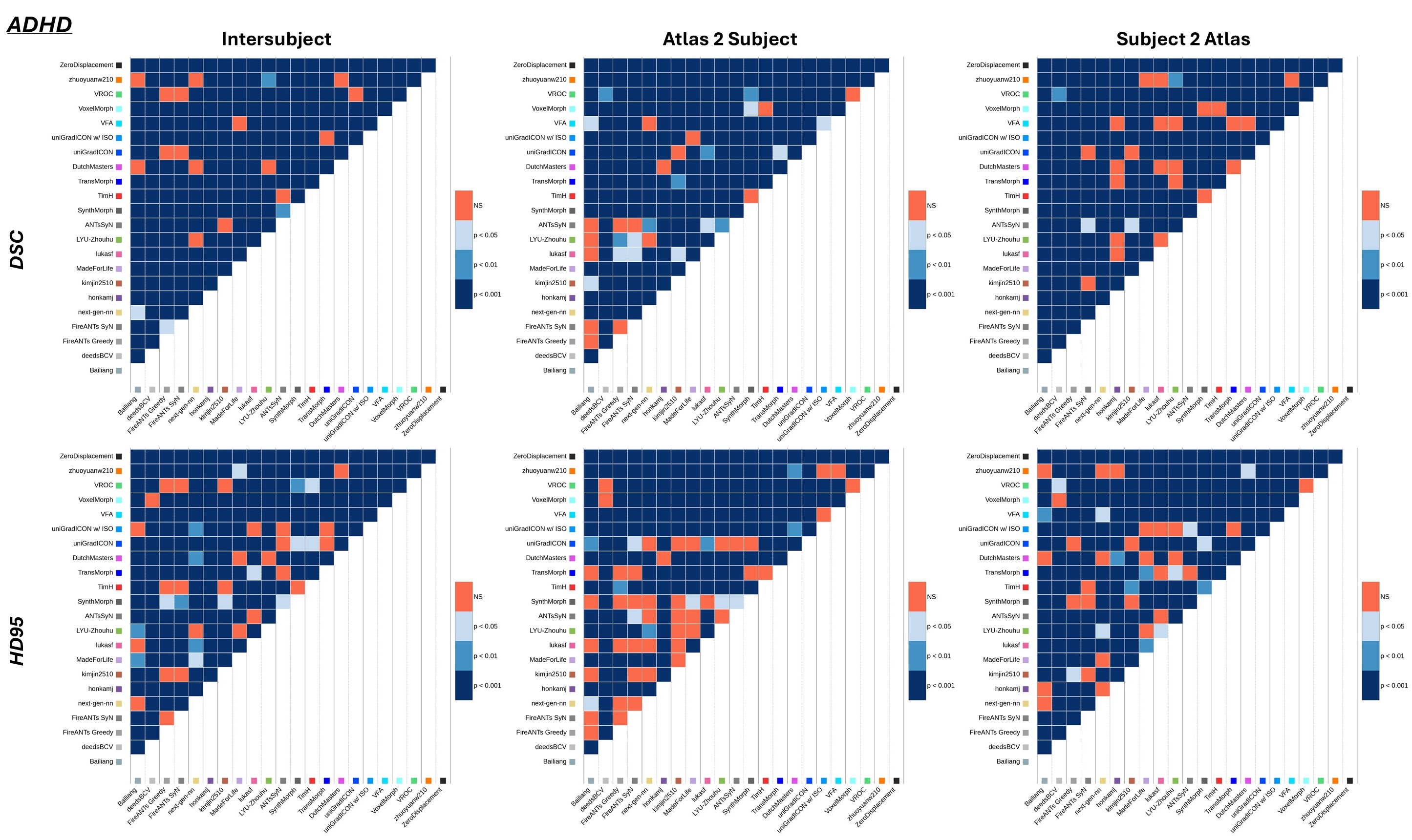}
    \caption{\textcolor{black}{Significance matrix for pairwise method comparisons on the ADHD dataset. The upper row shows results for DSC, and the lower row shows results for HD95. Columns correspond to three registration tasks: Intersubject, Atlas to Subject~(Atlas2Subject), and subject to atlas~(Subject2Atlas) registration. Each cell indicates the statistical significance of performance differences between a pair of methods, assessed using Wilcoxon signed-rank post-hoc tests with Bonferroni correction. Color encodes the significance level, ranging from non-significant~(NS) to $p < 0.001$. Only the upper triangular matrix is shown to avoid redundancy.}}
    \label{fig:adhd_sig}
\end{figure*}
\begin{figure*}[!tbh]
    \centering
    \includegraphics[width = 0.9\linewidth]{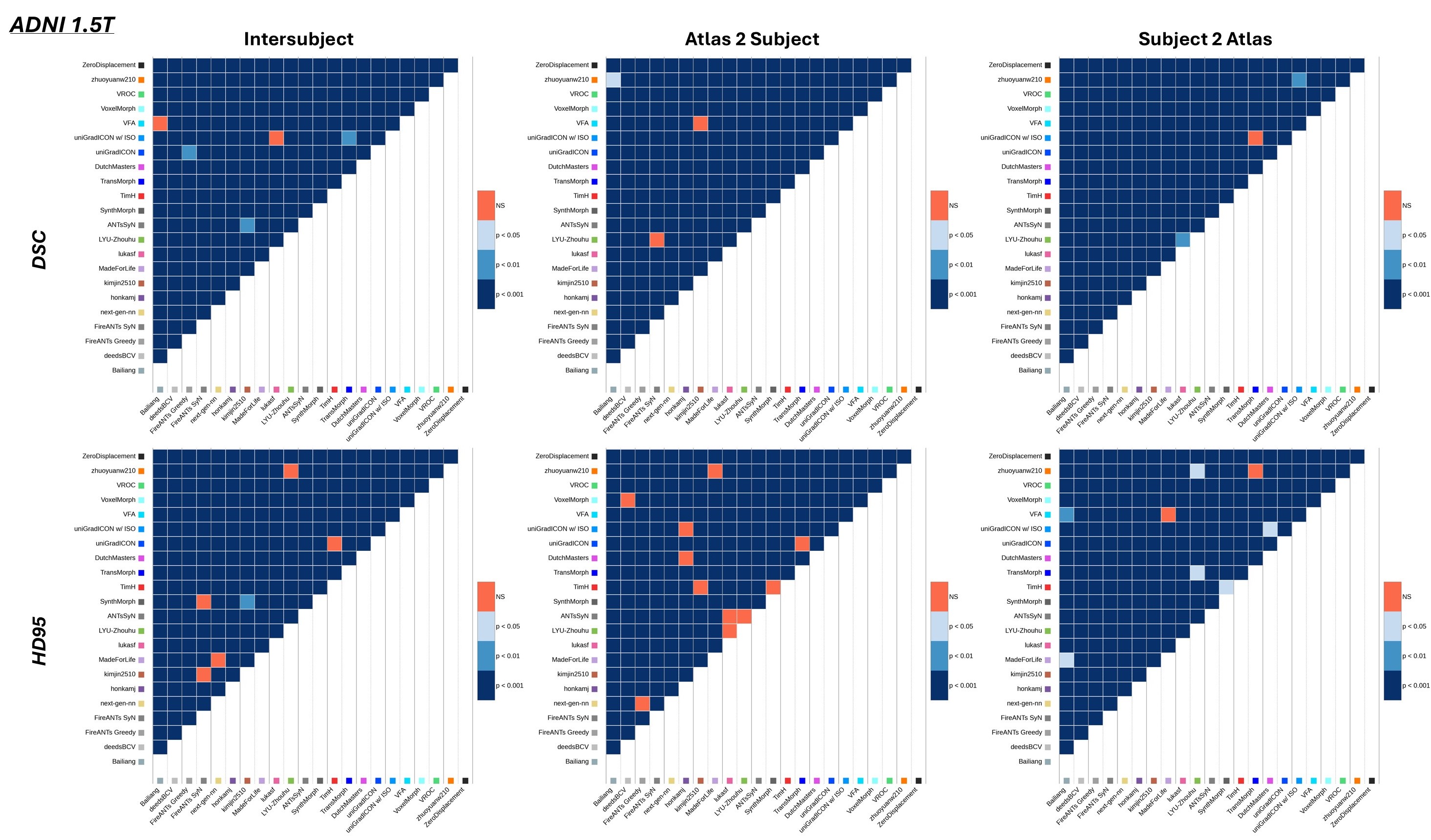}
    \caption{\textcolor{black}{Significance matrix for pairwise method comparisons on the ADNI~1.5T dataset. The upper row shows results for DSC, and the lower row shows results for HD95. Columns correspond to three registration tasks: Intersubject, Atlas to Subject~(Atlas2Subject), and subject to atlas~(Subject2Atlas) registration. Each cell indicates the statistical significance of performance differences between a pair of methods, assessed using Wilcoxon signed-rank post-hoc tests with Bonferroni correction. Color encodes the significance level, ranging from non-significant~(NS) to $p < 0.001$. Only the upper triangular matrix is shown to avoid redundancy.}}
    \label{fig:adni1_5T_sig}
\end{figure*}

\clearpage
\begin{figure*}[!tbh]
    \centering
    \includegraphics[width = 0.9\linewidth]{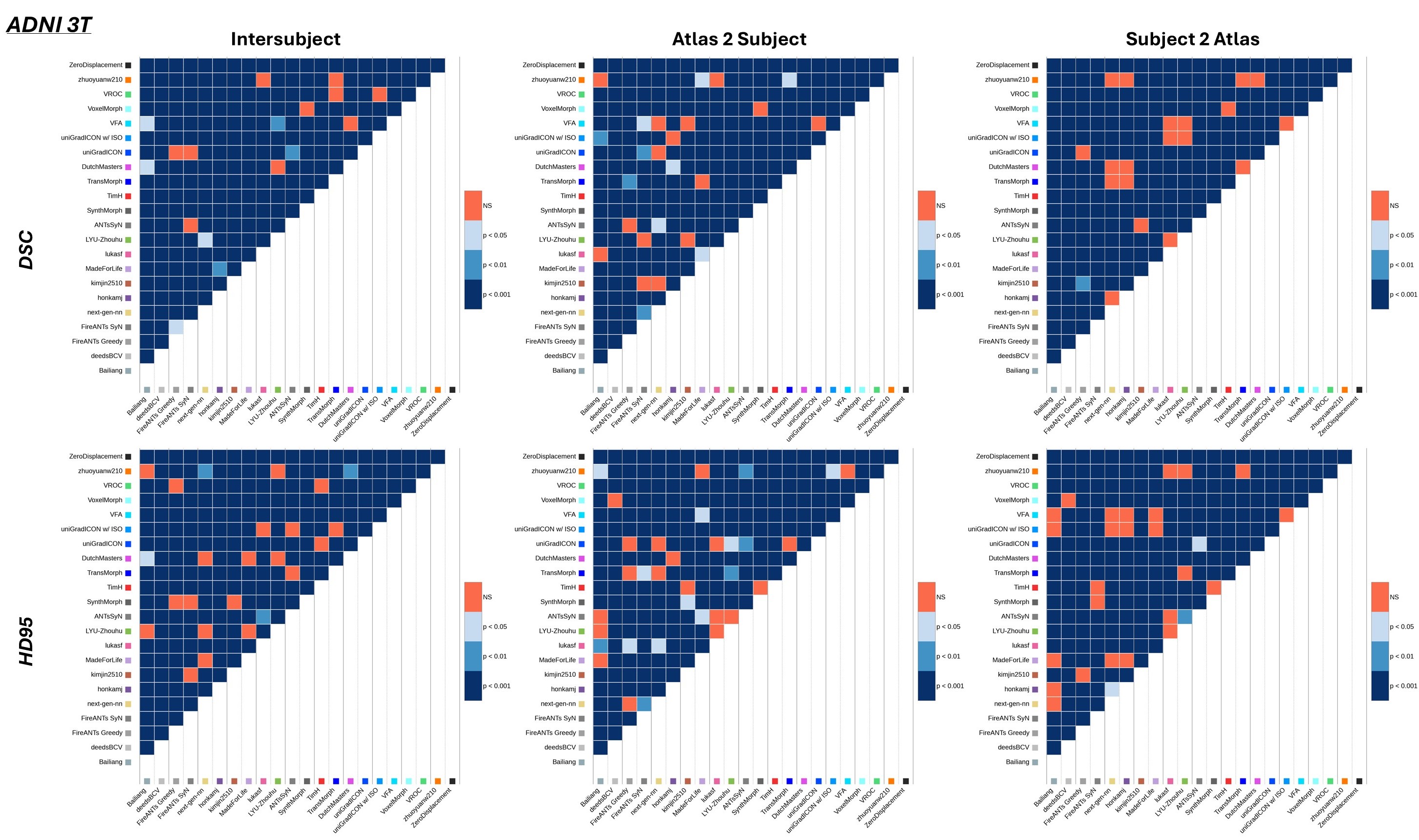}
    \caption{\textcolor{black}{Significance matrix for pairwise method comparisons on the ADNI~3T dataset. The upper row shows results for DSC, and the lower row shows results for HD95. Columns correspond to three registration tasks: Intersubject, Atlas to Subject~(Atlas2Subject), and subject to atlas~(Subject2Atlas) registration. Each cell indicates the statistical significance of performance differences between a pair of methods, assessed using Wilcoxon signed-rank post-hoc tests with Bonferroni correction. Color encodes the significance level, ranging from non-significant~(NS) to $p < 0.001$. Only the upper triangular matrix is shown to avoid redundancy.}}
    \label{fig:adni3T_sig}
\end{figure*}
\begin{figure*}[!tbh]
    \centering
    \includegraphics[width = 0.9\linewidth]{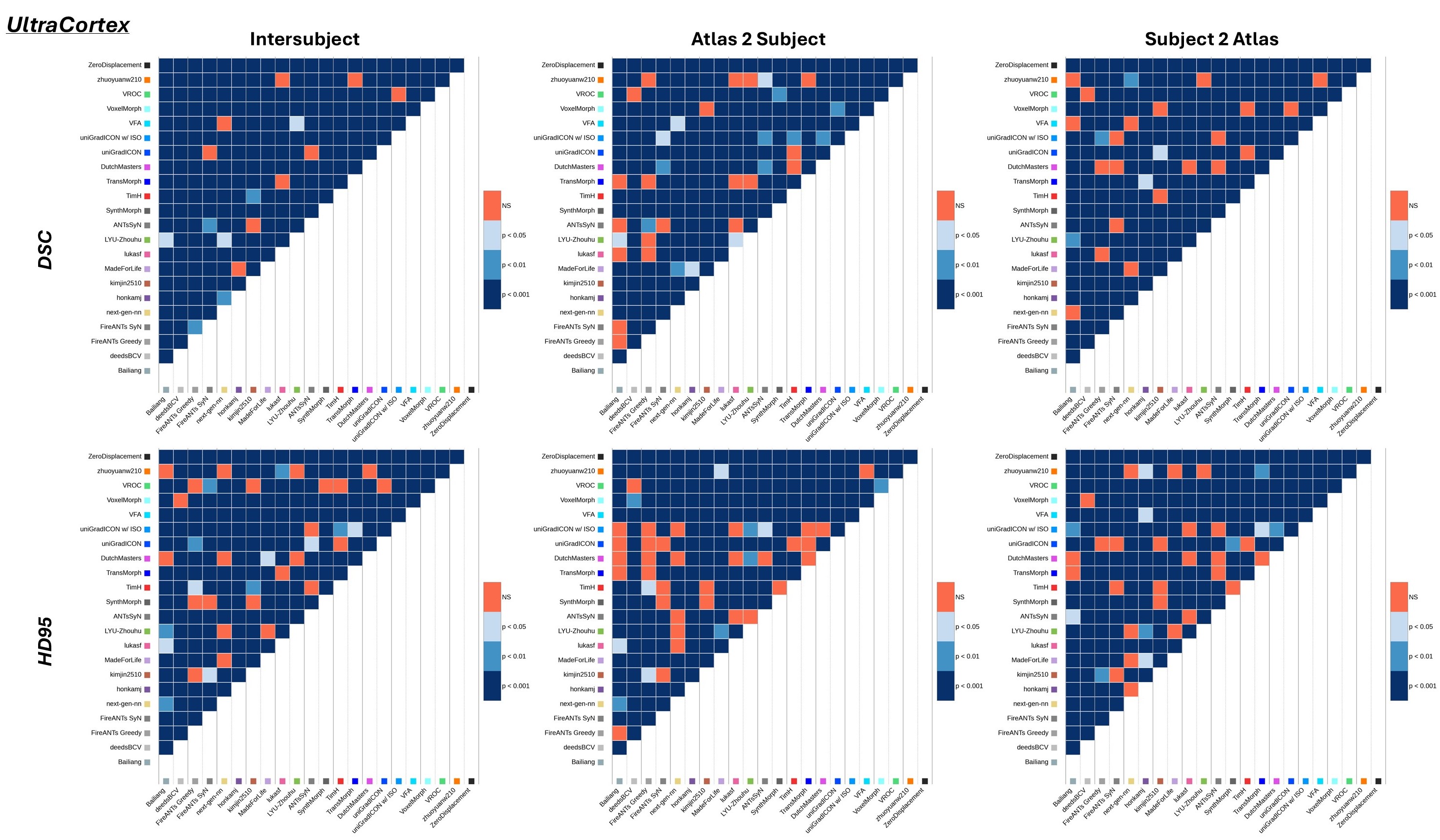}
    \caption{\textcolor{black}{Significance matrix for pairwise method comparisons on the UltraCortex dataset. The upper row shows results for DSC, and the lower row shows results for HD95. Columns correspond to three registration tasks: Intersubject, Atlas to Subject~(Atlas2Subject), and subject to atlas~(Subject2Atlas) registration. Each cell indicates the statistical significance of performance differences between a pair of methods, assessed using Wilcoxon signed-rank post-hoc tests with Bonferroni correction. Color encodes the significance level, ranging from non-significant~(NS) to $p < 0.001$. Only the upper triangular matrix is shown to avoid redundancy.}}
    \label{fig:ultracortex_sig}
\end{figure*}

\clearpage
\begin{figure*}[!tbh]
    \centering
    \includegraphics[width = 0.9\linewidth]{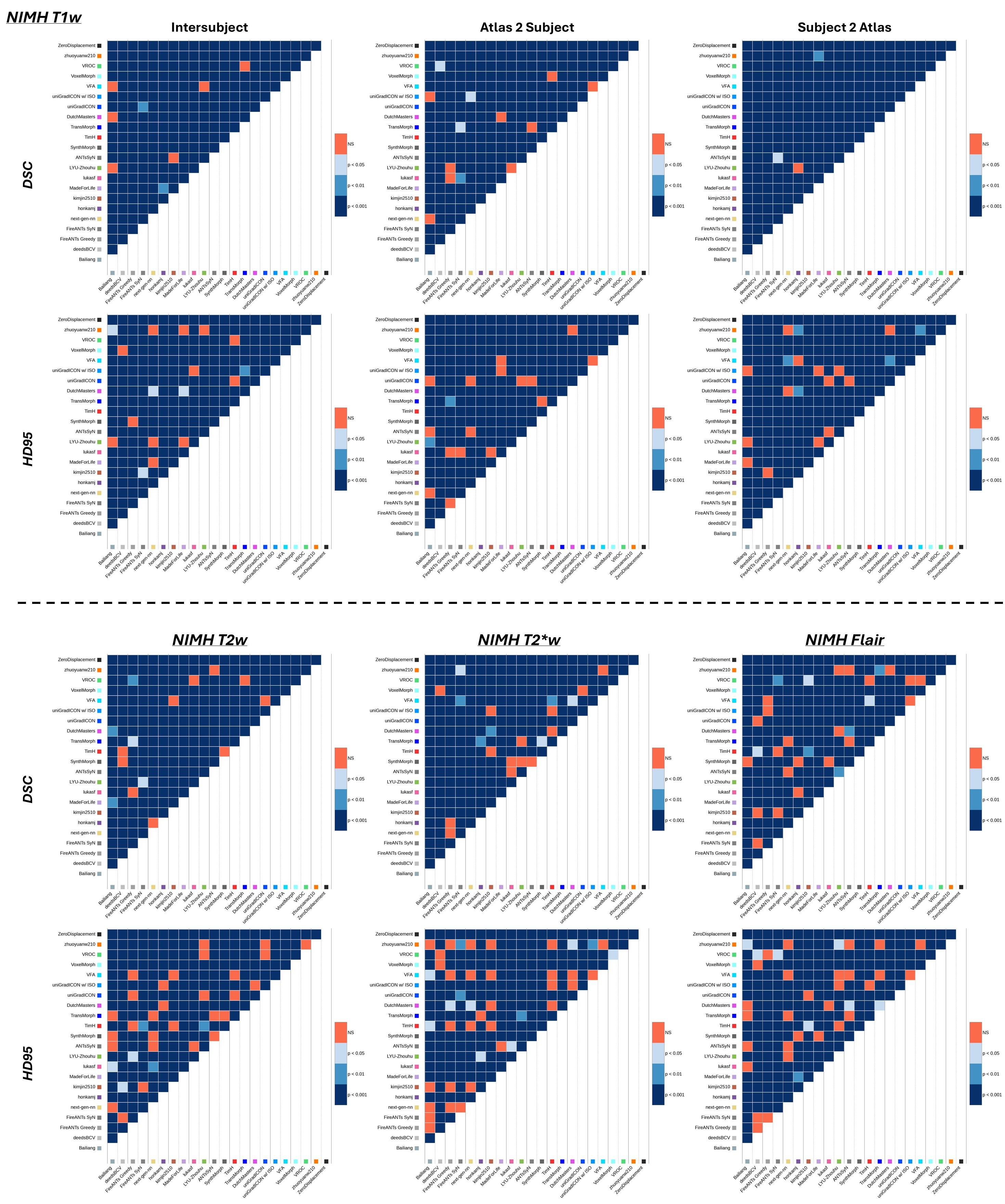}
    \caption{\textcolor{black}{Significance matrices for pairwise method comparisons on the NIMH dataset. For T1-weighted images, results are shown for three registration tasks, Intersubject, Atlas to Subject~(Atlas2Subject), in the upper panel separated by the dashed line. For the T2-weighted, T2*-weighted, and FLAIR images, results are shown for Intersubject registration only in the lower panel. For each panel, the upper row shows results for DSC, and the lower row shows results for HD95. Each cell represents the statistical significance of performance differences between a pair of methods, assessed using Wilcoxon signed-rank post-hoc tests with Bonferroni correction. Color encodes the significance level, ranging from non-significant~(NS) to $p < 0.001$. Only the upper triangular matrices are shown.}}
    \label{fig:nimh_sig}
\end{figure*}
\begin{figure*}[!tbh]
    \centering
    \includegraphics[width = 0.9\linewidth]{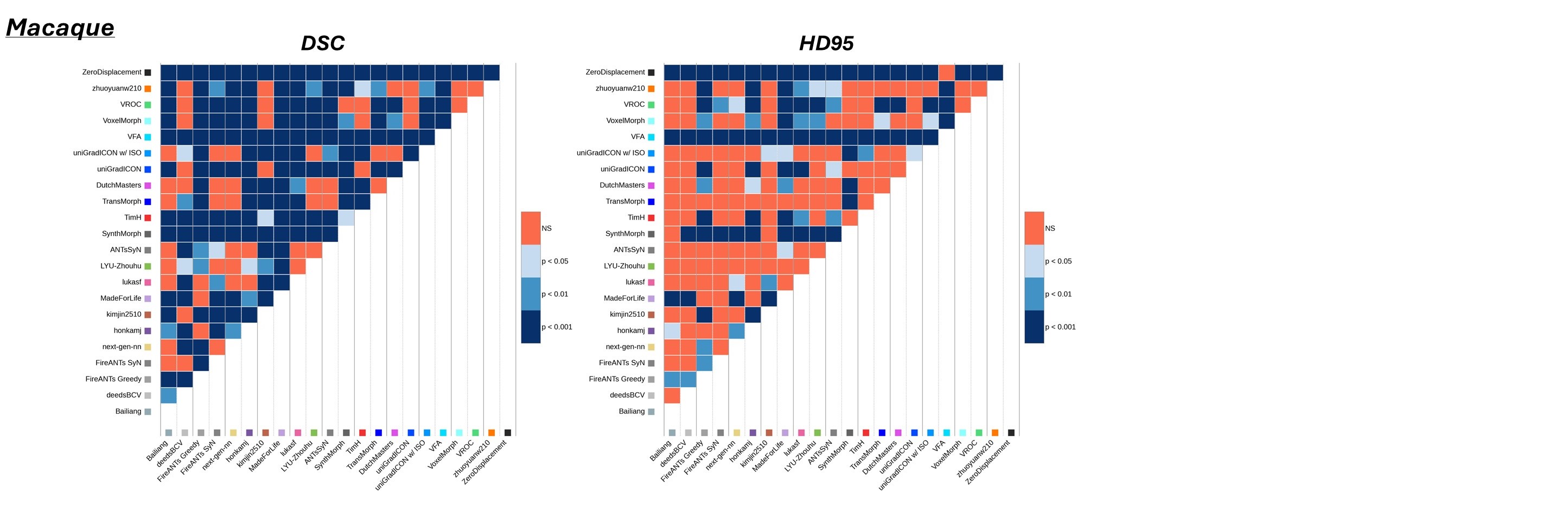}
    \caption{\textcolor{black}{Significance matrix for pairwise method comparisons on the macaque dataset. Results are shown for Intersubject registration only. The left plot shows results for DSC, and the right plot shows results for HD95. Each cell indicates the statistical significance of performance differences between a pair of methods, assessed using Wilcoxon signed-rank post-hoc tests with Bonferroni correction. Color encodes the significance level, ranging from non-significant~(NS) to $p < 0.001$. Only the upper triangular matrix is shown to avoid redundancy.}}
    \label{fig:macaque_sig}
\end{figure*}

\clearpage
\section{Challenge Uploads}
\label{a:uploads}
\setcounter{table}{0} 
In Table~\ref{tab:submission_summary}, we provide a summary of the total submissions to the challenge website through August 31,~2025.

\begin{table}[!h]
\caption{\textcolor{black}{Summary of submission attempts by team (as of August 31, 2025). Organizers denotes the LUMIR organizers. Others denotes teams that submitted results via the Grand Challenge platform but did not participate in the final challenge testing and competition.}}
\label{tab:submission_summary}
\begin{center}
    \begin{tabular}{lccc}
    \toprule
    & & & \textbf{Total}\\
    \textbf{Team} & \textbf{\# Failed} & \textbf{\# Successful} & \textbf{Attempts} \\
    \cmidrule(lr){1-4}
    \rowcolor{Gray}\twBox{honkamj} & 0 & 18  & 18 \\
    \twBox{MadeForLife} & 0 & 13 & 13 \\
    \rowcolor{Gray}\twBox{LYU1}  & 0 & 18 & 18 \\
    \twBox{next-gen-nn} & 0 & 21 & 21 \\
    \rowcolor{Gray}\twBox{zhuoyuanw210} & 0 & 14 & 14 \\
    \twBox{DutchMasters} & 0 & 18 & 18  \\
    \rowcolor{Gray}\twBox{lukasf} & 0 & 10 & 10\\
    \twBox{Bailiang} & 0 & 16 & 16 \\
    \rowcolor{Gray}\twBox{VROC}  & 0 & 50  & 50 \\
    \twBox{TimH} & 4 & 5 & 9 \\
    \rowcolor{Gray}\twBox{LoRA-FT} & 0 & 3 & 3\\
    \cmidrule(lr){2-4}
    Organizers & 0 & 3 & 3 \\
    \rowcolor{Gray}Others   & 10 & 116 & 126 \\
    \cmidrule(lr){1-4}
    \textbf{Total}  & \textbf{14} & \textbf{305} & \textbf{319}  \\
    \bottomrule
    \end{tabular}
\end{center}
\end{table}

\section*{Disclosures}
The authors have no relevant financial interests or conflicts of interest to disclose related to the content of this paper. 

AI (OpenAI ChatGPT-4o) was used for grammar and language refinement. All scientific content, ideas, and references were developed by the authors and carefully reviewed to ensure accuracy and integrity.

\section*{Acknowledgments}
We thank Lin Tian and Rohit Jena for their helpful discussions and guidance in the implementation of several of the baseline methods.
We also thank Jiafeng Zhu for valuable input on statistical analysis.

This work is supported by grants from the National Institutes of Health~(NIH), United States, R01CA297470~(PI: Y.~Du), R01EB031023~(PI: Y.~Du), U01EB031798~(PI: G.~Sgouros), P01CA272222~(PI: G.~Sgouros), R01CA253923~(PI: F.~Maldonado), R01CA275015~(PI: 
M.E.~Lenburg), R01HL169944~(PI: E.~Tkaczyk), U24AG074855~(PI: T.~Hohman; PI: M.~Cuccaro; PI: A.~Toga), R01MH121620~(PI: W.~Taylor), and R03CA286693~(PI: B.~Kimia; PI: H.~Bai). The views expressed in this publication and by the authors do not necessarily reflect the official policies of the NIH; nor does mention by trade names, commercial practices, or organizations imply endorsement by the U.S. Government.

\textcolor{black}{Data used in this paper were obtained in part from the Alzheimer's Disease Neuroimaging Initiative (ADNI) database (\url{adni.loni.usc.edu}).
Accordingly, the investigators within ADNI contributed to the design and implementation of ADNI and/or provided data, but did not participate in the analysis or writing of this report.
A complete list of ADNI investigators is available at this \href{http://adni.loni.usc.edu/wp-content/uploads/how_to_apply/ADNI_Acknowledgement_List.pdf}{link}.}

\bibliographystyle{model2-names.bst}
\biboptions{authoryear}
\bibliography{registration}

\end{document}